\documentclass[varenna]{cimento}

\usepackage{color}
\usepackage{latexsym}
\usepackage{amstext}
\usepackage{amsxtra}
\usepackage{epsfig}
\usepackage{graphicx}

\title{Basic theory tools for degenerate Fermi gases}
\author{Yvan Castin}
\institute{Laboratoire Kastler Brossel, \'Ecole normale sup\'erieure \\
24 rue Lhomond, 75231 Paris Cedex 5, France}
\def\be{\begin{equation}}
\def\ee{\end{equation}}
\def\bea{\begin{eqnarray}}
\def\eea{\end{eqnarray}}

\begin{document}
\maketitle
\section{The ideal Fermi gas}

We consider in this section a non-interacting Fermi gas in a single spin component,
at thermal equilibrium in the grand canonical ensemble. We review basic
properties of such a gas. We concentrate on the degenerate regime 
$\rho\lambda^3\gg 1$, with
$\rho$ the density and $\lambda$ the thermal de Broglie wavelength defined as
\be
\lambda^2 = \frac{2\pi\hbar^2}{m k_B T}
\label{eq:lambda}
\ee

\subsection{Basic facts}

In first quantization, the Hamiltonian of the non-interacting system is the sum of
one-body terms,
\be
H= \sum_{i=1}^{\hat{N}} \hat{h}_0(i)
\ee
where $\hat{N}$ is the operator giving the number of particles, and 
$\hat{h}_0$ is the one-body
Hamiltonian given here by
\be
\hat{h}_0=\frac{p^2}{2m} + U(\vec{r}\,)
\label{eq:obh}
\ee
where $\vec{p}$ is the momentum operator, $m$ is the mass of a particle
and $U(\vec{r}\,)$ is a position dependent external potential.
We shall consider two classes of external potential. Either a zero external potential, 
supplemented by periodic boundary conditions in a box of size $L$,
defined by $\vec{r}\in [0,L[^d$, or a harmonic
potential mimicking the traps used in experiments:
\be
U(\vec{r}\,) = \sum_{\alpha=1}^{d} \frac{1}{2} m\omega_\alpha^2 r_\alpha^2
\ee
where $\omega_\alpha$ is the oscillation frequency of a particle in the trap along
one of the $d$ dimensions of space.

The system is made of indistinguishable fermions, all in the same spin state.
To implement the antisymmetry of the many-body wavefunction, we move to second quantization.
Introducing the orthonormal basis of the eigenmodes $\phi_\lambda$ of the one-body
Hamiltonian, with eigenenergy $\epsilon_\lambda$, we expand the field operator as
\be
\hat{\psi}(\vec{r}\,) = \sum_\lambda \phi_\lambda(\vec{r}\,) a_\lambda
\ee
where $a_\lambda$ annihilates one particle in the mode $\phi_\lambda$ and
obeys anticommutation relations with the creation and annihilation operators:
\bea
\label{eq:anticom1}
\{ a_\lambda,a_{\lambda'}\} &=& 0 \\
\{ a_\lambda,a^\dagger_{\lambda'}\} &=& \delta_{\lambda\lambda'}.
\label{eq:anticom2}
\eea
The Hamiltonian then reads in second quantized form:
\be
H = \sum_\lambda \epsilon_\lambda a_\lambda^\dagger a_\lambda,
\ee
and the operator giving the number of particles:
\be
\hat{N} = \sum_\lambda  a_\lambda^\dagger a_\lambda.
\ee

The system is assumed to be at thermal equilibrium in the grand-canonical ensemble,
with a many-body density operator given by
\be
\hat{\sigma} =  \Xi^{-1} \, e^{-\beta(H-\mu \hat{N})}
\ee
where the factor $\Xi$ ensures that $\hat{\sigma}$ has unit trace, $\beta=1/k_B T$
and $\mu$ is the chemical potential. The choice of the grand-canonical ensemble
allows to decouple one mode from the other and makes the density operator Gaussian
in the field variables, so that Wick's theorem may be used to calculate expectation values,
which are thus (possibly complicated) functions of the only non-zero quadratic moments
\be
\langle a_\lambda^\dagger a_\lambda\rangle = 
n_\lambda = n(\epsilon_\lambda)\equiv 
\frac{1}{\exp[\beta(\epsilon_\lambda-\mu)]+1}.
\label{eq:fd}
\ee
This is the famous Fermi-Dirac law for the occupation numbers $n_\lambda$.
An elementary derivation of Wick's theorem can be found in an appendix
of \cite{Houches03}.

We note that, because of the fermionic nature of the system, the grand-canonical ensemble
cannot be subject to unphysically large fluctuations of the number of particles, contrarily
to the ideal Bose gas case.
Using Wick's theorem we find that the variance of the particle number is
\be
\mbox{Var}\, \hat{N} = \sum_{\lambda} n_\lambda (1-n_\lambda),
\ee
which is indeed always below the Poisson value $\langle\hat{N}\rangle$. 
In the zero temperature limit, for $\mu$ not coinciding
with an eigenmode energy $\epsilon_\lambda$, one finds a vanishing variance of the particle
number, so that the grand-canonical ensemble becomes equivalent to the canonical one.

\begin{figure}[htb]
\centerline{
\includegraphics[width=11cm,clip]{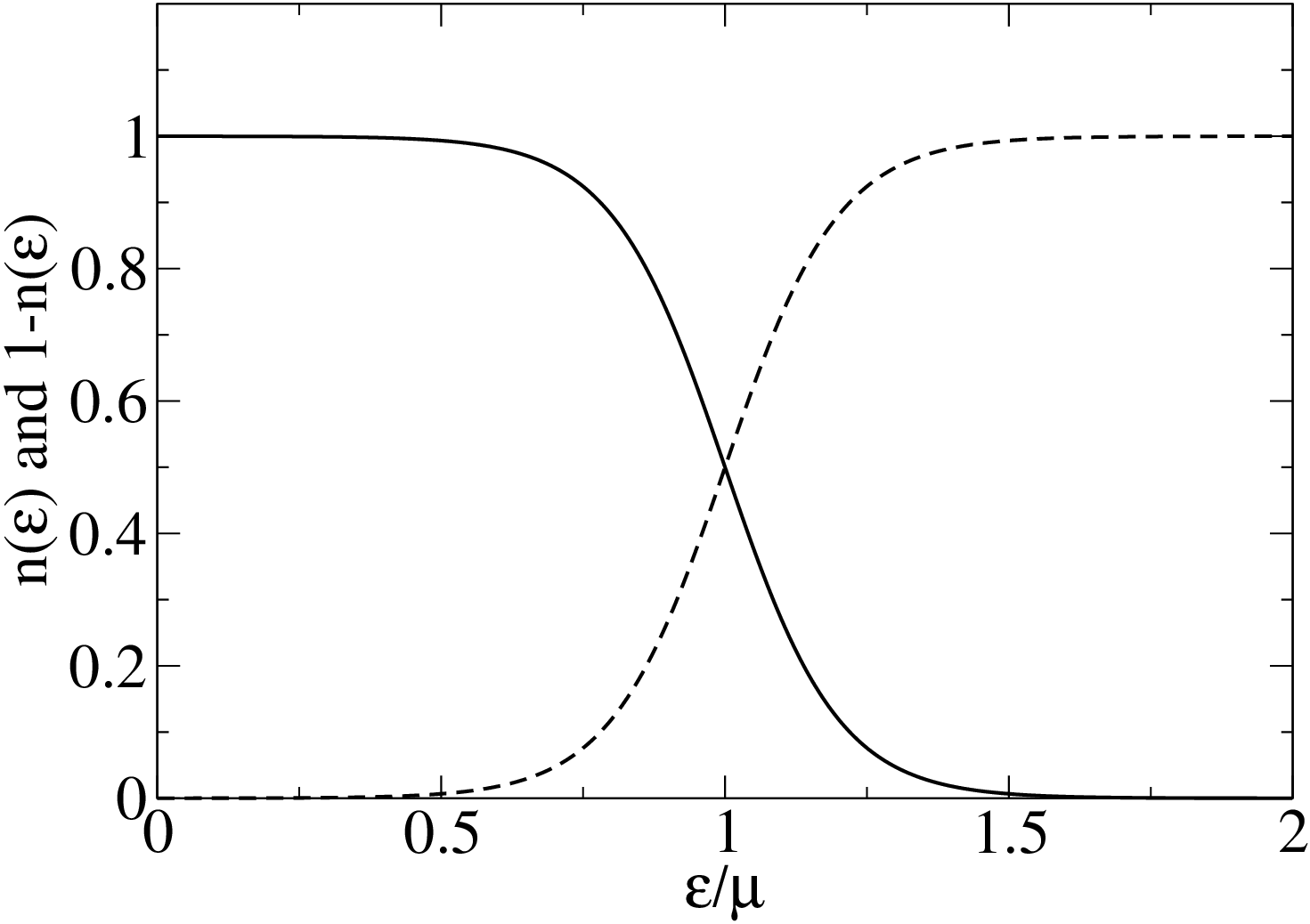}}
\caption{Fermi-Dirac distribution in the degenerate regime. Solid line: occupation
numbers $n(\epsilon)$ for the particles. Dashed line: occupation numbers
$1-n(\epsilon)$ for the holes. The temperature is $k_B T=\mu/10$.}
\label{fig:FD}
\end{figure}

In this lecture, we shall be concerned with the degenerate limit, where the temperature
is much smaller than the chemical potential:
\be
k_B T \ll \mu.
\ee
The Fermi-Dirac distribution has then the shape presented in Fig.~\ref{fig:FD}. We shall
repeatedly use the fact that the occupation number of holes $1-n_\lambda$ for $\epsilon_\lambda
< \mu$ and the occupation number of particles $n_\lambda$
for $\epsilon_\lambda > \mu$  are narrow functions of $\epsilon_\lambda-\mu$, of
energy width $\sim k_B T$.

More precisely, a standard low-$T$ expansion technique proceeds as follows \cite{Diu}.
One writes the Fermi-Dirac formula as
\be
n(\epsilon)=n_{T=0}(\epsilon)+ \frac{\mbox{sign}\,(\epsilon-\mu)}{\exp(\beta|\epsilon-\mu|)+1},
\label{eq:tech}
\ee
singling out as a second term a narrow function of $\epsilon-\mu$.
In the case of a continuous density of states $\rho(\epsilon)$, an integral over $\epsilon$
appears in thermodynamic quantities.
One then extends the integration over $\epsilon$ to $]-\infty,+\infty[$ for the second
term of the right-hand side, neglecting a contribution $O[\exp(-\mu/k_B T)]$,
after having expanded in powers of $\epsilon-\mu$ the
functions of $\epsilon$ multiplicating this second term, such as
$\rho(\epsilon)$.

We illustrate this technique for the calculation of the density $\rho$ in a spatially
homogeneous system in the thermodynamic limit:
\be
\rho_{\rm hom}(\mu,T) = \int \frac{d^d k}{(2\pi)^d} n_k = \int_0^{+\infty}  d\epsilon\, \rho(\epsilon)
n(\epsilon)
\label{eq:hom}
\ee
where we introduced the free space density of states
\be
\rho(\epsilon) = \int \frac{d^d k}{(2\pi)^d}  \delta(\epsilon-\hbar^2k^2/2m)=
\frac{d}{2\epsilon} (2\pi)^{-d} V_d(\sqrt{2m\epsilon}/\hbar)
\ee
with $V_d(u)$ is the volume enclosed by 
the sphere of radius $u$ in dimension $d$.
We go to the limit $T\to 0$ for a fixed chemical potential,
and we take $\mu>0$ in order to have a non-zero density at zero temperature.
From the rewriting Eq.(\ref{eq:tech}) of the Fermi-Dirac formula
we obtain
\be
\rho_{\rm hom}(\mu,T)= \rho_{\rm hom}(\mu,0) + \int_0^{\mu} d(\delta\epsilon)\,
\frac{\rho(\mu+\delta\epsilon)-\rho(\mu-\delta\epsilon)}
{\exp(\beta\delta\epsilon)+1}
 + O\left(e^{-\beta \mu}\right),
\ee
where we cut the integration over $\epsilon$ to $2\mu$, introducing an 
exponentially small error.
The zero temperature contribution is readily evaluated by the integral in $k$ space:
\be
\rho_{\rm hom}(\mu,0)= 
\int_{\hbar^2 k^2/2m < \mu} \frac{d^dk}{(2\pi)^d} = \frac{V_d(1)}{(2\pi)^d}
\left(\frac{2m\mu}{\hbar^2}\right)^{d/2}.
\label{eq:rho0_hom}
\ee
The thermal contribution is obtained as a low-$T$ expansion by expanding
$\rho(\mu\pm\delta\epsilon)$ in powers of $\delta\epsilon$: only odd powers of
$\delta\epsilon$ contribute; in the resulting integrals over $\delta\epsilon$,
we extend the upper bound from $\mu$ to $+\infty$, paying the price of 
an exponentially small error. Finally a series expansion
with even powers of $T$ is obtained:
\be
\rho_{\rm hom}(\mu,T)=\rho_{\rm hom}(\mu,0) \left[1+ \frac{\pi^2}{24} d(d-2) 
\left(\frac{k_B T}{\mu}\right)^2+ O\left((k_B T/\mu)^4\right)\right]
\ee
The bidimensional case $d=2$ deserves a particular discussion. One finds
that all the powers of $T$ in the expansion have vanishing coefficients,
since the density of states is constant, $\rho(\epsilon)=m/(2\pi\hbar^2)$. 
The integral over $\epsilon$ in Eq.(\ref{eq:hom}) can be performed exactly,
\be
\rho_{\mathrm{hom}}^{d=2}(\mu,T) 
= \frac{m k_B T}{2\pi\hbar^2} \ln\left(1+e^{\beta\mu}\right).
\ee
For a fixed and negative chemical potential, one sees that the density tends exponentially
rapidly to zero when $T\to 0$.
For $\mu>0$ a more convenient rewriting in the low temperature limit is
\be
\rho_{\mathrm{hom}}^{d=2}(\mu,T) 
= \frac{m\mu}{2\pi\hbar^2} \left[1+\frac{k_B T}{\mu} \ln\left(1+e^{-\beta\mu}\right)\right].
\ee
This explicitly shows that the deviation of the spatial density from the zero temperature
value $\rho_{\mathrm{hom}}(\mu,0)$ is exponentially small for $T\to 0$ if $d=2$.

\subsection{Coherence and correlation functions of the homogeneous gas}

We consider here a gas in a box with periodic boundary conditions.
The eigenmodes are thus plane waves $\phi_{\vec{k}}(\vec{r}\,)=\exp(i\vec{k}\cdot\vec{r}\,)/L^{d/2}$,
with a wavevector equal to $2\pi/L$ times a vector with integer components.
We shall immediately go to the thermodynamic limit, setting $L$ to infinity
for a fixed $\mu$ and $T$.

The first-order coherence function of the field is defined as the following thermal average:
\be
g_1(\vec{r}\,) = \langle \hat{\psi}^\dagger(\vec{r}\,) \hat{\psi}(\vec{0}\,)\rangle.
\ee
In the thermodynamic limit, it is given by the Fourier transform of the momentum distribution
of the gas, which is an isotropic function:
\begin{equation}
g_1(\vec{r}\,) = \phi(r) \equiv \int \frac{d^d k}{(2\pi)^d} n_k \,
e^{i\vec{k}\cdot\vec{r}}.
\label{eq:phi}
\end{equation}

It is easy to see that no long-range order exists for this coherence function,
even at zero temperature. At $T=0$, $n_k=1$ for $k<k_0$ and $n_k=0$ for
$k>k_0$, where the wavevector $k_0$ is defined by
\be
\frac{\hbar^2 k_0^2}{2m} = \mu
\ee
and is simply the Fermi wavevector, in the present case of zero temperature.
In the dimensions for $d=1$ to $d=3$ this leads to
\bea
\phi_{1D}(r)&=& \frac{\sin k_0 r}{\pi r}
\label{eq:phi1D} \\
\phi_{2D}(r)&=& \frac{k_0}{2\pi r}\, J_1(k_0 r)
\label{eq:phi2D} \\
\phi_{3D}(r)&=& -\frac{1}{2\pi r} \partial_r \phi_{1D}(r), 
\label{eq:phi3D}
\eea
where $J_1(x)$ is a Bessel function, behaving as $-\cos(x+\pi/4) (2/\pi x)^{1/2}$
for $x\rightarrow +\infty$. One then gets an algebraic decay of $\phi$ at large
distances, in all dimensions, because of the discontinuity of the momentum distribution.
Note that Eq.(\ref{eq:phi3D}) holds at all temperatures.

At low but non-zero temperature, the Fermi surface is no longer infinitely sharp but
has a smooth variation with an energy width $\sim k_B T$. This corresponds to a momentum width
$\delta k_0$ such that
\be
\epsilon_{k_0+\delta k_0}-\epsilon_{k_0} \simeq \frac{\hbar^2 k_0 \delta k_0}{m} \equiv  k_B T.
\label{eq:dk0}
\ee
We then expect a decreasing envelope for the oscillating function $\phi(r)$ with
a decay faster than algebraic with a length scale $1/\delta k_0$.
This can be checked by the following approximate calculation in 1D: we restrict to $k>0$
using parity, we single out from 
$n_k$ the zero-temperature contribution $\theta(k_0-k)$, where $\theta$ is the Heaviside function,
and we linearize the kinetic energy dispersion relation around $k_0$, writing
$\epsilon_{k_0+\delta k} \simeq \mu+ \hbar k_0 \delta k/m$.
Extending the integration over $\delta k$ to $-\infty$ we obtain
\be
\phi_{1D}(r)-\phi_{1D}^{T=0}(r) \simeq -\frac{2 \delta k_0}{\pi} \sin(k_0 r)
\int_0^{+\infty} du\, \frac{\sin(u \delta k_0 r)}{\exp(u)+1}
\ee
where $u$ originates from the change of variable $u=|\delta k|/\delta k_0$.
Integrating by parts over $u$, taking the integral of the sine function, gives
a fully integrated term exactly compensating the $T=0$ contribution, so that one
is left with
\be
\phi_{1D}(r) \simeq \frac{\sin(k_0 r)}{2\pi r} 
\int_0^{+\infty} du\, \frac{\cos(u \delta k_0 r)}{\cosh^2(u/2)}.
\ee
After extension to an integral over the whole real line (thanks to the parity
of the integrand), one may close the integration contour at infinity and sum
the residues over all the poles with $\mbox{Im}\, u>0$ to get:
\be
\phi_{1D}(r) \simeq \frac{\delta k_0\, \sin(k_0 r)}{\sinh(\pi \delta k_0 r)}.
\ee
We have recovered a result derived in \cite{Efetov}.
This approximate formula can also be used to calculate the 3D case, by virtue of
Eq.(\ref{eq:phi3D}).

The correlation function of the gas, also called the pair distribution function, 
is defined as the following thermal average:
\be
g_2(\vec{r}\,) = \langle \hat{\psi}^\dagger(\vec{r}\,) \hat{\psi}^\dagger(\vec{0}\,)
\hat{\psi}(\vec{0}\,) \hat{\psi}(\vec{r}\,)\rangle.
\ee
By virtue of Wick's theorem, $g_2$ is related to the function $\phi$ as
\be
g_2(\vec{r}\,) = \phi^2(0)-\phi^2(r)
\ee
where $\phi(0)$ is simply the mean particle density $\rho$. A key feature is
that $g_2$ vanishes in the limit $r\rightarrow 0$ (quadratically in $r$): this is
a direct consequence of the Pauli exclusion principle,
and expresses the fact that one cannot find (with a finite probability
density) two fermions with the same spin state at the same location.

This spatial antibunching is intuitively expected to lead to reduced
density fluctuations for an ideal degenerate Fermi gas, as compared to
an ideal Bose gas. This is quantified in the next subsection.

\subsection{Fluctuations of the number of fermions in a given spatial zone}
\label{subsec:fluc}

We define the operator giving the number of particles within the sphere of radius $R$:
\be
\hat{N}_R = \int_{r<R} d^dr\, \hat{\psi}^\dagger(\vec{r}\,) \hat{\psi}(\vec{r}\,),
\ee
the gas being spatially homogeneous and taken in the thermodynamic limit.
We now show that the number $\hat{N}_R$ has subpoissonian fluctuations for a 
degenerate ideal Fermi gas.

The mean value of the number of particles within the sphere is simply
\be
\langle \hat{N}_R\rangle = \rho V_d(R) 
\ee
where $\rho=\phi(0)$ is the mean density and $V_d(R)$ the volume enclosed
by the sphere of
radius $R$ in dimension $d$.
From Wick's theorem we find a variance
\be
\mbox{Var}\, \hat{N}_R = \langle \hat{N}_R\rangle - \int_{r<R} d^dr\,
\int_{r'<R} d^dr'\, \phi^2(|\vec{r}-\vec{r}\,'|).
\ee
This immediately reveals the subpoissonian nature of the fluctuations.

We quantify this subpoissonian nature in the limit of a large radius $R$.
Replacing $\phi$ by its expression as a Fourier transform of $n_k$, see Eq.(\ref{eq:phi}),
we obtain
\be
\mbox{Var}\, \hat{N}_R = \langle \hat{N}_R\rangle - \int \frac{d^dk}{(2\pi)^d}\,
\int \frac{d^dk'}{(2\pi)^d}\, n_k n_{k'} F(\vec{k}-\vec{k}')
\ee
where $F$ is the modulus squared of the Fourier transform of 
the characteristic function of the sphere:
\be
F(\vec{q}\,) = \left|\int_{r<R} d^dr\, e^{i\vec{q}\cdot\vec{r}}\right|^2.
\ee
One can show that $F$ is an integrable function peaked in $q=0$ and with a width
$\sim 1/R$. From the Parseval-Plancherel identity the integral over the whole momentum
space of $F$ is $(2\pi)^d V_d(R)$. When $R \delta k_0 \gg 1$, where $\delta k_0$ is the
momentum width of the finite temperature Fermi surface, see Eq.(\ref{eq:dk0}),
we can replace $F$ by a Dirac distribution:
\be
F(\vec{q}\,) \simeq (2\pi)^d V_d(R) \delta^d(\vec{q}\,).
\ee
This leads to
\be
\mbox{Var}\, \hat{N}_R \simeq V_d(R) \int \frac{d^dk}{(2\pi)^d} n_k(1-n_k).
\ee
To easily calculate the resulting integral to leading order in $T$, we note that
\be
\partial_\mu n_k = \beta n_k (1-n_k)
\ee
so that the $T=0$ value of the density only is required, Eq.(\ref{eq:rho0_hom}).
We finally obtain
\be
\mbox{Var}\, \hat{N}_R \simeq \langle \hat{N}_R\rangle \frac{d\, k_B T}{2\mu}.
\ee

The above result does not apply when $R \delta k_0 \ll 1$. In the range of radius
$R \ll (\delta k_0)^{-1}$ we may simply perform a $T=0$ calculation.
The technique to take advantage of the translational and rotational invariance
of $\phi^2(|\vec{r}-\vec{r}\,'|)$ is to introduce the purely geometrical quantity
\be
K(X) = \int_{r<R} d^dr \int_{r'<R} d^dr' \delta(|\vec{r}-\vec{r}\,'|-X),
\ee
so that
\be
\mbox{Var}\, \hat{N}_R =\rho V_d(R)-\int_0^{2R} dX\, K(X) \phi^2(X).
\ee
An exact calculation of $K(X)$ is possible: for $X\leq 2R$, one obtains
\bea
K_{1D}(X) &=& 4(R-X/2) \\
K_{2D}(X) &=& \pi X \left[4R^2\mbox{arccos}\, \frac{X}{2R}-X\,(4R^2-X^2)^{1/2}\right]\\
K_{3D}(X) &=& \frac{\pi^2}{3} X^2(X+4R)(X-2R)^2.
\eea
When combined with Eqs.(\ref{eq:phi1D},\ref{eq:phi3D}), this leads to exact expressions
for the variance in 1D and 3D, in terms of the complex integral $Ci$ and $Si$ functions.
The large $k_0 R$ expressions are then readily obtained. The asymptotic expansion
of the 2D has to be worked out by hand, singling out the contribution of the low
$X$ quadratic expansion of $K(X)$. We obtain:
\bea
\label{eq:var_1d}
d=1: \ \ \mbox{Var}\, \hat{N}_R & = &  \frac{\gamma+1+\ln 4 k_0 R}{\pi^2} +O(1/(k_0 R)^2) \\
d=2: \ \ \mbox{Var}\, \hat{N}_R & = &  \frac{k_0 R}{\pi^2} \left[\ln(4 k_0 R) +  C\right]
+O(\ln(k_0R)/k_0 R)\\
d=3: \ \ \mbox{Var}\, \hat{N}_R & = &  \frac{1}{2\pi^2} \ln(A_3 k_0 R)\, (k_0 R)^2
-\frac{1}{24\pi^2} \ln(B_3 k_0 R) + O(1/k_0 R)
\eea
with 
\bea
C &=& \lim_{u\rightarrow +\infty} \frac{\pi}{12}u^3\, {}_2F_{3}(\frac{3}{2},\frac{3}{2};
2,\frac{5}{2},3;-u^2) -\ln u \simeq 0.656657 \\
A_3 &=& \exp(\gamma +2 \ln 2 -1/2) = 4.321100\ldots \\
B_3 &=& \exp(\gamma +2 \ln 2 -5/12) = 4.696621\ldots
\eea
and ${}_2F_{3}$ is a hypergeometric function, $\gamma=0.5772156649\ldots$ is Euler's constant.

\subsection{Application to the 1D gas of impenetrable bosons}

Consider a 1D homogeneous gas of bosons interacting {\it via} a delta potential $g \delta(x_1-x_2)$
in the limit $g\rightarrow +\infty$. In this so-called impenetrable boson limit,
the interaction potential can be replaced by the contact condition that the many-body
wavefunction vanishes when two bosons are in the same point. On the fundamental domain
of ordered positions of the $N$ bosons, $x_1<\ldots < x_N$, one then realizes that
the eigenwavefunction for bosons coincides with an eigenwavefunction of $N$ non-interacting
fermions, with the same eigenenergy.  Out of the fundamental domain, the bosonic
and fermionic wavefunctions may differ by a sign.

As a consequence, the spatial distribution of the impenetrable bosons, being sensitive
to the modulus squared of the wavefunction, will be the same as for the ideal Fermi
gas.  The discussion on the variance of the number of particles in an interval
$[-R,R]$ in subsection \ref{subsec:fluc} immediately applies.

On the contrary, the first order coherence function for the impenetrable bosons $g_1^B(x)$
is sensitive to the phase of the wavefunction and will differ from the $g_1$ fermionic
function. More details, obtained with the Jordan-Wigner transformation,
are given in \cite{Houches03} and references therein. At zero temperature, 
we simply recall the mathematical fact, showing the absence of condensate,
that at large distances \cite{Tracy}
\be
g_1^B(x) \sim \frac{A \rho}{|k_F x|^{1/2}}
\ee
where $k_F =\pi \rho$, $\rho$ being the density of bosons, and $A=0.92418\ldots$.
We also use the fact (that one can check numerically \cite{Houches03}) 
that the power law
behavior of $g_1^B(x)$ is the same as the one of the simpler function
\be
G(x) = \langle e^{i \pi \hat{N}_{[0,x]}}\rangle
\ee
where $\hat{N}_{[0,x]}$ is the operator giving the number of fermions in the interval
from $0$ to $x$ (here $x>0$) and the expectation value is taken in the ground state of
the ideal Fermi gas.  In this way, one relates the long range behavior of
$g_1^B$ to the counting statistics of an ideal Fermi gas: $G(x)$ is the difference
between the probabilities of having an even number and an odd number of fermions in the 
interval $[0,x]$.
One may assume
that the probability distribution of the number $n$ of fermions in the interval
has a Gaussian envelope  \cite{bosonisation}.
Using the Poisson formula
\be
\sum_{n=-\infty}^{+\infty} f(n) =\sum_{n=-\infty}^{+\infty} \tilde{f}(2\pi n)
\ee
where $\tilde{f}(k)=\int dx\, f(x)\exp(-ikx)$ is the Fourier transform of the arbitrary
function $f$, and restricting to the leading terms $n=0$ and $n=1$, we obtain
\be
G(x) \simeq 2 \cos(\pi\rho x) e^{-\pi^2\,(\mbox{\scriptsize Var}\,\hat{N}_{[0,x]})/2}.
\ee
From the asymptotic expression Eq.(\ref{eq:var_1d}) we obtain
\be
G(x) \propto \frac{\cos(\pi \rho x)}{(\rho x)^{1/2}}
\ee
which explains the $1/x^{1/2}$ decay of $g_1^B(x)$.

\subsection{In a harmonic trap}

In a harmonic trap, simple approximate results can be obtained in the limit where
the chemical potential is much larger than the trapping frequencies 
$\omega_\alpha$ times $\hbar$,
in a temperature range to be specified,
in the so-called semi-classical approximation.
We illustrate this semi-classical approximation for two quantities, a thermodynamic one,
the entropy, and a local observable, the density.

\subsubsection{Semi-classical calculation of the entropy}
The entropy $S$ is a physically appealing way to evaluate to which extent a system is cold,
as it is a constant in thermodynamically adiabatic transformations, contrarily to the
temperature. 

To calculate thermodynamic quantities in the grand-canonical ensemble, 
it is convenient to start from the grand-potential
\be
\Omega(\mu,T)= -k_B T \ln \mbox{Tr}\left[e^{-\beta(H-\mu \hat{N})}\right].
\ee
Then one uses the fact that 
\be
d\Omega =-S \, dT - N\, d\mu
\ee
where $N=\langle \hat{N}\rangle$ is the mean number of particles.
Since the various field eigenmodes decouple in the grand-canonical ensemble, one
finds
\be
\Omega(\mu,T)= k_B T \int d\epsilon \rho(\epsilon) \ln[1-n(\epsilon)]
\label{eq:Omega}
\ee
where $n(\epsilon)$ is the Fermi-Dirac formula and $\rho(\epsilon)$ is the density of states
for a particle in the trap:
\be
\rho(\epsilon) = \sum_{\vec{n}} \delta\left[\epsilon-
\sum_{\alpha=1}^{d} (n_\alpha+1/2)\hbar\omega_\alpha\right]
\ee
where the sum is over all vectors with non-negative integer coordinates.
In the semi-classical limit one replaces the sum by an integral 
\be
\rho(\epsilon)\simeq \rho_{\rm sc}(\epsilon) = 
\int_{[0,+\infty[^d} d^dn\, \delta(\epsilon-\sum_\alpha n_\alpha \hbar\omega_\alpha)
\ee
where we dropped the ground mode energy $\epsilon_0$ under the assumption $\epsilon\gg
\epsilon_0$. By the change of variables $n_\alpha=u_\alpha \epsilon/\hbar\omega_\alpha$
one obtains
\be
\rho_{\rm sc}(\epsilon) = \frac{\epsilon^{d-1}}{\prod_\alpha \hbar\omega_\alpha}
\int_{[0,+\infty[^d} d^d u\, \delta\left(1-\sum_\alpha u_\alpha\right).
\ee
The integral can be calculated by integrating over $u_d$ and by performing 
the change of variables $\tau_1=u_1, \tau_2=u_1+u_2,\ldots, \tau_{d-1}=u_1+\ldots+
u_{d-1}$. We finally obtain
\be
\rho_{\rm sc}(\epsilon) = \frac{\epsilon^{d-1}}{(d-1)! (\hbar\bar{\omega})^d}
\ee
where $\bar{\omega}=(\prod_\alpha \omega_\alpha)^{1/d}$ is the geometric
mean of the trap frequencies.

We replace $\rho$ by $\rho_{\rm sc}$ in Eq.(\ref{eq:Omega}), and we integrate by parts,
taking the derivative of 
the $\ln[1-n(\epsilon)]$, to obtain the semi-classical approximation:
\be
\Omega\simeq \Omega_{\rm sc} = -(\hbar\bar{\omega})^{-d}
\int_0^{+\infty} d\epsilon\, \frac{\epsilon^d}{d!}\, n(\epsilon).
\ee
In the degenerate regime, we use the technique sketched around Eq.(\ref{eq:tech}), based
on the fact that $n(\epsilon)-n_{T=0}(\epsilon)$ is a narrow function of the energy, to
obtain the low-$T$ expansion:
\be
\Omega_{\rm sc}(\mu,T) = -\frac{\mu^{d+1}}{(d+1)! (\hbar\bar{\omega})^d}
\left[1+\frac{\pi^2}{6}d(d+1)\left(\frac{k_B T}{\mu}\right)^2+\ldots
\right]
\ee
Calculating $S=-\partial_T \Omega$ and $N=-\partial_\mu\Omega$,  we obtain the semi-classical
approximation for the number of particles and for the entropy per particle:
\bea
\label{eq:Ntrap}
N &\simeq & \frac{\mu^d}{d!(\hbar\bar{\omega})^d} \\
S/N &\simeq& \frac{d\pi^2}{3} k_B T/T_F
\label{eq:approx_entropy}
\eea
with the Fermi energy 
\be
k_B T_F \equiv \hbar\bar{\omega} (N d!)^{1/d}.
\ee

\begin{figure}[htb]
\centerline{
\includegraphics[width=11cm,clip]{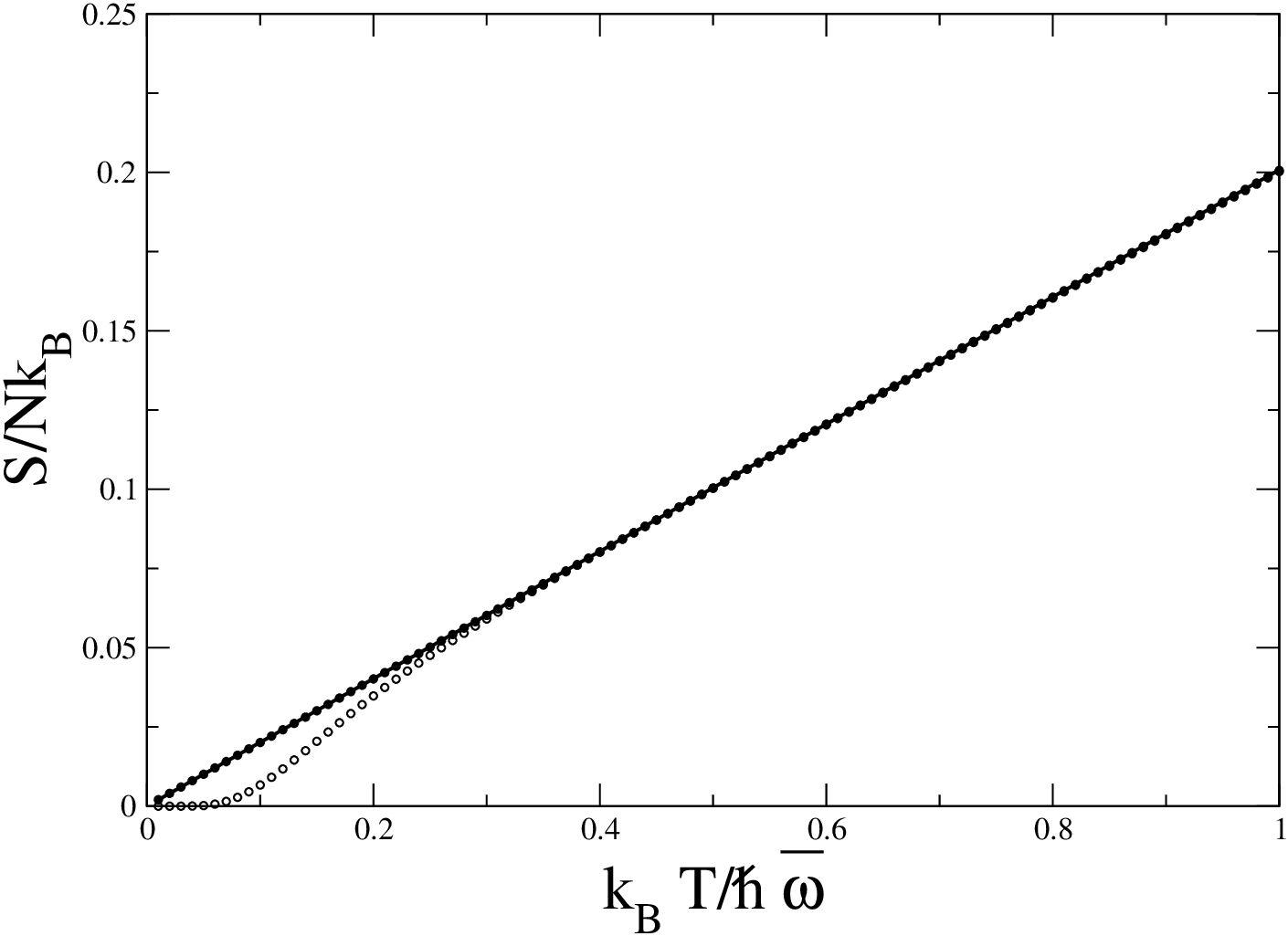}}
\caption{Entropy $S$ of a spin-polarized ideal Fermi gas in a 3D harmonic trap.
Black disks: exact result for trap frequencies $\omega_\alpha=\bar{\omega}
(0.5981188\ldots,1.2252355\ldots,1.36440128\ldots)$. Empty circles: exact result for an isotropic
trap $\omega_\alpha=\bar{\omega}(1,1,1)$. Solid line: approximate formula
Eq.(\ref{eq:approx_entropy}). 
The chemical potential is fixed to $\mu=\epsilon_0+47.6\hbar\bar{\omega}$,
where $\epsilon_0$ is the trap ground mode energy.
In order to test the accuracy of Eq.(\ref{eq:approx_entropy}), in a situation
where 
the number of particles $N$ rather than the chemical potential $\mu$ is
known, as it is the case in experiments with atomic gases, 
we use the exact, numerical value of $N$ to evaluate
Eq.(\ref{eq:approx_entropy}) rather than its approximation
Eq.(\ref{eq:Ntrap}). Here $N$ is about $2\times 10^4$.
}
\label{fig:entropy}
\end{figure}

What are the validity conditions of this approximate formula for the entropy? 
The temperature should be $T\ll T_F$ but, contrarily to the thermodynamic
limit case, the temperature should remain high enough, to allow 
the approximation of the density of states (which is a sum of Dirac peaks) 
by a smooth function. The typical distance between the Dirac peaks at the Fermi
energy should be smaller than the thermal energy width $\sim k_B T$ of the Fermi
surface. In the case of an isotropic harmonic trap, $\omega_\alpha=
\omega$ for all $d$ dimensions, the Dirac peaks are regularly
spaced by $\hbar\omega$ so that we impose
\be
k_B T > \hbar \omega.
\ee
In the case of incommensurable trap frequencies, the peaks are irregularly spaced
and we estimate a typical spacing from the inverse of the smoothed density of states at the
Fermi surface: we require
\be
k_B T > \frac{1}{\rho_{\rm sc}(\mu)},
\ee
with $\mu\simeq k_B T_F$. This leads to the condition
\be
k_B T \gg \frac{k_B T_F}{dN} = \hbar\bar{\omega} \frac{(d!)^{1/d}}{d N^{1-1/d}}.
\ee
One sees that this is a much weaker condition than in the isotropic case,
for $d>1$, since we are here in the large $N$ limit. 
This fact is illustrated by a numerical example in 3D, calculating
the entropy for an isotropic trap and an anisotropic trap, see Fig.~\ref{fig:entropy}.

\subsubsection{Semi-classical calculation of the density}

The semi-classical approximation for the density profile in the external
potential $U(\vec{r}\,)$ is strictly equivalent to
the so-called local density approximation (LDA), as can be revealed by the writings:
\bea
\rho_{\rm sc}(\vec{r}\,) &=& \int \frac{d^dp}{(2\pi\hbar)^d} n[U(\vec{r}\,)+p^2/2m] \\
\rho_{\rm LDA}(\vec{r}\,) &=& \rho_{\mathrm{hom}}[\mu-U(\vec{r}\,),T] 
\eea
where $\rho_{\rm hom}$ is defined in Eq.(\ref{eq:hom}) and $n(\epsilon)$
is the Fermi-Dirac formula Eq.(\ref{eq:fd}).
The philosophies of the two approximations of course differ.
In the semi-classical case, one relies on a phase space density of the particles,
which is an approximation to the Wigner representation 
$W(\vec{r},\vec{p}\,)$ of the one-body density operator
of the gas \cite{Wigner}:
\be
\hat{\rho} = \frac{1}{\exp[\beta(\hat{h}_0-\mu)]+1},
\ee
where the one-body Hamiltonian $\hat{h}_0$ is defined in
Eq.(\ref{eq:obh}).  One assumes
\be
W(\vec{r},\vec{p}\,) \equiv  
\int \frac{d^du}{(2\pi\hbar)^d} \langle \vec{r}-\frac{\vec{u}}{2}|\hat{\rho}|\vec{r}+\frac{\vec{u}}{2}\rangle
e^{i\vec{p}\cdot\vec{u}/\hbar}
\simeq \frac{1}{(2\pi\hbar)^d}\,\frac{1}{\exp\{\beta[U(\vec{r}\,)+\frac{p^2}{2m}-\mu]\}+1}.
\ee
It remains to integrate this approximation of $W(\vec{r},\vec{p}\,)$ over
$\vec{p}$ to obtain $\rho_{\mathrm{sc}}(\vec{r}\,)$.

In the local density approximation, the gas is considered as a collection of 
quasi-macros\-co\-pic pieces
that have the same properties as the homogeneous gas of temperature $T$
and chemical potential $\mu_{\rm LDA}=\mu-U(\vec{r}\,)$. Note that this last
equation intuitively expresses the fact that the chemical potential is uniform
in a gas at thermal equilibrium.

These approximations assume that the external potential 
varies weakly over the correlation length of the homogeneous gas,
$\sim 1/k_0$  where $k_0(\vec{r}\,)$ is the local Fermi wavevector, such
that $\hbar^2 k_0^2/2m = \mu_{\rm LDA}$.
In a harmonic trap, the scale of variation of the potential may be taken
as the spatial extension of the gas, the so-called Thomas-Fermi length
$R_\alpha$ along direction $\alpha$ such that
\be
\mu = \frac{1}{2} m \omega_\alpha^2 R_\alpha^2.
\ee
The condition $k_0(\vec{r}\,) R_\alpha \gg 1$, and correspondingly the 
semi-classical approximation
and the LDA, will fail close to the borders of the cloud where $k_0$ diverges.

At $T=0$, as is apparent from Eq.(\ref{eq:rho0_hom}),
these approximations predict the density profiles:
\be
\rho(\vec{r}\,) = \frac{V_d(1)}{(2\pi)^d}\left[\frac{2m(\mu-U(\vec{r}\,))}{\hbar^2}\right]^{d/2}.
\ee
Integrating this density profile over the domain $|r_\alpha|\leq R_\alpha$
gives an expression of the total number of particles as a function of the chemical
potential: this expression coincides with the righthand side
of Eq.(\ref{eq:Ntrap}), whatever the dimension of space $d$,
which illustrates the consistency of the various approaches.

It is instructive to see if the LDA for the density profile can be used
at $T=0$ strictly, or if a lower bound on $T$ is required, as was the case 
for the entropy. An answer to this question is obtained by calculating
exactly the density of the gas and its second order derivatives
in the trap center and comparing to the LDA.
The value of $\rho(\vec{0}\,)$ is simple to obtain from a numerical summation
over the harmonic oscillator quantum numbers $n_\alpha$, since the value 
in $x=0$ of the 1D harmonic oscillator eigenstate wavefunction of quantum
number $n\geq0$ is known exactly: for $n$ odd, $\phi_n(0)=0$, and for
$n$ even,
\be
\phi_n(0) = (-1)^{n/2}\frac{(n!)^{1/2}}{2^{n/2} (n/2)! \pi^{1/4}} a_{\rm ho}^{-1/2}
\ee
where the harmonic oscillator length is $a_{\rm ho}=(\hbar/m\omega)^{1/2}$.
This may be obtained from properties of the Hermite polynomials $H_n$, using
$\phi_n(x) = \mathcal{N}_n 
H_n(x/a_{\rm ho}) \exp(-x^2/2 a_{\rm ho}^2)$ with a positive
normalisation factor such that
$\mathcal{N}_n^{-2}=a_{\rm ho}\pi^{1/2}2^{n}n!$,
or from the
recursion relation $\phi_n(0)=-(1-1/n)^{1/2}\phi_{n-2}(0)$, obtained
from the $a$ and $a^\dagger$ formalism.
The second order derivatives of the density in $\vec{r}=\vec{0}$ are also easily
obtained: they involve $\phi_n'(0)=\sqrt{2n}\phi_{n-1}(0)/a_{\rm ho}$ and the second order 
derivative of $\phi_n$, which is related to $\phi_n$ by Schr\"odinger's equation,
$\phi_n''(0)=-(2n+1)\phi_n(0)/a_{\rm ho}^2$.

We apply this check in 1D.  At zero temperature, explicit expressions can be obtained.
For a chemical potential $n+1/2< \mu/\hbar\omega < n+3/2$, where $n$ is an even
non-negative integer, we obtain
\bea
\rho(0) &=& (n+1) \phi_n^2(0) = \frac{1}{\pi^{1/2}a_{\rm ho}} \frac{(n+1)!}{2^n[(n/2)!]^2} \\
\rho''(0) &=& - 2 a_{\rm ho}^{-2} \rho(0).
\eea
For a chemical potential $n+3/2< \mu/\hbar\omega < n+5/2$, where $n$ is still even,
the $x=0$ density assumes the same value (since $\phi_{n+1}(0)=0$) whereas
the second order derivative changes sign:
\bea
\rho(0) &=& \frac{1}{\pi^{1/2}a_{\rm ho}} \frac{(n+1)!}{2^n[(n/2)!]^2} \\
\rho''(0) &=& 2 a_{\rm ho}^{-2} \rho(0).
\eea

\begin{figure}[htb]
\begin{center}
\includegraphics[height=4.5cm,clip]{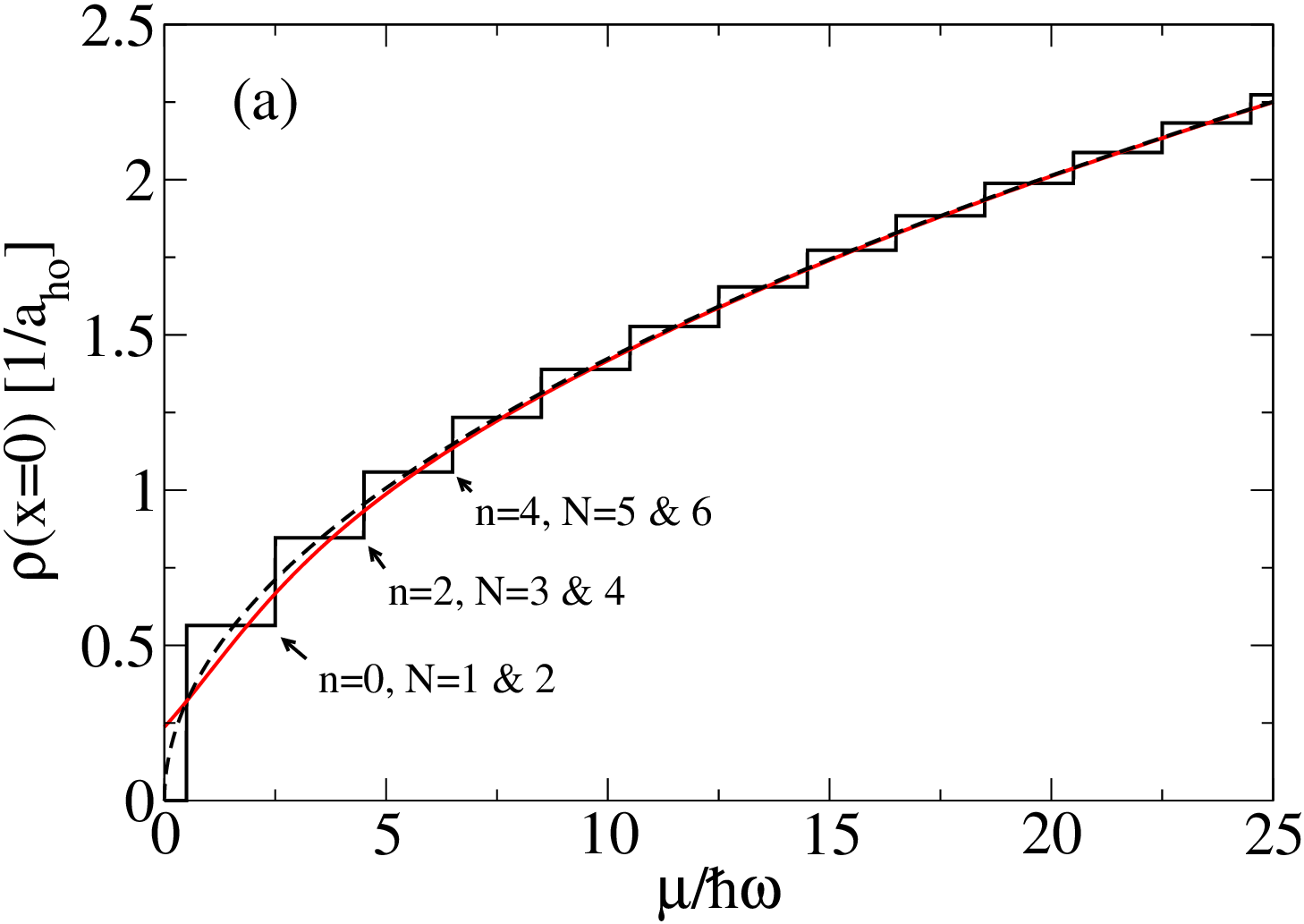}
\includegraphics[height=4.5cm,clip]{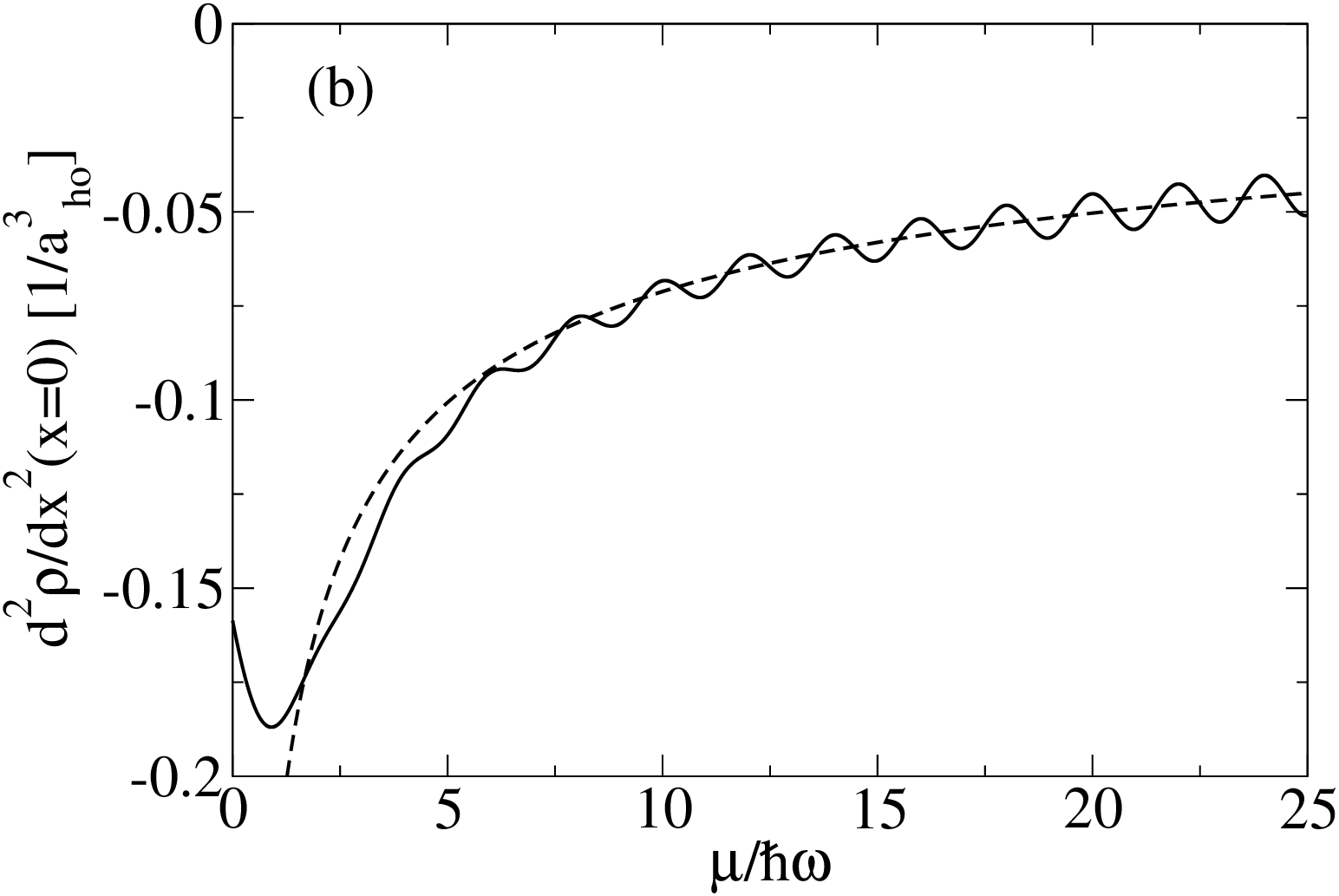}
\end{center}
\caption{
Central density of a 1D harmonically trapped ideal Fermi gas, in the grand canonical
ensemble, as a function of the chemical potential $\mu$. (a) Density in the trap
center $x=0$: exact result for $T=0$ (black solid line with steps), exact numerical sum for 
$k_B T=\hbar\omega$ (red smooth solid line), vs the $T=0$ LDA (dashed line).
(b) Second order derivative of the density in the trap center, $d^2\rho/dx^2(x=0)$,
for $k_B T=\hbar\omega$: exact numerical sum
(solid line) vs the $T=0$ LDA (dashed line).}
\label{fig:dens}
\end{figure}

The $T=0$ exact value of the density as a function of the chemical potential
is compared to the LDA in Fig.~\ref{fig:dens}a: the LDA
does not reproduce the steps but nicely interpolates between them.
By using Stirling's formula for the exact result, one finds
that the relative error in the LDA is $O(1/N)$, where $N$ is the particle number.
If one cleverly approximates the value of a step in $\rho(0)$ 
by taking the median value of the
chemical potential, $\mu=(n+3/2)\hbar\omega$, one even finds that the relative error
in the LDA vanishes as $O(1/N^2)$.
The LDA, on the contrary, gives a totally incorrect prediction for $\rho''(0)$
(not shown in the figure): one has
\be
\rho_{\rm LDA}''(0)= -\frac{1}{\pi^2 a_{\rm ho}^4 \rho_{\rm LDA}(0)}
\ee
which tends to zero in the large $N$ limit, whereas the exact result oscillates
with a diverging amplitude.
All this can be understood from the fact that the exact density profile
at $T=0$ is an oscillating function of $x$, oscillating at smaller and smaller
scales in the large
$N$ limit. This fact can be revealed numerically \cite{Minguzzi}, but more
elegantly results from the exact summation formula \cite{Baranov_pc}:
\be
\sum_{k=0}^{m} \phi_k^2(x) = (m+1)\phi_m^2(x)-\sqrt{m(m+1)}\phi_{m+1}(x)\phi_{m-1}(x).
\ee
The LDA of course misses these oscillations.

What happens at finite temperature~? We expect that, at $k_B T \sim \hbar\omega$,
these oscillations are sufficiently reduced to make the LDA more accurate, 
while the temperature is low enough to 
allow the use of the $T=0$ limit of the LDA. 
This is confirmed by the $k_B T=\hbar\omega$ data in Fig.~\ref{fig:dens}:
the plateaus in $\rho(0)$ as a function of $\mu$ are hardly visible (see a),
so are the oscillations of $\rho(x)$  as a function of $x$ \cite{Minguzzi2}, 
and the oscillations of $\rho''(0)$ are so strongly reduced that the LDA
approximation becomes acceptable (see b).

\section{Two-body aspects of the interaction potential}

In real gases, there are interactions among the particles. In fermionic atomic
gases it is even possible to reach the maximally interacting regime allowed by
quantum mechanics in a gas, the so-called unitary regime,
without altering the stability and lifetime of the gas.
This section reviews possible models that can be used to represent the interaction
potential and recalls basic facts of two-body scattering theory.
We restrict to a three-dimensional 
two spin-state, single species Fermi gas; the $p$-wave interactions
among atoms of a common spin state are neglected, usually an excellent
approximation away from $p$-wave Feshbach resonances. The $s$-wave interactions
are on the contrary supposed to be strong, the scattering length being much larger than
the potential range, in the vicinity of a Feshbach resonance.

\subsection{Which model for the interaction potential ?}

The detailed description of the interaction of two atoms is involved.
At large enough interatomic distances, in particular 
much larger than the size of the electronic orbitals of an atom, 
one can hope to represent this interaction by
a position dependent potential $V(\vec{r}\,)$, which includes the van der Waals interaction
term
\be
V(r) \simeq - \frac{C_6}{r^6},
\ee
a simple formula that actually neglects retardation effects and long range dipole-dipole
magnetic interactions. Forgetting about these complications, we can use Schr\"odinger's
equation and the
$C_6$ coefficient to construct a length $b$, called the van der Waals
length, that we shall consider as the `true'
range of the interaction potential:
\be
\frac{\hbar^2}{m b^2} = \frac{C_6}{b^6}.
\ee
For alkali atoms, $b$ is in the nanometer range.

At short interatomic distances, however, this simple picture of a scalar interaction
potential has to be abandoned, and one has to include the various Born-Oppenheimer
potentials curves coming from the QED Hamiltonian for two atomic nuclei at fixed distance,
including all the electronic and nuclear spin states,
and the motional couplings among the Born-Oppenheimer curves due to the finite mass of an atom.

Fortunately we are dealing with gaseous systems: the mean interparticle distance
is much larger than the potential range $b$:
\be
\rho^{1/3} b \ll 1
\ee
where $\rho$ is the mean density. We shall also consider atomic dimers, but these
dimers shall be very weakly bound, with a size on the order of the scattering length $a$,
which is much larger than $b$ in the vicinity of a Feshbach resonance.
As a consequence, the central postulate for the theory of quantum gases is that the
short-range details of the atomic interactions are unimportant, only the low-momentum
behavior of the scattering amplitude between two atoms is relevant.
For fermions of equal masses, this postulate has proved robust in its confrontation
to experiments, even in the unitary regime. 

A practical consequence of this postulate is that any short range model for the interaction
leading to almost the same scattering amplitude $f_k$ as the true interaction, in the typical
relative momentum range $k$ of atoms in a quantum gas, is an acceptable model.
We then put the constraint on any acceptable model for the interaction:
\be
f_k^{\rm model} \simeq f_k
\label{eq:philo}
\ee
for the relevant values of the relative momenta $k$ populated in the gas.
We insist here that we impose similar scattering {\it amplitudes} over some
momentum range, not just equal scattering lengths.
Typical values of $k$ can be the following ones:
\be
k_{\rm typ} \in \{a^{-1}, k_F, \lambda^{-1} \}
\label{eq:k_typ}
\ee
where $a$ is the s-wave scattering length between opposite spin fermions, $k_F=(6\pi^2\rho_\uparrow)^{1/3}$
 is the Fermi momentum expressed in terms of the density of a single spin component
$\rho_\uparrow=\rho_\downarrow$, and $\lambda$ is the thermal de Broglie wavelength Eq.(\ref{eq:lambda}).
The choice of the correct value of $k_{\rm typ}$ is left to the user and depends on the physical
situation. The first choice $k_{\rm typ}\sim a^{-1}$ is well suited to the case
of a condensate of dimers ($a>0$) since it is the relative momentum of two atoms forming
the dimer. The second choice $k_{\rm typ}\sim k_F$ is well suited to a degenerate
Fermi gas of atoms (not dimers). The third choice $k_{\rm typ} \sim \lambda^{-1}$
is relevant for a non-degenerate Fermi gas, a case not discussed in this lecture.

A condition for the gas not to be sensitive to the microscopic details
of the potential is then that
\be
k_{\rm typ} b \ll 1.
\label{eq:nice_regime}
\ee
As we shall see, the scattering amplitude in this momentum range should be described
by a very limited number of parameters, which allows to use very simple models
in the many-body problem.

The situation is more subtle for
bosons, or for mixtures of fermions with widely different masses, where the Efimov
phenomenon takes place (see the lecture by Gora Shlyapnikov in this volume): at
the unitary limit, an effective 3-body attractive interaction occurs and
accelerates the atoms; the wavevector $k$ can then, for some type of
Feshbach resonances (the broad Feshbach resonances to be defined below),
reach the range $\sim b^{-1}$, in which case the details of the true interaction 
may become important. We shall not consider this case here.

\subsection{Reminder of scattering theory}

Suppose for simplicity that two particles of mass $m$ interact {\sl via} the short potential
$V(\vec{r}_1-\vec{r}_2)$ that tends to zero at infinity faster than
$1/r_{12}^3$. Moving to the center of mass frame, with a center of mass
with vanishing momentum, one gets
the Schr\"odinger equation at energy $E$:
\be
E \phi(\vec{r}\,) = \left[-\frac{\hbar^2}{m} \Delta_{\vec{r}} + V(\vec{r}\,) \right]
\phi(\vec{r}\,).
\ee
At negative energies $E<0$, one looks for the discrete values of $E$ such that
$\phi$ is square integrable: these eigenstates correspond to two-body bound states, that
we call dimers.
At positive energies $E>0$, one sets
\be
E= \frac{\hbar^2 k_0^2}{m}
\ee
with $k_0>0$,
one looks for scattering states that obey the boundary conditions
suited to describe a scattering experiment:
\bea
\label{eq:boundary1}
\phi(\vec{r}\,) &=& \phi_0(\vec{r}\,) + \phi_{s}(\vec{r}\,)  \\
\label{eq:boundary2}
\phi_0(\vec{r}\,) &=& \langle\vec{r}\,|\vec{k}_0\rangle \equiv e^{i \vec{k}_0\cdot\vec{r}}  \\
\label{eq:boundary3}
\phi_s(\vec{r}\,) & \sim &  f_{\vec{k}_0}(\vec{n}\,) \frac{e^{i k_0 r}}{r}
\ \ \ \ \mbox{for} \ \ \ r\rightarrow + \infty.
\eea
The wavefunction $\phi_0$ represents the free wave coming from infinity, taken
here to be a plane wave of wavevector $\vec{k}_0$ of modulus $k_0$.
The function $\phi_{s}(\vec{r}\,)$ represents the scattered wave which, at infinity,
depends on the distance as an outgoing spherical wave and possibly depends
on the direction of observation $\vec{n}\equiv \vec{r}/r$ through the
scattering amplitude $f_{\vec{k}_0}$.

It is important to keep in mind that the Schr\"odinger equation with the
above boundary conditions is formally solved by
\be
|\phi\rangle = [1+G(E+i0^+)V]|\phi_0\rangle
\ee
where $G(z)=(z-H)^{-1}$ is the resolvent of the Hamiltonian, $z$ a complex number.
Using the identity $G=G_0+G_0 V G$, where $G_0(z)=(z-H_0)^{-1}$ is the resolvent
of the free, kinetic energy Hamiltonian, one gets
\be
|\phi\rangle = [1+G_0(E+i0^+) T(E+i0^+)]|\phi_0\rangle
\label{eq:in_k_space}
\ee
where we have introduced the $T$ operator (or $T$-matrix)
\be
T(z)= V+ V G(z) V.
\ee
Eq.(\ref{eq:in_k_space}) allows to obtain the equivalence in momentum space
of the asymptotic conditions Eqs.(\ref{eq:boundary1},\ref{eq:boundary2},\ref{eq:boundary3}) in real space:
\be
\langle\vec{k}\,|\phi\rangle = (2\pi)^3 \delta^3(\vec{k}-\vec{k}_0)
+\frac{1}{E+i0^+-\hbar^2 k^2/m} \,
\langle\vec{k}| T(E+i0^+)|\vec{k}_0\rangle,
\label{eq:asymptk}
\ee
where the matrix element of $T$ is an {\sl a priori} unknown smooth, regular function of $\vec{k}$.
The link between the $k$-space and the $r$-space points of view is completed
by the following relation:
\be
f_{\vec{k}_0}(\vec{n}\,) = -\frac{m}{4\pi\hbar^2} \langle k_0\vec{n}|T(E+i0^+)|\vec{k}_0\rangle,
\label{eq:f_vs_T}
\ee
which follows from the expression of the 
kinetic energy Green's function for $E>0$:
\be
\langle \vec{r}\,| G_0(E+i0^+) |\vec{r}\,'\rangle =
-\frac{m}{4\pi\hbar^2} \frac{e^{i k_0|\vec{r}-\vec{r}\,'|}}{|\vec{r}-\vec{r}\,'|}.
\label{eq:green}
\ee

Clearly, the central object is the scattering amplitude. It obeys the optical theorem
\be
\sigma_{\rm scatt}(k_0)= \int d^2n\, |f_{\vec{k}_0}(\vec{n}\,)|^2 =
\frac{4\pi}{k_0} \mbox{Im}\, f_{\vec{k}_0}(\vec{k}_0/k_0),
\ee
where $\sigma_{\rm scatt}$ is the total scattering cross section for distinguishable particles. 
This theorem is a direct
consequence of the conservation of probability: the matter current 
$\vec{\mbox{\j}}$ in the
stationary state $\phi$ has a vanishing divergence, that is a zero flux through a sphere
of radius $r\rightarrow + \infty$: using the asymptotic form of $\phi$, one gets
the announced theorem. Here we shall consider model potentials that
scatter only in the $s$-wave: $\phi_s$ is isotropic and so is the scattering
amplitude, which reduces to an energy dependent complex number:
\be
f_{\vec{k}_0}(\vec{n}\,) = f_{k_0}.
\ee
The optical theorem simplifies to
\be
|f_{k_0}|^2 = \frac{\mbox{Im}\, f_{k_0}}{k_0}.
\ee
Using $\mbox{Im}\, z^{-1}=-\mbox{Im}\, z/|z|^2$, for a complex number $z$, one realizes
that this simply imposes:
\be
f_{k_0} = -\frac{1}{u(k_0) + i k_0}
\label{eq:u}
\ee
where $u(k_0)$ is at this stage an arbitrary {\sl real} function. The scattering amplitude takes
its maximal modulus when $u$ is negligible as compared to $k_0$, this constitutes
the unitary limit
\be
f_{k_0} \simeq -\frac{1}{i k_0}.
\ee

A last point concerns the two-body bound states: their eigenergies correspond to
poles of the resolvent $G$, which are also poles of the $T$ matrix.
A simple way to find the dimer eigenenergies is therefore to look for poles
in the analytic continuation of the scattering amplitude to negative energies,
setting $k_0=i q_0$, $q_0>0$, so that $E=-\hbar^2 q_0^2/m <0$.
Note that the determination $+i q_0$ is chosen for $k_0=\sqrt{m E}/\hbar$ so that
the matrix elements of $G_0$, see Eq.(\ref{eq:green}), assume the correct value
(that tends to zero at infinity). 
Furthermore one can easily have access to the asymptotic
behavior of the dimer wavefunctions $\phi_j$, 
including the correct normalization factor.
One uses the closure relation
\be
\int \frac{d^3k_0}{(2\pi)^3}  |\phi_{\vec{k}_0}\rangle\langle \phi_{\vec{k}_0}|+
\sum_j |\phi_j\rangle \langle \phi_j| = I
\ee
where $I$ is the identity operator.
One takes the matrix elements of this identity between $\langle\vec{r}\,|$ and
$|\vec{r}\,'\rangle$, for large values of $r$ and $r'$ so that the asymptotic
expression Eq.(\ref{eq:boundary3}) may be used. After angular average (for a $s$-wave
scattering), one gets factors $e^{\pm ik_0(r+r')}$ and $e^{ik_0(r-r')}$.
Using the optical theorem one shows that the coefficient of $e^{ik_0(r-r')}$
vanishes. Assuming that $u(k)$ is an {\sl even} analytical function of $k$
one can extend the integral over $k_0$ to $]-\infty,+\infty[$, setting
$f_{k_0}^*=f_{-k_0}$, and use a contour integral technique by closing the contour
with an infinite half-circle in the $\mbox{Im}\, z>0$ part of the complex plane.
From Eq.(\ref{eq:u}) the residue of $f_k$ in the pole $k=i q_j$ is obtained
in terms of $u'(i q_j)$. We finally obtain the large $r$ behavior of any
$s$-wave correctly normalized dimer wavefunction:
\be
\phi_j(r) \sim \left(\frac{q_j}{1-i u'(i q_j)}\right)^{1/2}
\frac{e^{-q_j r}}{\sqrt{2\pi}r}.
\label{eq:dimer_asymp}
\ee

\subsection{Effective range expansion and various physical regimes}

For the quantum gases, it is very convenient to introduce the usual low-$k$ expansion
of the scattering amplitude:
\be
f_{k_0} = -\frac{1}{a^{-1} + i k_0 -k_0^2 r_e/2+ \ldots}
\label{eq:fk0}
\ee
where the parameter $a$ is called the scattering length,
the parameter $r_e$ is called the effective range of the interaction,
not to be confused with the true range $b$. As discussed
below, this expansion is interesting if the
omitted terms $\ldots$ in the denominator of Eq.(\ref{eq:fk0}) 
are indeed negligible when Eq.(\ref{eq:nice_regime}) is satisfied.

We take the opportunity to point out that the general allowed values for the
effective range can go from $-\infty$ to $+\infty$. This can be demonstrated on a specific
example, the square well potential, with $V(r)=-V_0$ for $r<b$, $V(r)=0$ otherwise,
the true range $b$ being fixed and $V_0$ being a variable positive quantity. Setting $\kappa_0=(m V_0/\hbar^2)^{1/2}$,
we obtain the exact formulas:
\bea
a &=& b -\frac{\tan \kappa_0 b}{\kappa_0} \\
r_e &=& b- \frac{b^3}{3 a^2} - \frac{1}{\kappa_0^2 a}.
\eea
This easily shows that for all {\it non-zero} values of $\kappa_0$ such that the
scattering length vanishes, the effective range $r_e$ tends to $-\infty$.
When $\kappa_0$ tends to zero, the scattering length tends to zero (more precisely, $0^-$), quadratically
in $\kappa_0^2$, as is evident also from the Born approximation; in the expression for $r_e$, 
the last two terms diverge to infinity, with opposite sign; a systematic series expansion in powers
of $\kappa_0 b$ leads to
\be
r_e = -\frac{2 b^2}{5a} + \frac{176}{175} b + O(\kappa_0^2 b^3) \ \ \ \ \mbox{for} \ \kappa_0\rightarrow 0^+.
\label{eq:reinf}
\ee
One then sees that $r_e$ tends to $+\infty$ for $\kappa_0\rightarrow 0^+$ since $a\to 0^-$ in this limit.
Note that the leading term in Eq.(\ref{eq:reinf}) can also be obtained in the Born approximation.

As already mentioned,
a very favorable case is when the $\ldots$ in the denominator of Eq.(\ref{eq:fk0})
is negligible as compared to the sum of the first three terms, in the low
momentum regime $k_0 b \ll 1$: one can then characterize the true interaction
by only two parameters, the scattering length and the effective range.
In what follows, it is implicitly assumed that this favorable case is obtained,
otherwise more realistic modeling of the interaction should be performed.
One can then identify interesting limiting cases.

\noindent {\it The zero-range limit:}
This is a limit where 
the only relevant parameter of the true interaction
is the scattering length $a$, i.e.\ we shall assume that the typical momentum
$k_{\rm typ}$ satisfies
\be
k^2_{\rm typ} |r_e| \ll |a^{-1}+i k_{\rm typ}|,
\label{eq:zero_range}
\ee
where $r_e$ is the effective range of the true interaction.
In practice, in present experiments and for typical atomic densities, 
this is true for the so-called broad Feshbach resonances,
where $r_e \sim b$: the condition Eq.(\ref{eq:zero_range}) is then a direct consequence
of Eq.(\ref{eq:nice_regime}). For narrow Feshbach resonances, $|r_e|$ can be considerably
larger than $b$ and Eq.(\ref{eq:zero_range}) may be violated for typical densities.

What type of model can be chosen in this zero-range limit ?
A natural idea is to use a strictly zero-range model, having both a vanishing
true range and effective range, and characterized by the scattering
length as the unique parameter. This `ideal world' idea corresponds to the Bethe-Peierls
model of subsection \ref{subsec:Bethe-Peierls}, where the interaction is replaced
by contact conditions. 

In practice, the contact conditions are not always convenient to implement, in an approximate
or numerical calculation. It is therefore also useful to introduce a finite range
model, usually with a true range of the order of its effective range $r_e^{\rm model}$.
One then usually puts the model range to zero at the end of the calculation. Note that
this is not always strictly speaking necessary: it may be sufficient that the model
is also in the zero-range limit,
\be
k^2_{\rm typ} |r_e^{\rm model}| \ll |a^{-1}+i k_{\rm typ}|.
\label{eq:zero_range_for_model}
\ee

In section \ref{sec:BCS} on BCS theory 
we shall restrict to the zero-range limit.
We shall then approximate the true interaction by an on-site interaction in
a Hubbard lattice model, see
subsection \ref{subsec:lat_mod}: the effective range of the model is then of the order
of the lattice spacing, and one may easily take the limit of a vanishing lattice spacing
in the BCS theory.

\noindent{\it The unitary limit:}
What is the condition to reach the unitary limit in a gas, in a regime where
the low-$k$ expansion holds ?  One should have
\be
|a^{-1}|, k_{\rm typ}^2 |r_e| \ll k_{\rm typ}.
\ee
The condition on $a$ is in principle possible
to fulfill, as $a$ diverges close to a Feshbach resonance.
The condition on $r_e$ shows that  the zero-range limit Eq.(\ref{eq:zero_range}) 
is a necessary condition to reach the unitary limit. This is why the broad Feshbach
resonances are favored in present experiments (over the narrow ones) to reach the strongly
interacting regime. To make it simple, one may say that the unitary limit
is simply the zero-range limit with infinite scattering length.

\subsection{A two-channel model}

\begin{figure}[htb]
\centerline{
\includegraphics[width=11cm,clip=]{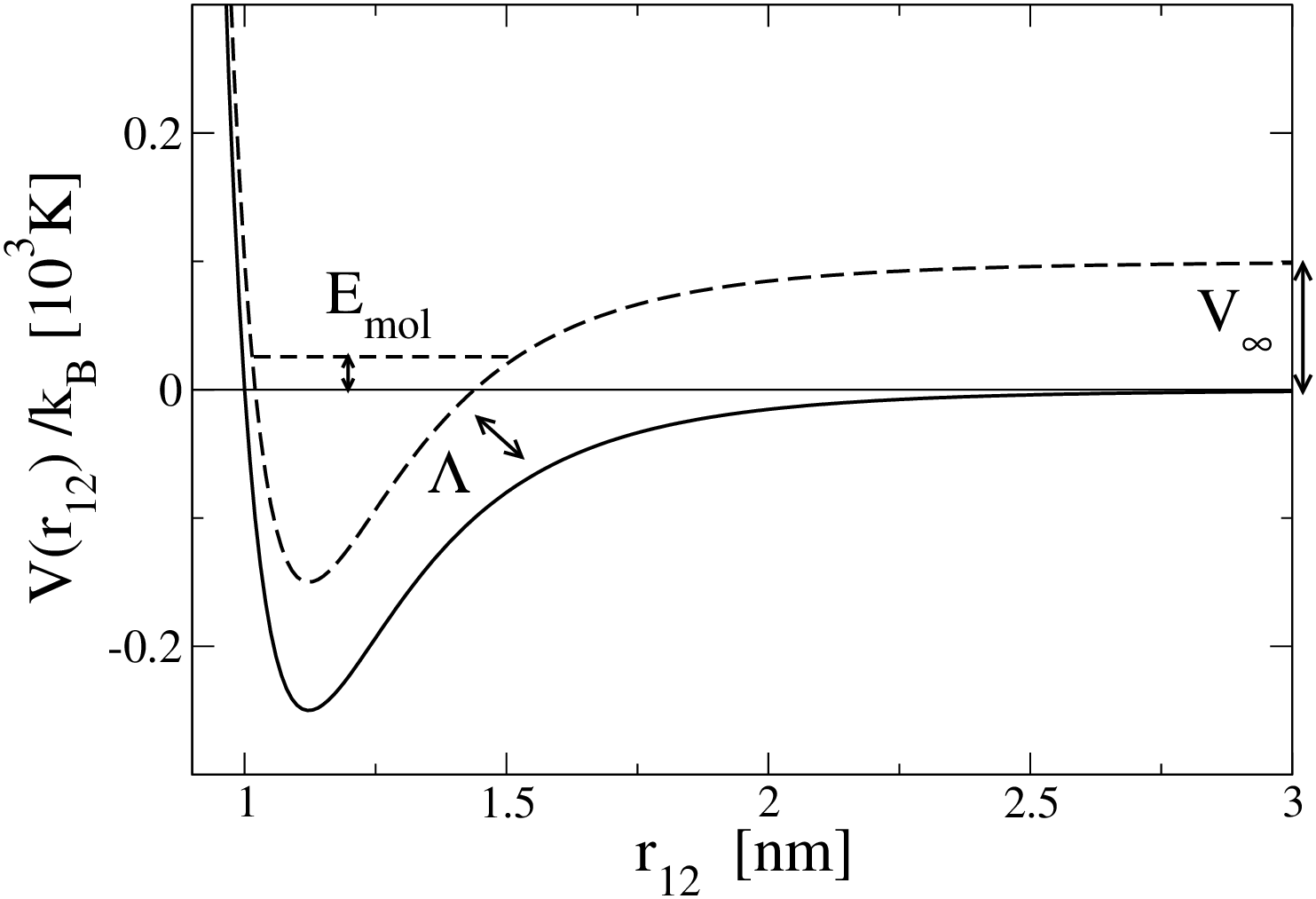}}
\caption{Oversimplified view of a magnetically induced Feshbach resonance.
The interaction between two atoms is described by two Born-Oppenheimer curves.
Solid line: open channel potential curve. Dashed line: closed channel potential curve.
When one neglects the coupling $\Lambda$ between the two curves, the closed channel
has a molecular state of energy $E_{\rm mol}$ with respect to the dissociation
limit of the open channel. Note that the energy spacing between the solid curve 
and the dashed curve was greatly exaggerated, for clarity.}
\label{fig:feshbach}
\end{figure}

We have the oversimplified view of a magnetically induced Feshbach resonance,
see Fig.~\ref{fig:feshbach}:
two atoms interact {\sl via} two potential curves, $V_{\rm open}(r)$ and
$V_{\rm closed}(r)$. 
At short distances these two curves correspond to the electronic
spin singlet and triplet, respectively, for atoms of electronic spin 1/2,
so that they experience different shifts in presence of an external magnetic field.
At large distances, $V_{\rm open}(r)$ tends to zero whereas 
$V_{\rm closed}(r)$ tends to a strictly positive value $V_{\infty}$.

The atoms are supposed to come from infinity in the internal state corresponding
to the curve $V_{\rm open}(r)$, the so-called open channel. 
The atoms are ultracold and have a low kinetic energy,
\be
E \ll V_{\infty}.
\ee
In real life, the two curves are actually weakly coupled. 
Due to this coupling, the atoms can have access to the internal state corresponding
to the curve $V_{\rm closed}(r)$, but only at short distances; at long distances, 
the external atomic wavefunction in this so-called closed channel
will be an evanescent wave that decays exponentially with $r$ since $E<V_{\infty}$.

Now assume that, in the absence of coupling between the channels, the closed channel
supports a bound state of energy $E_{\rm mol}$, denoted in this text as {\sl the molecular state}.
Assume also that, by applying a judicious magnetic field, one sets the energy of
this molecular state close to zero, that is to the dissociation limit of the
open channel. In this case one may expect that the scattering amplitude of
two atoms may be strongly affected, by a resonance effect, when a weak coupling
exists between the two channels.

A quite quantitative discussion is obtained thanks to a very simple two-channel model.
This model is the most realistic of the models presented here, and we shall
consider it as a `reality' against which the other models should be confronted.
Assuming that the particles are free in the open channel, and can populate only
the molecular state in the closed channel, the only non-trivial ingredient is
a coupling between the molecule with center of mass momentum
zero, $|\mbox{mol}\rangle$,
and a pair of atoms of opposite momenta and spin states $||\vec{k}\rangle$, 
with
\be
||\vec{k}\rangle \equiv  a_{\vec{k}\uparrow}^\dagger a_{-\vec{k}\downarrow}^\dagger
|0\rangle,
\ee
where the creation and annihilation operators obey the free space anticommutation relation
\be
\{a_{\vec{k}\sigma},a^\dagger_{\vec{k}'\sigma'}\}
= (2\pi)^3\delta(\vec{k}-\vec{k}')
\delta_{\sigma \sigma'}.
\ee
This coupling in the momentum representation is defined as
\bea
V|\mbox{mol}\rangle &=& \int \frac{d^3k}{(2\pi)^3} \Lambda \chi(k) ||\vec{k}\rangle  \\
V||\vec{k}\rangle &=& \Lambda \chi(k) |\mbox{mol}\rangle.
\eea
We have introduced a real coupling constant $\Lambda$, and a real and isotropic
cut-off function $\chi(k)$, equal to $1$ for $k=0$
and to 0 for $k=\infty$, to avoid ultraviolet divergencies. It is convenient to take
\be
\chi(k) = e^{-k^2\sigma^2/2}
\label{eq:gauss}
\ee
where the length $\sigma$ is {\sl a priori} of the order of the true potential range $b$.
The two-channel model is thus defined by the three parameters
$E_{\rm mol}, \Lambda$ and $\sigma$. It scatters in the $s$-wave channel only,
since $\chi$ is isotropic.

An eigenstate of energy $E$ of the two-channel Hamiltonian is of the form
\be
|\phi\rangle =  \beta |\mbox{mol}\rangle+
\int \frac{d^3k}{(2\pi)^3}\, \alpha(\vec{k}\,) ||\vec{k}\rangle.
\ee
Including the molecular state energy and the kinetic energy of a pair of atoms, the
stationary Schr\"odinger equation leads to
\bea
\label{eq:alpha}
\frac{\hbar^2 k^2}{m} \alpha(\vec{k}\,)  + \Lambda \chi(k) \beta &=&  E \alpha(\vec{k}\,) \\
E_{\rm mol}\, \beta + \int \frac{d^3k}{(2\pi)^3} \Lambda \chi(k) \alpha(\vec{k}\,) &=&
E \beta.
\label{eq:beta}
\eea
We restrict to a scattering problem, $E>0$.
Instructed by the previous discussion around Eq.(\ref{eq:asymptk}), we correctly
solve for $\alpha$ in terms of $\beta$ in Eq.(\ref{eq:alpha}):
\be
\alpha(\vec{k}\,) = (2\pi)^3\delta(\vec{k}-\vec{k}_0) + \frac{\Lambda \chi(k)}{E+i0^+ -\hbar^2 k^2/m}\beta,
\ee
where $\vec{k}_0$ is the wavevector of the incident wave, and $E=\hbar^2 k_0^2/m$.
Insertion of this solution in Eq.(\ref{eq:beta}) leads to a closed
equation for $\beta$ that is readily solved.
From Eq.(\ref{eq:asymptk}) and  Eq.(\ref{eq:f_vs_T}) we obtain the scattering amplitude
\be
f_{k_0} = \frac{m}{4\pi\hbar^2} 
\frac{\Lambda^2 \chi^2(k_0)}{E_{\rm mol}-E +\int \frac{d^3k}{(2\pi)^3} 
\frac{\Lambda^2 \chi^2(k)}{E+i0^+-\hbar^2 k^2/m}}.
\ee

Let us work a little bit on this scattering amplitude.
First, using the identity
\be
\frac{1}{X+i0^+} = {\mathcal P}\frac{1}{X} -i\pi \delta(X)
\ee
where $\mathcal P$ denotes the principal value, one can check that
$f_{k_0}$ is indeed of the form Eq.(\ref{eq:u}), so that it obeys the optical theorem.
Second, the scattering length is readily obtained by taking the $k_0\rightarrow 0$
limit:
\be
-a^{-1} = \frac{4\pi\hbar^2}{m} \left[ \frac{E_{\rm mol}}{\Lambda^2}
-\int \frac{d^3k}{(2\pi)^3}\, \frac{m \chi^2(k)}{\hbar^2 k^2}
\right].
\label{eq:sl2c}
\ee
This shows that the location of the resonance (where $a^{-1}=0$)
is not $E_{\rm mol}=0$  but a positive value $E_{\rm mol}^0$, this shift
being due to the coupling between the two channels. For $E_{\rm mol}$ larger than this
value, $a<0$. For $E_{\rm mol}$ below this value, $a>0$.

Another general result, not specific to a Gaussian choice for $\chi(k)$
(but assuming that $\chi(k)$ varies quadratically around $k=0$), concerns the zero
range limit $\sigma\rightarrow 0$ for fixed values of $k_0$, $\Lambda$ and
the scattering length $a$ (note that the molecular energy is then not fixed):
\be
f_{\rm k_0} \rightarrow -\frac{1}{a^{-1}+ik_0 + \frac{4\pi \hbar^2}{m} \,
\frac{E}{\Lambda^2}}.
\label{eq:lim}
\ee
This can be seen formally as resulting from the identity 
$\int_0^{+\infty} dX\, \mathcal{P}1/(X^2-1)=0$.
This shows that the two-channel model has a well defined limit, in the zero true-range
limit, characterized only by the scattering length and
by a non-zero and negative effective range,
\be
r_e^0 = -\frac{8\pi \hbar^4}{m^2\Lambda^2}.
\ee
This limit may be described
by generalized Bethe-Peierls contact conditions \cite{Petrov}.
This illustrates dramatically
the difference between the true range and the effective range of an interaction.
This also shows that, in this limit, the two-channel model can support a weakly
bound state, that is a dimer with a size much larger than the true range $\sigma$.
Setting $k_0=i q_0$,  one finds that the right hand side of Eq.(\ref{eq:lim}) has
a pole with $q_0$ real and positive if and only if $a>0$:
\be
q_0 = \frac{1-\sqrt{1-2r_e^0/a}}{r_e^0}.
\ee

Finally, an exact calculation of the scattering amplitude for the Gaussian choice
Eq.(\ref{eq:gauss}) is possible: the trick put forward by Mattia
Jona-Lasinio is to calculate the analytic continuation
to imaginary $k_0=i q_0$, so that one no longer needs to introduce the principal part
distribution. One gets
\be
(f_{i q_0})^{-1} = \left[ -a^{-1}-\frac{1}{2} r_e^0 q_0^2\right] e^{-q_0^2\sigma^2}
+q_0\, \mbox{erfc}(q_0\sigma),
\ee
where erfc is the complementary error function, that tends to unity in zero,
$\mathrm{erfc}(x)=1-2\pi^{-1/2}x + O(x^3)$.
The scattering length is given by
\be
-a^{-1} = \frac{4\pi\hbar^2 E_{\rm mol}}{m\Lambda^2}-\frac{1}{\sigma \sqrt{\pi}}.
\ee
The calculation of the effective range for a finite $\sigma$
is then straightforward:
\be
r_e = r_e^0 + \frac{4\sigma}{\sqrt{\pi}} 
\left(1-\frac{\sqrt{\pi} \sigma}{2a}\right).
\ee

We shall then classify the Feshbach resonances as follows: for $a=\infty$, the
ones with $r_e \simeq r_e^0 < 0$, much larger than the true range
in absolute value, constitute {\sl narrow} Feshbach resonances.
The ones with $r_e \simeq \sigma$ constitute {\sl broad} Feshbach resonances.

This classification is not only theoretical but also reflects the magnetic
width $\Delta B$ of the resonance. If one assumes that $E_{\rm mol}$ is an affine
function of the magnetic field $B$, with a slope $-\mu$, one finds from Eq.(\ref{eq:sl2c})
that the scattering length $a$ of the two-channel model varies with $B$ as
\be
a = \frac{a_{\rm bg} \Delta B}{B_0-B},
\label{eq:aB}
\ee
where $B_0$ is the magnetic field value of the resonance center,
and we have singled out the value $a_{\rm bg}$ of the background scattering
length (that is the value of the scattering length of the true interaction far from
the resonance), motivated by the fact that $a=a_{\rm bg} [1+\Delta B/(B_0-B)]$
in a more complete theory including the direct interaction in the open channel
\cite{Moerdijk}.
This defines the width $\Delta B$, such that
\be
\mu \Delta B = \frac{2\hbar^2}{m a_{\rm bg} r_e^0}.
\ee
So, when $a_{\rm bg}\sim \sigma\sim b$, comparing $\mu \Delta B$ to the natural
energy scale $\hbar^2/m b^2$ amounts to comparing $|r_e^0|$ to $b$.

\subsection{The Bethe-Peierls model}
\label{subsec:Bethe-Peierls}

In this model, the true interaction is replaced by contact conditions on the
wavefunction. For two particles in free space, 
in the center of mass frame, the contact condition
is that there exists a constant $A$ such that
\be
\phi(\vec{r}\,) = A\left[r^{-1}-a^{-1}\right]+ O(r)
\label{eq:bethe}
\ee
in the limit $r\rightarrow 0$. For $r>0$ the wavefunction is an eigenstate of
the free Hamiltonian:
\be
E \phi(\vec{r}\,) = -\frac{\hbar^2}{m} \Delta_{\vec{r}}\,\phi(\vec{r}\,).
\label{eq:free}
\ee
In mathematical terms, this defines the domain of the Hamiltonian. The Hilbert
space remains however the same, with the usual scalar product, because a $1/r$
divergence in 3D is square integrable.

An equation valid for all values of $r$ can be written using the theory of distributions:
\be
E \phi(\vec{r}\,) = -\frac{\hbar^2}{m} \Delta_{\vec{r}}\phi(\vec{r}\,) + 
\frac{4\pi\hbar^2a}{m}\,\phi_{\rm reg}\, \delta(\vec{r}\,) 
\ee
where the so-called regular part of $\phi$ is
\be
\phi_{\rm reg} =-\frac{A}{a} =\partial_r\left[r\phi(\vec{r}\,)\right]_{r=0} =
-a^{-1}\lim_{r\rightarrow 0} r\phi(\vec{r}\,). 
\ee
This establishes the equivalence with the regularized pseudo-potential model
\cite{Weisskopf,VarennaJean,Houches99,CdF}. The last form of $\phi_{\rm reg}$ is convenient to use
in the limit $a=\infty$, since $a^{-1}$ simplifies with the factor $a$
in front of the Dirac $\delta(\vec{r}\,)$.

The solution of Eq.(\ref{eq:free}) with the boundary conditions Eq.(\ref{eq:boundary3})
and the contact condition is simply:
\be
\phi(\vec{r}\,) = e^{i \vec{k}_0\cdot\vec{r}}
+f_{k_0} \frac{e^{i k_0 r}}{r}
\ee
with the scattering amplitude
\be
f_{k_0} = -\frac{1}{a^{-1}+i k }.
\ee
This shows that the model has both a zero true range and zero effective range.
It may be applied to mimic the effect of the true interaction potential
when the condition Eq.(\ref{eq:zero_range}) is satisfied.

Following the discussion above Eq.(\ref{eq:dimer_asymp}), one finds that the model
supports a dimer if and only if $a>0$. The pole of $f_{i q_0}$ is then $q_0=1/a$,
resulting in a dimer energy
\be
E_0 = -\frac{\hbar^2}{m a^2}.
\ee
Since the model has a zero true range, it turns out that the dimer wavefunction 
coincides everywhere with its asymptotic form Eq.(\ref{eq:dimer_asymp}),
\be
\phi_0(r) = \frac{1}{\sqrt{2\pi a}} \frac{e^{-r/a}}{r}.
\label{eq:phi0}
\ee

The advantage of the Bethe-Peierls model for the $N$-body problem
is that it introduces the minimal number
of parameters (the scattering length). In the unitary limit,
it is a zero-parameter model, which is ideal to describe universal states.
This advantage actually really exists if one restricts to analytical solutions
of the model (numerical solutions tend to introduce extra parameters like
a discretization step or a momentum cut-off).
Analytical solutions in free space are available up to $N=3$
in the unitary limit \cite{Efimov1,Efimov2}. 
For particles trapped in harmonic potentials,
analytical solutions are available for two particles at any $a$ for isotropic traps
\cite{Wilkens} and cylindrically symmetric traps \cite{Calarco}.
For three particles in an isotropic harmonic trap, the problem was solved
analytically only in the unitary limit $a=\infty$ \cite{Pethick,Werner}.
In the many-body problem, for the universal unitary gas in an isotropic
harmonic trap, a scaling solution can be found for the many-body wavefunction
in the time dependent case \cite{Castin} and an exact mapping to free space zero energy
eigenstates can be constructed \cite{Tan,Werner_v1,Werner2}.

It is worth mentioning that, for $N=3$ {\it bosons}, the Bethe-Peirls
model does not define a self-adjoint operator and has to be supplemented
by extra contact conditions; this is related to the Efimov
phenomenon \cite{Efimov1,Efimov2}; the problem of course persists for
$N>3$, making the model non satisfactory. Fortunately this
lecture is devoted to equal mass spin 1/2
fermions, for which no Efimov effect appears; the corresponding
unitary gas is then called {\it universal}, to emphasize this aspect
that no extra parameter has to be added to the Bethe-Peirls model.
Note that this absence of Efimov effect is not simply due to
the fact that the Pauli principle prevents one
from having three fermions of spin 1/2 in the same point of space:
if the spin up and spin down fermions have widely different masses,
an Efimov effect may appear \cite{Petrov_fermions}.

We take the opportunity to mention another trap to avoid, even for fermions.
With the formulation of the model in terms of binary interactions
with the pseudo-potential $V_{\rm PP}(\vec{r}\,) = g \delta(\vec{r}\,)
\partial_r(r\,\cdot)$, where $g=4\pi\hbar^2a/m$,
it is easy to `forget' that contact conditions are
imposed on the many-body wavefunction. One may then be tempted to
perform a variational calculation with a $N$-body trial wavefunction
which does not satisfy the contact conditions Eq.(\ref{eq:bethe}), that is which is not
in the domain of the Hamiltonian \cite{math_domain}.
Whereas in the low $a$-limit the corresponding prediction for the energy
may make sense as a {\it perturbative} prediction, it may become totally wrong
in the large $a$ limit.

To illustrate this statement, let us consider two identical particles in a harmonic isotropic
trap interacting {\it via} the pseudo-potential. After separation of the center of mass
coordinates, one is left with the following Hamiltonian for the relative motion
\be
H = -\frac{\hbar^2}{2\mu}\Delta_{\vec{r}}
+\frac{1}{2} \mu \omega^2 r^2 + g \delta(\vec{r}\,) \partial_r(r\,\cdot),
\label{eq:do}
\ee
where $\mu=m/2$ is the reduced mass.
Let us take as a trial wavefunction $|\phi_t\rangle$ the Gaussian of the ground state
of the harmonic oscillator, 
\be
\phi_t(r) = \frac{e^{-r^2/(4 a_{\rm ho}^2)}}{(2\pi)^{3/4} a_{\rm ho}^{3/2}},
\ee
where $a_{\rm ho}=\sqrt{\hbar/m\omega}$. This trial wavefunction does not obey
the contact conditions, so it is not in the domain of the Hamiltonian.
It is therefore mathematically incorrect to calculate the mean energy 
of $\phi_t$ by representing the action of $H$ on $\phi_t$ by the
differential operator Eq.(\ref{eq:do}).
One rather has to calculate the
overlap $\langle \psi_n|\phi_t\rangle$ of $|\phi_t\rangle$
with each exact eigenstate $\psi_n$ of $H$ of eigenenergy $E_n$, and
calculate the sum $\sum_n E_n |\langle \psi_n|\phi_t\rangle|^2$.
From \cite{Wilkens} and for a finite $a$
one finds that the overlap squared scales
as $1/n^{3/2}$ whereas the energy $E_n$ scales as $n$, for large $n$.
The correct mean energy of $\phi_t$ is thus infinite.

If one however falls in the trap, one gets the wrong mean energy
\be
E_{\rm wrong} =\langle \phi_t|H|\phi_t \rangle = \frac{3}{2} \hbar \omega +
\left(\frac{2}{\pi}\right)^{1/2} \hbar \omega \frac{a}{a_{\rm ho}}.
\ee
In the low $a$ limit this can be compared to the expansion of the exact minimal
positive eigenenergy \cite{Wilkens}
\be
E_{\rm exact} = \frac{3}{2} \hbar\omega
+ \hbar \omega \left[\left(\frac{2}{\pi}\right)^{1/2} \frac{a}{a_{\rm ho}}
+ \frac{2}{\pi} (1-\ln 2) \left(\frac{a}{a_{\rm ho}}\right)^2+ \ldots\right].
\label{eq:exact_expan}
\ee
One sees that the wrong variational calculation gives the correct 
linear correction in $a$, that is the correct perturbative result.
One may then be tempted to use the wrong variational calculation to predict
the sign of the coefficient of $a^2$; one would naively assume
$E_{\rm exact} \leq E_{\rm wrong}$ and one would wrongly predict
a negative sign for the coefficient of $a^2$ \cite{doubly_naive}.
In the opposite limit of an infinite scattering length, one directly realizes
the absurdity of the assumption $E_{\rm exact} \leq E_{\rm wrong}$: for
$a\to -\infty$, one finds $E_{\rm wrong}\to -\infty$, whereas
$E_{\rm exact}\to \hbar\omega/2$, with the exact wavefunction
$\propto e^{-r^2/4 a_{\rm ho}^2}/r$.

Note that the same pathology occurs for the many-body problem.
Consider for example a two spin-component spatially homogeneous
Fermi gas, with interactions modeled by the pseudo-potential.
If one uses the Hartree-Fock variational ansatz, which is not
in the domain of the Hamiltonian, one correctly obtains, in
the weakly attractive regime $k_F a\to 0^{-}$, the mean field shift
of the ground state energy, but one wrongly predicts a negative sign
for the coefficient of $a^2$ in the low $k_F a$ expansion of the 
ground state energy, by wrongly arguing that the exact ground state
energy has to be below the `variational' Hartree-Fock energy.
As can be checked on the systematic expansion \cite{Heiselberg}
the coefficient of $a^2$ is actually positive, because $11> 2\ln 2$.

One should not leave this section with the impression that any variational
calculation is impossible within the Bethe-Peierls model. To be
safe, one simply has to take a variational ansatz which is in the
domain of the Hamiltonian. An example of
a successful variational calculation is the derivation of the virial
theorem for a unitary gas in a non necessarily isotropic
harmonic trap; this derivation, due to Fr\'ed\'eric Chevy, is reported in
\cite{Werner2}, and uses the fact, for the universal unitary gas $1/a=0$,
that the function of the rescaled coordinates
$\Psi(\vec{r}_1/\lambda,\ldots,\vec{r}_N/\lambda)$, where $\lambda>0$,
is in the domain of the Hamiltonian if the wavefunction
$\Psi(\vec{r}_1,\ldots,\vec{r}_N)$ is in the domain of the Hamiltonian.

\subsection{The lattice model}
\label{subsec:lat_mod}

The last model we consider is at the intersection of two very popular
classes of models in condensed matter physics, the separable potentials
and the lattice models. It is slightly simpler than the two-channel model, but
it does not apply to narrow Feshbach resonances in a situation where the effective range
term cannot be neglected in the scattering amplitude.
It is more tractable than the Bethe-Peierls model for variational
treatments (like the BCS method, see section \ref{sec:BCS}) or for exact numerical treatments
since the domain of the Hamiltonian is the one of usual quantum mechanics.
In particular, since it is a sort of Hubbard model, the machinery of quantum Monte Carlo
methods applicable to the Hubbard model \cite{Svistunov,Bulgac,Lee,Juillet} 
could be applied to it.

In this model, the spatial coordinates $\vec{r}$ of the particles are discretized on a cubic
grid of step $l$. As a consequence, the components of the wavevector of a particle have 
a meaning modulo $2\pi/l$ only, since the function $\vec{r}\rightarrow \exp(i\vec{k}\cdot\vec{r}\,)$
defined on the grid is not changed if a component of $\vec{k}$ is shifted by an integer
multiple of $2\pi/l$. We shall therefore restrict the wavevectors to the first Brillouin
zone:
\be
\vec{k} \in {\mathcal D}\equiv \left[-\frac{\pi}{l},\frac{\pi}{l}\right[^3.
\ee
This also shows that the lattice structure in real space provides a cut-off in
momentum space.

The interaction between opposite spin particles takes place when two particles are
at the same lattice site, as in the Hubbard model. In first quantized form, 
it is represented by a discrete delta potential:
\be
V= \frac{g_0}{l^3} \delta_{\vec{r}_1,\vec{r}_2}.
\ee
The coupling constant $g_0$ is a function of the grid spacing $l$. It is adjusted
to reproduce the scattering length of the true interaction.
The scattering amplitude of two atoms on the lattice with vanishing total momentum
is given in \cite{Mora}, and a detailed discussion is presented in \cite{Houches03}.
We give here only the result, generalized to an arbitrary even dispersion relation 
$\vec{k}\rightarrow E_{\vec{k}}$ for the kinetic energy on the lattice:
\be
f_{k_0} = -\frac{m}{4\pi\hbar^2}
\frac{1}{g_0^{-1}-\int_{\mathcal D}\frac{d^3k}{(2\pi)^3}
\frac{1}{E+i0^+-2 E_{\vec{k}}}}.
\ee
Adjusting $g_0$ to recover the correct scattering length then gives
\be
\frac{1}{g_0}= \frac{1}{g} -\int_{\mathcal D} \frac{d^3k}{(2\pi)^3} \,
\frac{1}{2 E_{\vec{k}}},
\label{eq:g0_gen}
\ee
with $g=4\pi\hbar^2 a/m$.
This formula is reminiscent of the technique of renormalization of the coupling
constant. 

A natural case to consider is the one of the usual parabolic dispersion relation,
$E_{\vec k}=\hbar^2 k^2/2m$. A more explicit form of Eq.(\ref{eq:g0_gen}) is then
\be
g_0= \frac{4\pi\hbar^2 a/m}{1-Ka/l}
\label{eq:g0_exp}
\ee
with a numerical constant given by
\be
K=\frac{12}{\pi}\int_{0}^{\pi/4} 
d\theta\, \ln(1+1/\cos^{2}\theta)
= 2.442\ 749\ 607\ 806\ 335\ldots,
\ee
and that may be expressed analytically in terms of the dilog special function.
The effective range of the model is easily calculated with the complexification
technique $k_0=i q_0$; it is positive and proportional to $l$:
\be
r_e^{\rm model} = \frac{l}{2\pi^3} \int_{\vec{k}\notin [-1/2,1/2[^3} \frac{d^3k}{k^4}=
0.336\ 868 \ 47\ldots l.
\ee

In the $l\rightarrow 0$ limit, for a fixed $a$ (not fixed $g_0$)
the lattice model then reproduces the scattering amplitude
of the Bethe-Peierls model, 
\be
\lim_{l\rightarrow 0} f_{k_0} = -\frac{1}{a^{-1}+ik_0},
\ee
and admits a dimer for $a>0$.
Note that $g_0$ indeed assumes negative values that tend to $0^-$ linearly with $l$ in
this limit: the interaction is attractive, 
whatever the sign of the scattering length
\footnote{More generally, the lattice model admits a dimer if $g_0 <g_0^c$, where
$g_0^c$ is the $a\rightarrow \infty$ limit of Eq.(\ref{eq:g0_exp}). This 
condition is equivalent to $g_0<0$ and $a>0$.}.

\noindent {\sl Studying the weakly interacting regime:}
In the opposite case of $l \gg |a|$, the bare coupling constant
is
\be
g_0 \simeq \frac{4\pi \hbar^2 a}{m},
\ee
so that $g_0$ has the sign of the scattering length.
For $a>0$, the model interaction is repulsive, for $a<0$ it is attractive.
In both cases, it does not support any two-body bound state.
Its scattering length can be calculated in the Born approximation, since the term in $a$ in the
denominator of Eq.(\ref{eq:g0_exp}) is small as compared to unity.
This makes it an appropriate model to study
the weakly interacting regime, as was done for Bose gases in \cite{Mora}: the Hartree-Fock
theory, for example, which relies on the Born approximation, may be applied,
and its accuracy may be supplemented by keeping higher order terms in
a perturbative expansion of the interaction.
In the context of fermions, it may be used in the weakly attractive or weakly repulsive
regime; in the latter case, it may be an alternative to the celebrated hard sphere model
(with hard spheres of diameter $a>0$).

It is instructive to apply the philosophy around Eq.(\ref{eq:philo}) to understand
why the lattice model with $l\gg |a|$ is restricted to the weakly interacting regime,
for a quantum degenerate gas of atoms with $k_{\rm typ}= k_F$.
\begin{itemize}
\item In an effort to have $f_k^{\rm model}\simeq f_k$, one first adjusts the scattering length
of the model to the correct value $a$ by tuning $g_0$. 
\item If one further adjusts the lattice
spacing $l$ to have $r_e^{\rm model}=r_e$: this assumes a broad Feshbach resonance with
$r_e\sim b$, since the narrow Feshbach resonance has $r_e<0$; one then finds $l\sim b$;
since $k_{\rm typ} b \ll 1$ (otherwise microscopic details of the interaction will come in,
and the $\ldots$ in Eq.(\ref{eq:fk0}) is not negligible), having $|a| \ll l$ leads to
$k_F |a| \ll 1$, which is the weakly interacting regime.
\item Alternatively, one may consider the more favorable case of a true interaction
being in the zero-range limit Eq.(\ref{eq:zero_range}). 
The model should also be in the zero-range limit Eq.(\ref{eq:zero_range_for_model}), 
which can be written here
\be
k_F l \ll \left|\frac{1}{k_F a} + i\right|,
\ee
and this is sufficient ($r_e^{\rm model}\propto l$ may differ a lot from $r_e$).
Since we wish to have $|a|\ll l$, the above equation leads to $k_F |a|\ll |(k_F a)^{-1}+i|$,
a condition that may be satisfied only in the weakly interacting limit $k_F |a| \ll 1$.
\item The same result may be obtained in the following
much faster way: since we are dealing here with
fermions, there will be at most one fermion per spin component per lattice site, so that
$k_F l< (6\pi^2)^{1/3}$; the condition $|a|\ll l$ then immediately
leads to $k_F|a|<1$. This reasoning is however specific to the lattice model,
whereas the previous reasoning is easily generalized to the hard sphere model (where
$r_e^{\rm model}\propto a$).
\end{itemize}

\noindent {\sl The `true' Hubbard model}:
To be complete, we consider
a second case for the dispersion relation, the one of the true Hubbard model: this makes
the link with condensed matter physics, and this also shows that
a universal quantum gas may be described
by an attractive Hubbard model in the limit of a vanishing filling factor.
The Hubbard model is usually presented in terms of the tunneling
amplitude between neighboring sites, here $t=-\hbar^2/2ml^2$, and of the on-site
interaction $U=g_0/l^3$. The dispersion relation is then
\be
E_{\vec{k}} = \frac{\hbar^2}{ml^2} \sum_{\alpha} \left[1-\cos(k_\alpha l)\right]
\label{eq:disphub}
\ee
where the summation is over the three dimensions of space. Close to $\vec{k}=\vec{0}$,
the Hubbard dispersion relation by construction reproduces the free space
one $\hbar^2 k^2/2m$. The explicit version of Eq.(\ref{eq:g0_gen}) is 
obtained from Eq.(\ref{eq:g0_exp}) by replacing the numerical constant 
$K$ by $K^{\rm Hub}=3.175 911 \ldots$. At the unitary limit this leads to 
$U/t=7.913 552\ldots$, which corresponds to an attractive Hubbard model
since $t<0$, lending itself to a Quantum Monte Carlo analysis for equal
spin populations with no sign problem \cite{Svistunov,Bulgac,Lee}.
We have also calculated the effective range of the Hubbard model, which remarkably 
is negative:
\be
r_e^{\rm Hub} \simeq -0.3056 l.
\ee

\subsection{Application of Bethe-Peierls to a toy model: two macroscopic branches}

In this subsection, we move to the problem of $N$ interacting fermions,
$N/2$ with spin component $\uparrow$ and $N/2$ with spin component $\downarrow$.
In order to have a global view of what the BCS theory will be useful for, it is instructive to 
start with a purely qualitative model \cite{boite}.

Consider a matter wave of isotropic wavefunction $\phi(r)$
in a hard wall spherical cavity of radius $R$, in presence
of a point-like scatterer of fixed position in the center of the cavity.
The effect of the point-like scatterer is characterized by the scattering length
$a$ and is treated by imposing
the Bethe-Peierls contact condition Eq.(\ref{eq:bethe}).

\begin{figure}[htb]
\begin{center}
\includegraphics[height=4.5cm,clip=]{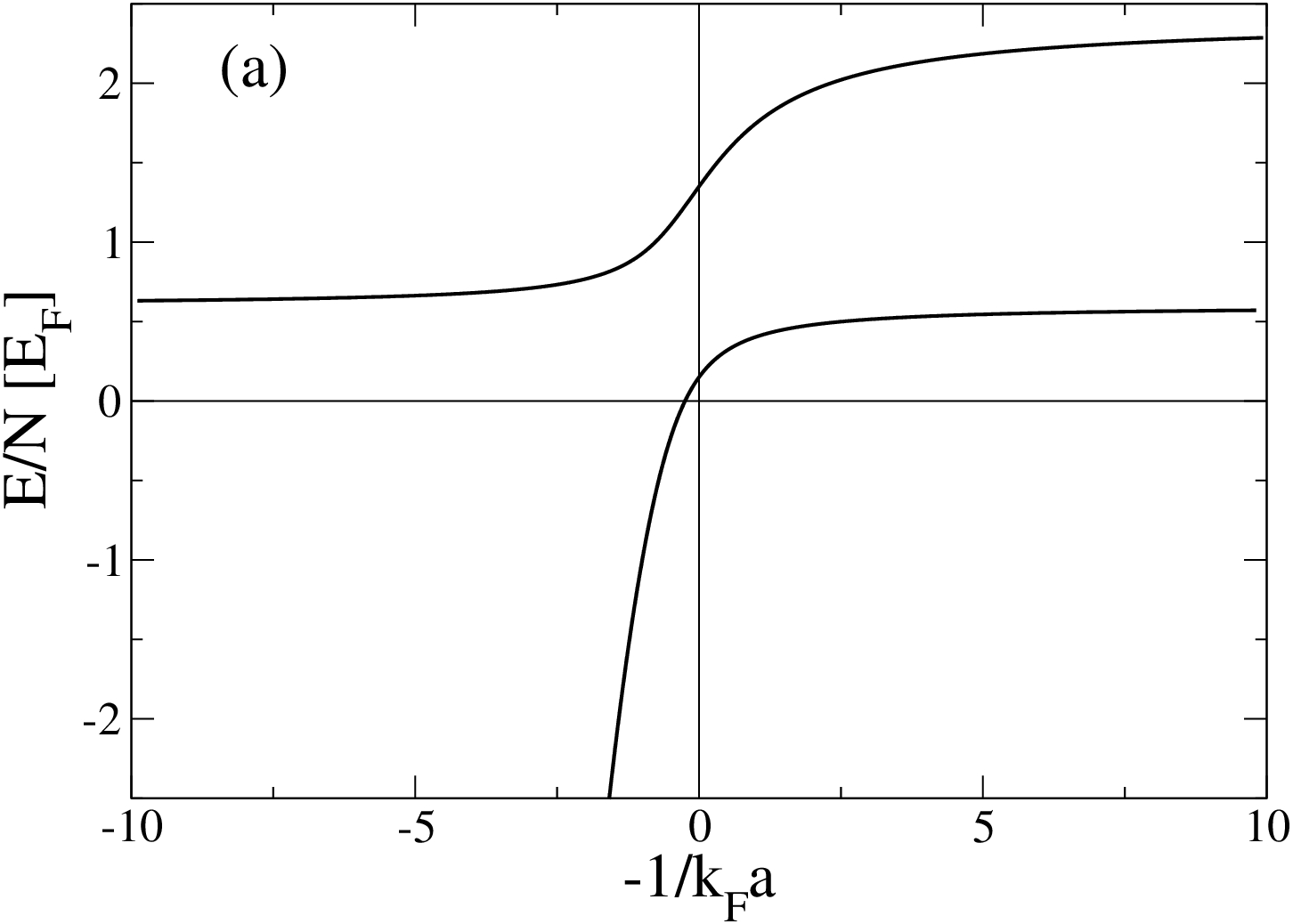}
\includegraphics[height=4.5cm,clip=]{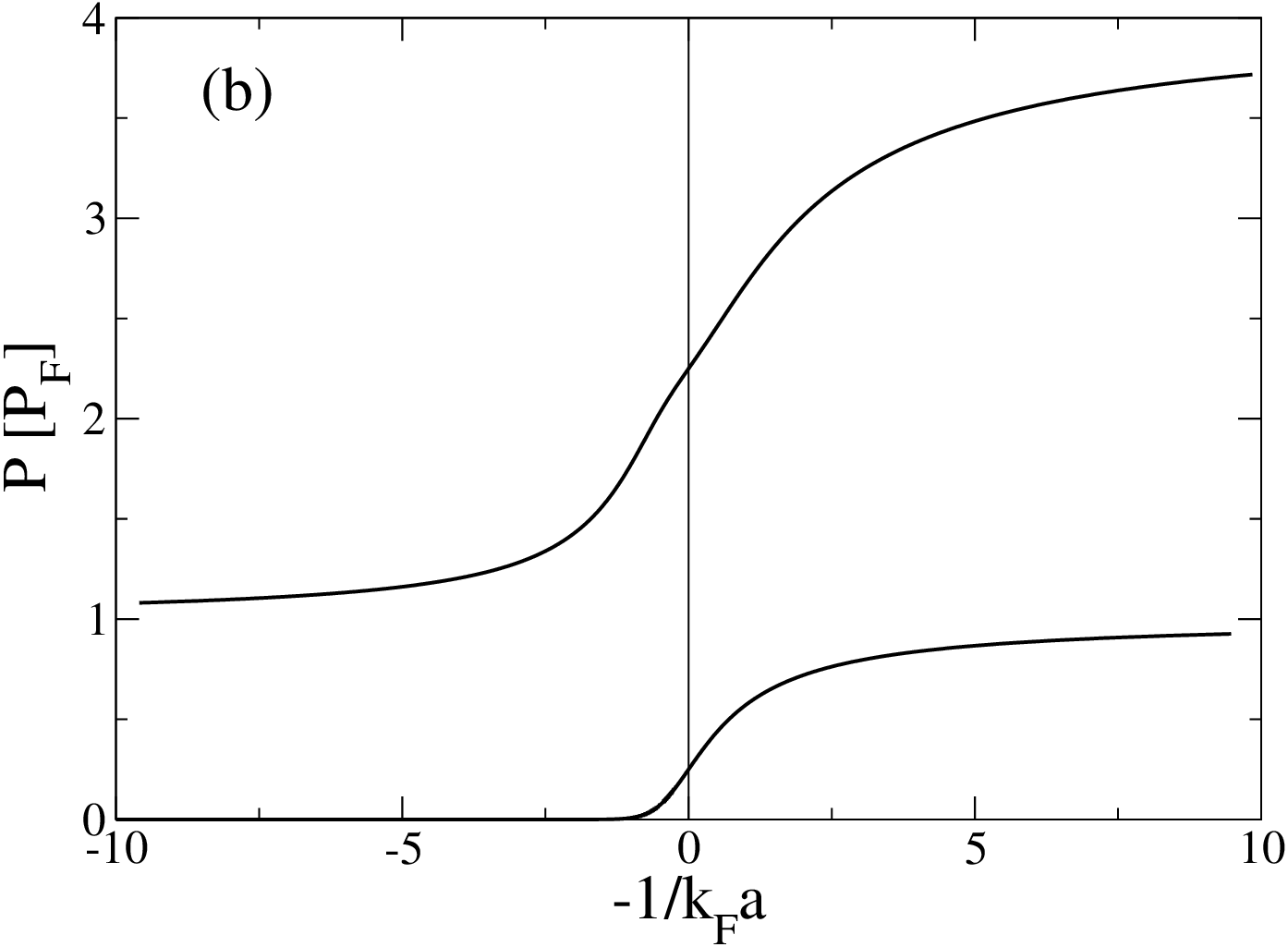}
\end{center}
\caption{In the toy model with a scatterer in a hard wall spherical cavity:
(a) energy per particle and (b) pressure of the gas in the first two branches, 
as functions of $-1/k_F a$. $k_F$, $E_F$ and $P_F$  are respectively the
Fermi wavevector, the Fermi energy and the pressure of the $T=0$ ideal Fermi
gas with the same density as the interacting gas.}
\label{fig:toy}
\end{figure}

What is the link between this model and the many-body problem
of $N$ interacting fermions~? The wavefunction $\phi(r)$ describes    
the relative motion of a, let us say, spin $\uparrow$ atom, with respect
to the nearest spin $\downarrow$ atom modelized by the point-like scatterer.
The cavity represents (i) the interaction
effect of the other $N/2-1$ spin $\downarrow$ atoms and (ii)
 the Pauli blocking effect of the other $N/2-1$ spin $\uparrow$ atoms.
Interaction effect (i): the radius of the cavity should then be of the
order of the mean
interparticle separation in the gas,
\be
R \propto \frac{1}{k_F}.
\label{eq:choixR}
\ee
Pauli blocking effect (ii): in the case $g=0$, the zero point
energy of $\phi$ should be of the order of the Fermi energy so that
one has also the choice (\ref{eq:choixR}).
This explains also the choice of hard walls for the cavity;
in the case of bosons, imposing that the derivative $\partial_r \phi$
vanishes in $r=R$ \cite{Pandha_bosons}, or taking
a cubic cavity with periodic boundary conditions would be appropriate.
Finally, the total energy of the gas is related to $\epsilon$ by
\be
E \propto N \epsilon.
\label{eq:choixE}
\ee
Specific but arbitrary proportionality coefficients are used  in
Eqs.(\ref{eq:choixR},\ref{eq:choixE}) to produce the
Fig.\ref{fig:toy} to come,
as detailed in \cite{boite}.

Let us proceed with the calculation of the eigenenergies of the matter
wave in the cavity.
An eigenmode of the cavity with energy
$\epsilon$ solves Schr\"odinger's equation for $0<r<R$
\be
-\frac{\hbar^2}{m} \Delta \phi(r) = \epsilon \phi(r)
\ee
with the Bethe-Peierls contact condition Eq.(\ref{eq:bethe}). If $\epsilon>0$, we
set $\epsilon=\hbar^2 k^2/m$ and $k$ solves
\be
\tan kR= ka.
\ee
If $\epsilon<0$, we set $\epsilon=-\hbar^2\kappa^2/m$ and $\kappa$
solves
\be
\tanh \kappa R = \kappa a.
\label{eq:tanh}
\ee

In the cavity, there is an infinite number of discrete modes. To each of this mode
we associate a distinct macroscopic state of the gas.
In figure~\ref{fig:toy}
we have plotted the energy per particle and the pressure
$P=-\partial_V E$, where $V=N/\rho$ is the total volume of the gas, for the first two branches,
as functions of $-1/k_F a$. We have taken $-1/k_F a$
as the abscissa because it allows an almost direct mapping with the $B$
field axis in a real experiment around the location $B_0$ of the Feshbach resonance,
see Eq.(\ref{eq:aB}).

The first excited branch is metastable. It starts with a weakly repulsive
Fermi gas on the extreme left of the figure
and has a larger energy than the ideal
Fermi gas, indicating effective repulsion. When $0<a\ll k_F^{-1}$, three-body
collisions (not included in the toy model) lead in a real gas to
the formation of dimers $\phi_0$: in the language of the toy model,
the system starts populating the ground branch. 
This suggests a first experimental way
to produce a gas of dimers.

The ground branch continuously connects the weakly attractive Fermi
gas (on the right) to a gas of dimers (on the left). This provides a second
experimental way to produce a condensate of dimers, by adiabatic following
of the ground branch. The sharp
decrease of the total energy on the left part of the figure reflects the
$1/a^2$ dependence of the dimer binding energy $\epsilon_0$.
The pressure is always less than the Fermi pressure of the ideal
Fermi gas, indicating effective attraction with respect to the ideal
Fermi gas. Note that $P$ drops very rapidly on the left side,
due to the absence of interaction between the molecules in the toy
model. In real life, this interaction occurs with a scattering
length that was calculated by solution of the 4-body problem
\cite{dim_dim}. As discussed in other lectures of this volume,
the molecular condensates have been obtained experimentally
\cite{Salomon_eta,Jin_mol,Grimm_mol,Ketterle_mol}.

The regime of infinite scattering length $|a|=+\infty$
is universal in the toy model: 
one finds that the energy of the corresponding
unitary gas has to be proportional to the ground state energy of the ideal
Fermi gas:
\be
E^{\rm unitary} = \eta E_0^{\rm ideal}
\label{eq:eta}
\ee
where $E_0^{\rm ideal}=3 N E_F/5$, $\eta$ is a numerical constant depending
on the branch, and $E_F$ is the Fermi energy of the ideal gas.
In the present state of the art of the field, it is accepted that this universality
property also holds for a real ground state Fermi gas. Approximate fixed node Monte Carlo
calculations \cite{Pandharipande, Giorgini} give for the ground branch $\eta \simeq 0.4$,
in disagreement with early experiments \cite{Thomas}
but in good agreement with recent experimental measures
\cite{Grimm_eta,Salomon_eta,Hulet_eta} and with recent exact Quantum Monte Carlo
results \cite{Juillet}.

The toy model allows to easily solve the following paradox:
\begin{itemize}
\item A common saying is that a gas with a positive scattering length $a>0$
experiences effective repulsion, and that it experiences effective attraction
for $a<0$.
\item Another saying is that for $|a|=+\infty$ the energy of the gas is universal and
does not depend on the sign of $a$. 
\item However, if $a\rightarrow +\infty$ (scenario 1), one expects that the universal state
has effective repulsion, whereas for $a\rightarrow -\infty$ (scenario 2),
one expects that the universal state experiences attraction. How can it be that there is
no dependence on the sign of $a$ in the unitary limit?
\end{itemize}
The answer provided by the toy model is simple: scenario 1 and scenario 2 are predicted
to lead in the
unitary limit to {\it two different} universal states, belonging respectively 
to the first excited branch and to the ground branch, one experiencing effective repulsion,
the other effective attraction.

\section{Zero temperature BCS theory: study of the ground branch}
\label{sec:BCS}

In this section, we consider the many-body problem of a two-component Fermi
gas with an interaction characterized only by the $s$-wave scattering length $a$.

To be able to use the BCS theory for all values of $k_F a$, while
getting sensible results, we restrict to the zero temperature case: the usual
static BCS theory is indeed unable to get a fair approximation to the critical temperature
on the bosonic side of the Feshbach resonance ($a>0, k_Fa\ll 1$), 
for reasons that will become clear
at the end of this section.
An improved finite temperature theory can be found in \cite{Nozieres,Randeria}.

For simplicity we also assume that the two spin states $\uparrow$ and $\downarrow$
have the same number of particles, so that they have a common chemical potential $\mu$.
The case of imbalanced chemical potentials and particle numbers 
is presently the object of experimental and theoretical studies,
and may lead to a variety of observed and/or predicted phenomena, as the
not yet observed 
non-demixed BCS phases with spatially modulated order parameter (the so-called LOFF phases)
\cite{LOFF,MoraLOFF}
or the already observed spatial demixing of the two spin components
\cite{Ketterle_unba,Hulet_eta,Hulet_unba,Chevy_unba}.

Going beyond the toy model of previous section, the zero temperature
BCS theory predicts that the particles arrange in pairs and gives several
properties of the gas of pairs:
\begin{itemize}
\item The existence of long range order in the pair coherence function:
\[
g_1^{\mathrm{pair}}(\vec{r}\,) = \langle \hat{\Psi}^\dagger(\vec{r}\,) 
\hat{\Psi}(\vec{0}\,)\rangle
\]
where $\hat{\Psi}(\vec{r}\,)=\hat{\psi}_\downarrow(\vec{r}\,) \hat{\psi}_\uparrow(\vec{r}\,)$
annihilates a pair of particles with opposite spin in $\vec{r}$.
In the thermodynamic limit, the zero temperature
BCS theory predicts that $g_1^{\mathrm{pair}}$
has a non-zero limit for large $r$: the pairs form a condensate. Except in the
regime $k_F a\rightarrow 0^+$, the pairs are not bosons,
so that this condensate is not simply a Bose-Einstein condensate.
\item The BCS theory predicts an energy required to break a pair.
\item The time-dependent BCS theory also predicts collective modes for the motion
of the pairs, like sound waves, associated to the famous Anderson-Bogoliubov phonon
branch \cite{Bogoliubov,Anderson_rpa}. In a trapped system, the time dependent BCS theory predicts the equivalent
of superfluid hydrodynamic modes.
\end{itemize}

In what follows, we shall use the lattice model of
subsection \ref{subsec:lat_mod}.
In second quantized form we therefore take the grand-canonical Hamiltonian
\be
H= \sum_{\vec{k}\in{\mathcal D}} \sum_{\sigma=\uparrow,\downarrow}
\left(\frac{\hbar^2 k^2}{2m}-\mu\right) a^\dagger_{\vec{k}\sigma} 
a_{\vec{k}\sigma}
+ l^3\sum_{\vec{r},\sigma}  U(\vec{r}\,) \hat{\psi}_\sigma^\dagger(\vec{r}\,) 
\hat{\psi}_\sigma(\vec{r}\,)
+g_0 l^3 \sum_{\vec{r}} 
\hat{\psi}_\uparrow^\dagger(\vec{r}\,)
\hat{\psi}_\downarrow^\dagger(\vec{r}\,)
\hat{\psi}_\downarrow(\vec{r}\,)
\hat{\psi}_\uparrow(\vec{r}\,).
\ee
The external potential $U(\vec{r}\,)$ may be accompanied by a rotational term
if one wishes to study vortices.
Here the $a_{\vec{k}\sigma}$'s obey the usual discrete anticommutation relations
Eqs.(\ref{eq:anticom1},\ref{eq:anticom2}). The field operators 
$\hat{\psi}_\sigma(\vec{r}\,)$
obey anticommutation relations mimicking the ones in continuous space in the limit
$l\rightarrow 0$:
\be
\{\hat{\psi}_\sigma(\vec{r}\,),\hat{\psi}^\dagger_{\sigma'}(\vec{r}\,')\}=
l^{-3}\,\delta_{\vec{r}\vec{r}\,'}\delta_{\sigma\sigma'}.
\ee
As discussed in subsection \ref{subsec:lat_mod}, this lattice model is very close
to the usual Hubbard model of condensed matter physics. What is unusual here
is that we take a quadratic dispersion relation, and more important the model
makes sense (to describe a real atomic gas with continuous positions) in the low
filling factor limit, an unusual limit in solid state physics.
We note that the original 
Hubbard model (with the usual cosine dispersion relation
Eq.(\ref{eq:disphub})
and with a filling factor of the order of unity) may be realized 
in a real experiment
with atoms in an optical lattice 
\cite{Zoller,Bloch,Esslinger_Fermi_lattice,Ketterle_Fermi_Hubbard}.

\subsection{The BCS ansatz}

\subsubsection{A coherent state of pairs}

Let us recall the Glauber coherent state of a bosonic field, 
\be
|\alpha\rangle = e^{-|\alpha|^2/2}\,e^{\alpha a^\dagger} |0\rangle,
\ee
where $\alpha$ is a complex number and $a^\dagger$ creates a boson in some
normalized one-body wavefunction $\phi(\vec{r}\,)$, 
\be
a^\dagger = \sum_{\vec{r}} l^3\, \phi(\vec{r}\,) \hat{\psi}^\dagger(\vec{r}\,).
\ee

By analogy, the BCS theory \cite{BCS}, which is a {\sl variational} theory,
takes as a trial state a coherent state of pairs:
\be
|\Psi_{\rm BCS}\rangle = {\mathcal N}\, e^{\gamma C^\dagger} |0\rangle
\label{eq:psi_bcs}
\ee
where ${\mathcal N}$ is a normalization factor, $\gamma$ is
a complex number and $C^\dagger$
creates two particles in the normalized pair wave function $ \phi(\vec{r}_1,\vec{r}_2)$:
\be
C^\dagger = \sum_{\vec{r}_1,\vec{r}_2} l^6 \phi(\vec{r}_1,\vec{r}_2)\, 
\hat{\psi}^\dagger_\uparrow(\vec{r}_1) \hat{\psi}^\dagger_\downarrow(\vec{r}_2).
\ee
Note that in general $C$ and $C^\dagger$ 
do not obey bosonic commutation relations.
For simplicity, we omit terms creating two particles in the same spin state.
This shall be sufficient for our purposes as we restrict here
to the case of balanced spin state populations.

\subsubsection{A more convenient form from the Schmidt decomposition}
\label{subsubsec:amcf}
To briefly review the main properties of this ansatz, it is convenient to introduce the 
Schmidt decomposition of $\gamma|\phi\rangle$, 
which is, in the physics of entanglement,
routinely applied
to the state vector of two arbitrary quantum particles:
\be
\gamma |\phi\rangle = \sum_\alpha \Gamma_\alpha |A_\alpha\rangle \otimes |B_\alpha\rangle
\ee
where the coefficients $\Gamma_\alpha$ are real numbers, the set of $|A_\alpha\rangle$
is an orthonormal basis here for a spinless particle, and the set of $|B_\alpha\rangle$
is also an orthonormal basis for a spinless particle. Note that in general
these two basis are distinct. As $|\phi\rangle$ is normalized to unity,
one has $\sum_\alpha \Gamma_\alpha^2=|\gamma|^2$. Then the pair creation operator
takes the simple expression
\be
\gamma \, C^\dagger  = \sum_\alpha \Gamma_\alpha \hat{c}_{\alpha\uparrow}^\dagger 
\hat{c}_{\alpha\downarrow}^\dagger
\ee
where $\hat{c}_{\alpha\uparrow}^\dagger$ is the creation operator of a particle
in the state $|A_\alpha\rangle |\uparrow\rangle$ and
$\hat{c}_{\alpha\downarrow}^\dagger$ is the creation operator of a particle
in the state $|B_\alpha\rangle |\downarrow\rangle$. Note that these creation
operators and the associated annihilation operators obey the usual
fermionic anticommutation relations.

The key advantage of the Schmidt decomposition is to allow a rewriting of the BCS
state, more familiar to the reader:
\be
|\Psi_{\rm BCS}\rangle = {\mathcal N}\, \left[\prod_\alpha \left(1+\Gamma_\alpha 
\hat{c}_{\alpha\uparrow}^\dagger \hat{c}_{\alpha\downarrow}^\dagger\right)
\right] |0\rangle.
\label{eq:bcs_exam}
\ee
To obtain this expression, we have used the fact that any binary product 
$\hat{c}_{\alpha\uparrow}^\dagger \hat{c}_{\alpha\downarrow}^\dagger$
commutes with any $\hat{c}_{\beta\uparrow}^\dagger \hat{c}_{\beta\downarrow}^\dagger$, 
and the fact that the series expansion
of $\exp(\gamma \hat{c}_{\alpha\uparrow}^\dagger\hat{c}_{\alpha\downarrow}^\dagger)$
terminates to first order in $\gamma$ since $(\hat{c}^\dagger)^2=0$ for fermions.
The form Eq.(\ref{eq:bcs_exam}) shows that one can consider 
each pair of modes $\{\alpha\uparrow,\alpha\downarrow\}$ as independent, since
each factor in the product commutes with any other factor.
The normalization factor is then readily calculated, 
${\mathcal N}=\prod_\alpha 1/\sqrt{1+\Gamma_\alpha^2}$. It can be absorbed in
the following rewriting, that may be the one directly familiar to the reader: 
\be
|\Psi_{\rm BCS}\rangle = \left[\prod_\alpha  \left(U_\alpha 
-V_\alpha \hat{c}_{\alpha\uparrow}^\dagger \hat{c}_{\alpha\downarrow}^\dagger\right)
\right] |0\rangle,
\label{eq:bcs_form}
\ee
where $U_\alpha$ and $V_\alpha$ are the real numbers 
\be
U_\alpha = \frac{1}{\sqrt{1+\Gamma_\alpha^2}} 
\ \ \ \ \ \ \ \mbox{and} \ \ \ \ \ \ \ 
V_\alpha = \frac{-\Gamma_\alpha}{\sqrt{1+\Gamma_\alpha^2}}.
\ee
Note that they satisfy $U_\alpha^2 + V_\alpha^2 = 1$, and the minus
sign in $V_\alpha$ was introduced to ensure consistency with coming notations.

\subsubsection{As a squeezed vacuum: Wick's theorem applies}
A third interesting rewriting of the BCS state
can be obtained from the identity
\be
e^{\theta \left(b^\dagger c^\dagger - c b\right)}|0\rangle
=\cos\theta |0\rangle + \sin \theta\ b^\dagger c^\dagger|0\rangle
\ee
where $\theta$ is a real number, $b$ and $c$ are two fermionic annihilation
operators with standard anticommutation relations
\footnote{This identity can be proved by a direct expansion of the exponential
in powers of $\theta$, or by obtaining a differential equation
satisfied by the left hand side considered as a function of $\theta$.}.
Then $|\Psi_{\rm BCS}\rangle = S |0\rangle$
where the unitary operator is
\be
S = \prod_\alpha e^{\theta_\alpha \left(\hat{c}_{\alpha \uparrow}^\dagger
\hat{c}_{\alpha \downarrow}^\dagger-\rm{h.c.}\right)}
\ee
and the angles $\theta_\alpha$ are such that
\be
U_\alpha = \cos\theta_\alpha \ \ \ \ \ \ V_\alpha=-\sin\theta_\alpha.
\ee
For bosons, $S$ would be interpreted as a squeezing operator \cite{ZollerGardiner}.
The BCS state is therefore the equivalent for fermions of the squeezed
vacuum for bosons.

Calculating the expectation value in the BCS state of a product of
operators $\hat{c}$ and $\hat{c}^\dagger$ is therefore equivalent
to calculating the expectation value in the vacuum state of
the product of the transformed operators $S^\dagger \hat{c} S$ and
$S^\dagger \hat{c}^\dagger S$. These transformed operators
have a linear expression in terms of the original $\hat{c}$ and
$\hat{c}^\dagger$, since one has
\bea
S^\dagger \hat{c}_{\alpha\uparrow}  S &=&  U_\alpha \hat{c}_{\alpha\uparrow} -
V_\alpha \hat{c}_{\alpha\downarrow}^\dagger \\
S^\dagger \hat{c}_{\alpha\downarrow}^\dagger  S &=& V_\alpha \hat{c}_{\alpha\uparrow}
+U_\alpha \hat{c}_{\alpha\downarrow}^\dagger.
\eea      
The linearity of these transformations is evident since $S$ can be formally considered
as the time evolution operator for a quadratic Hamiltonian, which generates
indeed linear equations of motion of the creation and annihilation operators.
As a consequence, Wick's theorem can be applied to calculate expectation
values in the BCS state, since it applies for the vacuum.

\subsubsection{Some basic properties of the BCS state}
Before moving to the energy minimization within the BCS ansatz, we calculate 
some interesting quantities. Since Wick's theorem applies, the expectation value
of any quantity is a function of the only non-zero two-point averages:
\bea
\langle \hat{c}_{\alpha\uparrow}^\dagger \hat{c}_{\alpha\uparrow}\rangle
= \langle \hat{c}_{\alpha\downarrow}^\dagger \hat{c}_{\alpha\downarrow}\rangle
&=& \frac{\Gamma_\alpha^2}{1+\Gamma_\alpha^2} \\
\langle \hat{c}_{\alpha\downarrow}\hat{c}_{\alpha\uparrow}\rangle =
\langle \hat{c}_{\alpha\uparrow}^\dagger \hat{c}_{\alpha\downarrow}^\dagger \rangle 
&=& \frac{\Gamma_\alpha}{1+\Gamma_\alpha^2},
\eea
where the expectation values are taken in the BCS state.

For the total number of particles in the BCS state, we obtain the mean value
and the variance
\bea
\langle \hat{N}\rangle &=& \sum_\alpha \frac{2\Gamma_\alpha^2}{1+\Gamma_\alpha^2} \\
\mbox{Var}\, \hat{N} &=& \sum_\alpha \frac{4\Gamma_\alpha^2}{(1+\Gamma_\alpha^2)^2}.
\eea
One then immediately obtains the inequality
\be
\mbox{Var}\, \hat{N} \leq 2 \langle \hat{N}\rangle,
\label{eq:var_bcs}
\ee
an inequality that becomes an equality in the limit where all $\Gamma_\alpha \ll 1$.
This shows that, in the large $N$ limit, the fact that the BCS state has not a well 
defined number of particles is not a problem for most practical purposes, since
the relative fluctuations are $O(1/\sqrt{N})$.

For the expectation value of the commutator of $C$ and $C^\dagger$ in the BCS state
we obtain
\be
\langle [C,C^\dagger]\rangle = 1 - \frac{\sum_\alpha 2\Gamma_\alpha^4/(1+\Gamma_\alpha^2)}
{\sum_\alpha \Gamma_\alpha^2}.
\ee
We see that, in the limit where all $\Gamma_\alpha^2 \ll 1$,
this expectation value is close to unity. In this limit, 
one may consider that $C^\dagger$ creates a
bosonic pair:  the BCS state becomes a Glauber coherent state 
for this bosonic pair, one obtains Poissonian fluctuations 
in the number $\hat{N}/2$ of pairs,
which explains the upper limit of Eq.(\ref{eq:var_bcs}).
This also shows that the BCS ansatz will have no difficulty to predict the
formation of a Bose-Einstein condensate of dimers in the $a\rightarrow 0^+$ limit \cite{Leggett}.

A last interesting property is that the vacuum of an arbitrary
quadratic Hamiltonian (quadratic in the quantum field) is actually 
a BCS state. We defer the proof of this statement to \S\ref{subsec:linham}.
A similar property is that the vacuum of a set of arbitrary operators
$b_\alpha$ obeying anticommutation relations is a BCS state: in other words,
the BCS state is a quasi-particle vacuum. We refer to \cite{Blaizot} for a
detailed description of this aspect.

\subsection{Energy minimization within the BCS family}
\label{subsec:emwtbf}

We now proceed with the minimization of the energy of the lattice model Hamiltonian
within the family of not necessarily normalized BCS states.
We define
\be
E[\Phi] = \frac{\langle \Psi_{\rm BCS}| H |\Psi_{\rm BCS}\rangle}
{\langle \Psi_{\rm BCS}| \Psi_{\rm BCS} \rangle}
\ee
where the BCS state $|\Psi_{\rm BCS} \rangle$ is of the form
Eq.(\ref{eq:psi_bcs}) parametrized by the unnormalized
pair wavefunction $\Phi=\gamma \phi$ and by the factor $\mathcal{N}$
now considered as an independent variable.
If one performs a variation of $\Phi$ around a minimizer of $E[\Phi]$,
$\Phi = \Phi_0 + \delta \Phi$,
this induces a variation $\delta|\Psi_{\rm BCS}\rangle$ of the BCS state around
$|\Psi_{\rm BCS}^0\rangle$,
\be
\delta|\Psi_{\rm BCS}\rangle = \sum_{\vec{r}_1,\vec{r}_2} l^6 \delta\Phi(\vec{r}_1,\vec{r}_2) 
\hat{\psi}^\dagger_\uparrow(\vec{r}_1) \hat{\psi}^\dagger_\downarrow(\vec{r}_2)
|\Psi_{\rm BCS}^0\rangle \equiv \delta\hat{X} |\Psi_{\rm BCS}^0\rangle.
\label{eq:vari}
\ee
The corresponding first order variation of the energy function $E[\Phi]$ 
has to vanish for all possible values of $\delta\Phi$,
\be
\delta E = (\delta\langle \Psi_{\rm BCS}|) (H-E[\Phi_0]) 
|\Psi_{\rm BCS}^0\rangle + \mbox{c.c.}=0 \ \ \ \ \ \ \forall \ \delta\Phi,
\label{eq:cond_ext}
\ee
where we assumed that the minimizer $|\Psi_{\rm BCS}^0\rangle$
is normalized to unity.

The second step is to introduce the quadratic Hamiltonian deduced from the full 
Hamiltonian $H$ by performing incomplete Wick's contractions. This recipe has to
be applied to the interaction term only, since the other terms of the Hamiltonian
are quadratic. The quartic on-site interaction is replaced by the quadratic expression:
\bea
g_0 \hat{\psi}_\uparrow^\dagger \hat{\psi}_\downarrow^\dagger 
\hat{\psi}_\downarrow \hat{\psi}_\uparrow
\rightarrow && \hat{\psi}_\uparrow^\dagger \hat{\psi}_\downarrow^\dagger 
\, g_0 \langle\hat{\psi}_\downarrow
\hat{\psi}_\uparrow\rangle + \mbox{h.c.} \nonumber \\
&& + \hat{\psi}_\uparrow^\dagger\hat{\psi}_\uparrow 
\, g_0 \langle \hat{\psi}_\downarrow^\dagger 
\hat{\psi}_\downarrow\rangle + \uparrow \leftrightarrow \downarrow,
\eea
where the expectation values are taken in the state $|\Psi_{\rm BCS}^0\rangle$.
The first line of the righthand side consists in a pairing term, 
that creates/annihilates
a pair of particles; it involves the following pairing field, also called `gap' for
historical reasons that will become clear:
\be
\Delta(\vec{r}\,) \equiv g_0 \langle \hat{\psi}_\downarrow(\vec{r}\,) 
\hat{\psi}_\uparrow(\vec{r}\,)\rangle.
\ee
The second line of the righthand side is simply the Hartree  mean field term:
each spin-up particle sees a mean field potential which is $g_0$ times the density
of spin-down particle. Note that, in the lattice model, this mean field term does
not diverge at the unitary limit, since $g_0$ remains bounded in this limit,
see Eq.(\ref{eq:g0_exp}).

The quadratic Hamiltonian associated by Wick's contractions to the full Hamiltonian
is the so-called BCS Hamiltonian, and it has the structure
\be
{\mathcal H} \equiv \sum_{\vec{r},\vec{r}\,', \sigma} l^3\, 
\hat{\psi}^\dagger_\sigma(\vec{r}\,)  h_{\vec{r},\vec{r}\,'} 
\hat{\psi}_\sigma(\vec{r}\,')
+  l^3 \sum_{\vec{r}} 
\left[\hat{\psi}^\dagger_\uparrow(\vec{r}\,) 
\hat{\psi}^\dagger_\downarrow(\vec{r}\,) \,
\Delta(\vec{r}\,) 
+\mbox{h.c.}\right] + E_{\rm adj}.
\label{eq:BCS_ham}
\ee
Here the matrix $h$ is the sum of the one-body part of the Hamiltonian $H$
and of the Hartree mean field terms,
\be
h_{\vec{r},\vec{r}\,'} = \left[\mbox{kin}\right]_{\vec{r},\vec{r}\,'} 
+\left[U(\vec{r}\,) -\mu+ 
g_0 \langle \hat{\psi}^\dagger_\uparrow(\vec{r}\,) \hat{\psi}_\uparrow(\vec{r}\,)\rangle
\right] \delta_{\vec{r},\vec{r}\,'}
\ee
where $\left[\mbox{kin}\right]$ is the matrix representing the kinetic energy
operator.
We have assumed for simplicity that the one-body
Hamiltonian is spin independent and that the two
spin density profiles are identical, 
so that the matrix $h$ does not depend on the spin
state.
Finally, the additive constant $E_{\rm adj}$ is adjusted to the value
\be
E_{\rm adj} = - g_0 \sum_{\vec{r}} l^3 \left[\langle \hat{\psi}^\dagger_\uparrow
\hat{\psi}_\uparrow\rangle^2 + |\langle \hat{\psi}_\downarrow
\hat{\psi}_\uparrow\rangle|^2 \right]
\label{eq:e_adj}
\ee
to ensure that the mean energy of $\mathcal{H}$ and of $H$ coincide
in the minimizer $|\Psi_{\rm BCS}^0\rangle$,
\be
\langle H\rangle = \langle \mathcal{H}\rangle.
\ee

Using Wick's theorem and the fact that the operator
$\delta\hat{X}$ is a quadratic function of the quantum
field, one can then show the marvelous property
\be
(\delta\langle\Psi_{\rm BCS}|) H |\Psi_{\rm BCS}^0\rangle 
= 
(\delta\langle\Psi_{\rm BCS}|) \mathcal{H} |\Psi_{\rm BCS}^0\rangle,
\ee
whatever the variation $\delta | \Psi_{\rm BCS} \rangle$ of
the BCS state {\it within} the BCS family.

There is the following key consequence. Considering the quadratic Hamiltonian
${\mathcal H}$  as a {\sl given} Hamiltonian, it turns out that the BCS state 
$|\Psi_{\mathrm{BCS}}^0\rangle$, obeying Eq.(\ref{eq:cond_ext}), also obeys the condition to be
a stationary state of the energy functional
\be
{\mathcal E}[\Phi] = 
\frac{\langle \Psi_{\rm BCS}| {\mathcal H} |\Psi_{\rm BCS}\rangle}{
\langle \Psi_{\rm BCS}| \Psi_{\rm BCS} \rangle},
\ee
noting that one has $E[\Phi_0] = {\mathcal E}[\Phi_0]$. Condition
Eq.(\ref{eq:cond_ext}) is thus equivalent to the condition
\be
\delta \mathcal{E} = (\delta\langle \Psi_{\rm BCS}|) 
(\mathcal{H}-\mathcal{E}[\Phi_0]) 
|\Psi_{\rm BCS}^0\rangle + \mbox{c.c.}=0 \ \ \ \ \ \ \forall \ \delta\Phi,
\label{eq:cond_ext_cal}
\ee
That is, to minimize $E[\Phi]$ within the BCS family, we shall take
the BCS state that minimizes the energy of $\mathcal H$.

We are left with the study of a quadratic Hamiltonian.
As reminded to the reader in \S\ref{subsec:linham}, the quadratic
Hamiltonian ${\cal H}$ may be diagonalized  by a canonical transformation,
the so-called Bogoliubov transformation. This allows to show that the ground
state $|\mbox{ground}\rangle$ of $\mathcal H$ is in fact a BCS state (see 
\S\ref{subsec:linham}), and to calculate 
the expectation values $\langle \hat{\psi}_\sigma^\dagger \hat{\psi}_\sigma\rangle_{\mathrm{ground}}$
and $\langle \hat{\psi}_\downarrow \hat{\psi}_\uparrow\rangle_{\mathrm{ground}}$
in $|\mbox{ground}\rangle$. Since $|\Psi_{\rm BCS}^0\rangle$ and
$|\mbox{ground}\rangle$ actually coincide, one is left with the self-consistency conditions:
\bea
\label{eq:sc1}
\langle \hat{\psi}_\sigma^\dagger \hat{\psi}_\sigma\rangle_{\mathrm{ground}} &=& 
\langle \hat{\psi}_\sigma^\dagger \hat{\psi}_\sigma\rangle \\
\langle \hat{\psi}_\downarrow \hat{\psi}_\uparrow\rangle_{\mathrm{ground}} &=& 
\langle \hat{\psi}_\downarrow \hat{\psi}_\uparrow\rangle.
\label{eq:sc2}
\eea
Since the expectation values in the ground state of $\mathcal H$ are
non-linear functionals of the coefficients $\langle \hat{\psi}_\sigma^\dagger \hat{\psi}_\sigma\rangle$
and $\langle \hat{\psi}_\downarrow \hat{\psi}_\uparrow\rangle$ of $\mathcal H$,
this constitutes a non-linear self-consistent problem, the so-called BCS
equations for the density and the gap function.

Another crucial consequence is to realize that one may write the self-consistency
conditions taking an excited eigenstate of the quadratic BCS Hamiltonian
${\cal H}$ rather than the ground state \cite{the_curious}: 
in this case, the variational BCS calculation
is used to predict excited states of the gas!
This shows that BCS theory immediately predicts elementary
excitations of the gas. As we shall see in the thermodynamic limit, 
these excitations correspond to pair breaking
and are characterized by an energy spectrum with a gap.

\subsection{Reminder on diagonalization of quadratic Hamiltonians}
\label{subsec:linham}

We consider here a quadratic Hamiltonian of the form Eq.(\ref{eq:BCS_ham}),
where, at this stage, $h$ is an arbitrary (spin independent)
hermitian matrix and $\Delta(\vec{r}\,)$
an arbitrary complex field.
We wish to put this Hamiltonian in canonical
form by performing a Bogoliubov transformation.

The key property of a quadratic Hamiltonian is that it leads in Heisenberg picture to
linear equations of motion. Collecting the field operators and their hermitian conjugate
at all points of the lattice 
in big vectors, we get
\be
i\hbar \partial_t \left(\begin{array}{c}\hat{\psi}_\sigma \\ \hat{\psi}_{-\sigma}^\dagger
\end{array}\right) = L^\sigma 
\left(\begin{array}{c}\hat{\psi}_\sigma \\ \hat{\psi}_{-\sigma}^\dagger
\end{array}\right) 
\label{eq:leom}
\ee
where $-\sigma$ stands for the spin component opposite to $\sigma$ and
where we have introduced two matrices
\be
L^\uparrow = \left( \begin{array}{cc} h & \mbox{diag}(\Delta) \\
\mbox{diag}(\Delta^*)  & - h^*\end{array}
\right)
\ \ \ \ \ \ \ \ \ \ \mbox{and}\ \ \ \ \ \ \ \ \ \  
L^\downarrow = \left( \begin{array}{cc} h & \mbox{diag}(-\Delta) \\
\mbox{diag}(-\Delta^*)  & - h^*\end{array}
\right).
\ee

These two matrices are hermitian: they can be diagonalized in an orthonormal basis.
Furthermore they obey the following symmetry property: if the vector $(u,v)$ 
is an eigenstate of $L^\uparrow$ with the energy $\epsilon$, then 
\begin{itemize}
\item the vector $(-v^*,u^*)$ is an eigenvector of $L^\uparrow$, with the eigenvalue $-\epsilon$, 
\item the vector $(u,-v)$ is an eigenstate of $L^\downarrow$, with the eigenvalue
$\epsilon$,
\item the vector $(v^*,u^*)$ is an eigenvector of $L^\downarrow$, with the eigenvalue $-\epsilon$. 
\end{itemize}

Assuming for simplicity that all the eigenenergies are non-zero, we can
collect the eigenvectors in pairs of opposite eigenenergies.
Expanding $(\hat{\psi}_\sigma, \hat{\psi}_{-\sigma}^\dagger)$ in the eigenbasis
of $L^\sigma$, and expressing the negative energy modes of $L^\uparrow$
and all the eigenmodes of $L^\downarrow$ 
in terms of the {\it positive} energy modes of $L^\uparrow$, 
noted as $(u_\alpha,v_\alpha)$ , we obtain
\bea
\label{eq:expan0}
\hat{\psi}_\uparrow(\vec{r}\,) &=& \sum_{\alpha} b_{\alpha\uparrow} u_\alpha(\vec{r}\,) -
b^\dagger_{\alpha\downarrow} v_\alpha^*(\vec{r}\,) \\
\hat{\psi}_\downarrow(\vec{r}\,) &=& \sum_{\alpha} b_{\alpha\downarrow} u_\alpha(\vec{r}\,) +
b^\dagger_{\alpha\uparrow} v_\alpha^*(\vec{r}\,)
\label{eq:expan}
\eea
where the positive energy eigenvectors of $L^\uparrow$ are normalized in
a way mimicking the continuous space limit,
with Dirac's notation $\langle \vec{r}\,|u_\alpha\rangle = u_\alpha(\vec{r}\,)$:
\be
\langle u_\alpha|u_\beta\rangle + \langle v_\alpha| v_\beta\rangle
=
l^3 \sum_{\vec{r}} u_\alpha^*(\vec{r}\,) u_\beta(\vec{r}\,)
+v_\alpha^*(\vec{r}\,) v_\beta(\vec{r}\,) = \delta_{\alpha\beta},
\label{eq:bon}
\ee
The coefficients in the expansion of the field operators are themselves operators that
are easy to express using the orthonormal nature of the 
eigenbasis of $L^\sigma$:
\bea
\label{eq:bau}
b_{\alpha\uparrow} &=&  l^3 \sum_{\vec{r}} u_\alpha^*(\vec{r}\,) \hat{\psi}_\uparrow(\vec{r}\,)
+ v_\alpha^*(\vec{r}\,) \hat{\psi}^\dagger_\downarrow (\vec{r}\,) \\
b_{\alpha\downarrow} &=& 
l^3 \sum_{\vec{r}} u_\alpha^*(\vec{r}\,) \hat{\psi}_\downarrow(\vec{r}\,)  
- v_\alpha^*(\vec{r}\,) \hat{\psi}^\dagger_\uparrow (\vec{r}\,).
\label{eq:bad}
\eea
Using Eq.(\ref{eq:bon}) again,
one can check that the $b$'s and
their hermitian conjugates obey the standard fermionic anticommutation 
relations, 

The last step is to express the Hamiltonian in normal form, using the $b_\alpha$'s.
One first realizes that the quadratic Hamiltonian can be expressed as a quadratic
form defined by the matrices $L^\sigma$,
\be
{\mathcal H} = \mbox{Tr}\, h + E_{\rm adj}+
\frac{1}{2}l^3\sum_\sigma  
\left(\hat{\psi}_\sigma^\dagger,\hat{\psi}_{-\sigma}\right)
L^\sigma 
\left(\begin{array}{c}\hat{\psi}_\sigma \\ \hat{\psi}_{-\sigma}^\dagger
\end{array}\right)
\ee
where the extra constant term, given by the trace of the matrix $h$,
originates from the anticommutator of
$\hat{\psi}^\dagger_\sigma$ with $\hat{\psi}_\sigma$.
Inserting the modal decomposition Eqs.(\ref{eq:expan0},\ref{eq:expan}) and using the fact that $(u_\alpha,v_\alpha)$
is an eigenvector of $L^\uparrow$, etc, we obtain the canonical form:
\be
{\mathcal H} = E_0 + \sum_{\alpha,\sigma}
\epsilon_\alpha b^\dagger_{\alpha\sigma} b_{\alpha\sigma},
\label{eq:can}
\ee
where we recall that all the $\epsilon_\alpha>0$ by construction.
The immediate expression for $E_0$ is $E_0=\mbox{Tr}\, h + E_{\mathrm{adj}}
-\sum_{\alpha} \epsilon_\alpha$;
it involves two quantities 
that are not well behaved in the zero lattice spacing limit. 
Expressing the matrix 
$L^\uparrow$ as a sum of $\pm\epsilon_\alpha$ times dyadics involving the eigenvectors
as ket and bra, and projecting over the upper left block leads to 
\be
h_{\vec{r},\vec{r}\,'} = l^3 \sum_{\alpha}  \epsilon_\alpha
\left[u_\alpha(\vec{r}\,)  u_\alpha^*(\vec{r}\,') -
v_\alpha^*(\vec{r}\,) v_\alpha(\vec{r}\,')\right],
\ee
where we used the orthonormal nature of the eigenbasis of $L^\uparrow$, 
as defined in Eq.(\ref{eq:bon}).
Taking the trace of this expression in the spatial basis of the lattice
leads to the more convenient form for $E_0$:
\be
E_0 = E_{\rm adj}
-2 \sum_{\alpha} \epsilon_\alpha 
\langle v_\alpha| v_\alpha\rangle,
\ee
where $E_{\mathrm{adj}}$ is given by Eq.(\ref{eq:e_adj}).
The canonical form Eq.(\ref{eq:can}), with all the $\epsilon_\alpha >0$
by construction,
immediately shows that the ground state $|\Psi_0\rangle$
of ${\mathcal H}$ is the vacuum of all the $b_\alpha$'s, with an eigenenergy $E_0$. Please remember that we are dealing here with the grand canonical 
Hamiltonian; this is why $E_0<0$ for the ideal gas $g_0=0$.

To show that this ground state is a BCS
state, we take for $|\Psi_0\rangle$ an ansatz of the BCS type
Eq.(\ref{eq:psi_bcs}) to solve the equations
\be
b_{\alpha\sigma} | \Psi_0\rangle =0
\label{eq:sys}
\ee
for all mode index $\alpha$ and spin component $\sigma=\uparrow, \downarrow$.
The commutator of a $b_{\alpha\sigma}$
with the pair creation operator $C^\dagger$ is a linear combination
of the $\hat{\psi}^\dagger$'s so it commutes with $C^\dagger$.
As a consequence,
\be
[b_{\alpha \sigma}, e^{\gamma C^\dagger}] =
\gamma e^{\gamma C^\dagger} [b_{\alpha \sigma},C^\dagger].
\label{eq:com}
\ee
Since $\exp(\gamma C^\dagger)$ is an invertible operator,
Eq.(\ref{eq:sys}) reduces to
\be
\left(b_{\alpha\sigma} + \gamma [b_{\alpha\sigma}, C^\dagger]\right)
|0\rangle =0.
\ee
From the expressions Eqs.(\ref{eq:bau},\ref{eq:bad}) we calculate
the commutator with $C^\dagger$ and obtain the linear system
\bea
v_\alpha^*(\vec{r}\,) + \gamma \sum_{\vec{r}\,'} l^3
\phi(\vec{r}\,',\vec{r}\,) u_\alpha^*(\vec{r}\,') &=& 0  \\
v_\alpha^*(\vec{r}\,) + \gamma \sum_{\vec{r}\,'} l^3
\phi(\vec{r},\vec{r}\,') u_\alpha^*(\vec{r}\,') &=& 0.
\eea
The solution may be written in matrix form as follows. Since the lattice
has discrete positions, we consider $\phi(\vec{r},\vec{r}\,')$
as the element of a square matrix $[\phi]$ on row $\vec{r}$ and column
$\vec{r}\,'$. Similarly we form the square matrix $[U]$ (respectively $[V]$)
such that the element on row $\vec{r}$ and column $\alpha$ is $u_\alpha(\vec{r}\,)$
(respectively $v_\alpha(\vec{r}\,)$). 
Then one has the explicit writing
\be
\gamma\, [\phi] = \gamma \, {}^t [\phi]= -l^{-3}\,  ([U]^\dagger)^{-1} [V]^\dagger,
\label{eq:gamma_phi}
\ee
where ${}^t M$ is the transposed matrix of a matrix $M$. 
Since the set of all the eigenvectors of $L^\uparrow$
forms an orthonormal basis, one has $l^3(U^\dagger U + V^\dagger V) 
= \mathrm{Id}$
and ${}^t U V = {}^t V U$, where $\mathrm{Id}$ is the identity matrix.
This ensures that the solution $[\phi]$ is indeed a symmetric matrix
\be
\phi(\vec{r}_1,\vec{r}_2) = \phi(\vec{r}_2,\vec{r}_1).
\ee
This symmetry property could be expected from the fact that
the quadratic Hamiltonian
$\mathcal{H}$ is invariant by an arbitrary rotation $R$ of the spin degree of
freedom, and that its ground state is not degenerate since we
have assumed $\epsilon_\alpha$ strictly positive.
The condition that $C^\dagger = R C^\dagger R^\dagger$ for all
spin rotations indeed imposes that $\phi$ is a symmetric function
of $\vec{r}_1$ and $\vec{r}_2$.

Using Wick's theorem one can easily calculate expectation values
of observables in $|\Psi_0\rangle$ using the modal decompositions
Eqs.(\ref{eq:expan0},\ref{eq:expan}). 
Two important examples are the mean density
and the anomalous average
\bea
\label{eq:rho_modal}
\rho_\uparrow(\vec{r}\,) = \rho_\downarrow(\vec{r}\,) &=&
\sum_{\alpha} |v_\alpha(\vec{r}\,)|^2, \\
\langle \hat{\psi}_\downarrow(\vec{r}\,)
\hat{\psi}_\uparrow(\vec{r}\,)\rangle
 &=& - \sum_{\alpha}  u_\alpha(\vec{r}\,) 
v_\alpha^*(\vec{r}\,).
\label{eq:delta_modal}
\eea

\subsection{Summary of BCS results for the homogeneous system}

\subsubsection{Gap equation in the thermodynamical limit}
In the case of a spatially homogeneous system (cubic box with periodic
boundary conditions) explicit predictions are easily extracted from the
BCS theory.
For a spatially uniform gap parameter $\Delta$, assumed to be real
non negative, and for a uniform density
profile  $\rho_\uparrow=\rho_\downarrow$
the spectral decomposition of $L^\uparrow$ is readily obtained: from
the translational invariance one seeks eigenvectors 
of the form
\be
(u_{\vec{k}}(\vec{r}\,),v_{\vec{k}}(\vec{r}\,))
=L^{-3/2} e^{i\vec{k}\cdot\vec{r}}(U_k,V_k).
\ee
Restricting to the positive part of the spectrum,
one finds the eigenenergies
\be
\epsilon_k = \left[\left(\frac{\hbar^2 k^2}{2m}- \tilde{\mu}\right)^2
+\Delta^2\right]^{1/2} = \left| \frac{\hbar^2 k^2}{2m} -\tilde{\mu} + i \Delta
\right|
\label{eq:spectre_bcs}
\ee
where $\tilde{\mu}= \mu - g_0 \rho_\uparrow$ is the chemical potential
minus the Hartree mean field term.
The amplitudes $U_k,V_k$, chosen to be real, are normalized as $U_k^2+V^2=1$
and are given by
\be
(U_k + i V_k)^2 = \frac{\frac{\hbar^2 k^2}{2m} -\tilde{\mu} + i \Delta}
{\epsilon_k}.
\label{eq:uiv}
\ee
To make the link with the quantities $\Gamma_\alpha$ of section
\ref{subsubsec:amcf}, we take for $|A_\alpha\rangle = |A_{\vec{k}}\rangle$
the plane wave of wavevector $\vec{k}$ and for
$|B_\alpha\rangle = |B_{\vec{k}}\rangle$ the plane wave of wavevector
$-\vec{k}$.
This gives
\be
\Gamma_k = -\frac{V_k}{U_k} = -\frac{\Delta}{\frac{\hbar^2 k^2}{2m} -\tilde{\mu}
+\epsilon_k},
\label{eq:gamma_homo}
\ee
and a pair wavefunction \cite{cafg}
\be
\gamma \phi(\vec{r}_1,\vec{r}_2) = \frac{1}{L^3} \sum_{\vec{k}\in\mathcal{D}}
\Gamma_k e^{i\vec{k}\cdot(\vec{r}_1-\vec{r}_2)}.
\label{eq:phi_homo}
\ee
Remember that the regime where {\it all} the $\Gamma_k^2\ll 1$
corresponds to the case where the BCS state is a condensate of bosonic
pairs.

Taking the real and imaginary parts of Eq.(\ref{eq:uiv}), and going
directly to the thermodynamic limit, one obtains
from Eqs.(\ref{eq:rho_modal},\ref{eq:delta_modal})
\bea
\label{eq:dens_homo}
\rho &=& \int_{\mathcal{D}} \frac{d^3k}{(2\pi)^3} 
\left[1- \frac{\frac{\hbar^2 k^2}{2m}-\tilde{\mu}}{\epsilon_k}\right] \\
\Delta &=& -g_0 \int_{\mathcal{D}} \frac{d^3k}{(2\pi)^3}
\frac{\Delta}{2\epsilon_k},
\label{eq:delta_homo}
\eea
where $\rho$ is the total density. This makes explicit the self-consistent
character of the conditions Eqs.(\ref{eq:sc1},\ref{eq:sc2}).
The last equation Eq.(\ref{eq:delta_homo}) is called the gap equation.

\subsubsection{In the limit of a vanishing lattice spacing}

An important point is to discuss the dependence of these predictions
with the lattice spacing $l$, since the relevant regime of a continuous
gas is described by the lattice model only in the $l\rightarrow 0$
limit. The integral giving $\rho$ converges at large $k$,
since the integrand is $O(1/k^4)$, so that one may directly set $l\to 0$
in Eq.(\ref{eq:dens_homo}). On the contrary the integral in Eq.(\ref{eq:delta_homo})
has an ultraviolet divergence when integrated over the whole momentum space.
However $g_0$ tends to zero in the $l\to 0$ limit, compensating
exactly this divergence. Dividing Eq.(\ref{eq:delta_homo})
by $g_0$ and expressing $1/g_0$ from Eq.(\ref{eq:g0_gen}) leads to 
a more convenient expression for the gap equation,
in which one may take directly $l\to 0$:
\be
\frac{\Delta}{g} = \Delta \int_{\mathcal{D}} \frac{d^3k}{(2\pi)^3}
\left[\frac{m}{\hbar^2 k^2} -\frac{1}{2\epsilon_k}\right].
\label{eq:gap_homo}
\ee
Furthermore, we note that the BCS theory becomes simpler
when $l\to 0$: since the coupling constant $g_0$ vanishes in this limit,
the Hartree mean field term also vanishes, so that $\tilde{\mu}\to \mu$
and the matrix $h$
in the quadratic Hamiltonian $\mathcal{H}$ reduces to the matrix 
of the kinetic energy minus $\mu$. The {\it a priori}
unknown coefficients of
$\mathcal{H}$ thus reduce to the anomalous average $\langle 
\hat{\psi}_\downarrow \hat{\psi}_\uparrow\rangle$, and
Eq.(\ref{eq:sc2}) remains as the only self-consistent equation,
the other one Eq.(\ref{eq:sc1}) giving {\it explicitly}
the density as a function of $\mu$ and of the anomalous average.

In general, one has in practice to use the BCS theory 
in this $l\to 0$ limit, to be
in the continuous gas regime. As we shall see in section \ref{subsubsec:lc},
a notable exception is the weakly attractive regime $k_F a \to 0^-$.

\subsubsection{BCS prediction for an energy gap}

As sketched in the last paragraph of section
\ref{subsec:emwtbf}, the excited states of the BCS Hamiltonian
$\mathcal{H}$ correspond to  
a variational BCS approximation for excited states of the 
full Hamiltonian of the gas. Here we come back to this statement,
and show how to derive one of the most celebrated predictions of
BCS theory, the minimum energy required to break a pair.

First we revisit briefly section \ref{subsec:emwtbf}: in the process
of energy minimization within the BCS family of states, we ended up
looking for a BCS state that is the ground state of a quadratic Hamiltonian
of the form Eq.(\ref{eq:BCS_ham}). Taking here for simplicity
the continuous limit $l\to 0$, for a spatially homogeneous solution,
we can omit the Hartree mean field term and perform a variational
calculation that {\it immediately} restricts to BCS states being the ground
state of the following Hamiltonian,
\be
\mathcal{H}_\lambda = \sum_{\vec{k}\in \mathcal{D}}\sum_\sigma
\left(E_k-\mu\right) a_{\vec{k}\sigma}^\dagger 
a_{\vec{k}\sigma}
+ l^3 \sum_{\vec{r}} \left[\lambda\hat{\psi}_\uparrow^\dagger(\vec{r}\,)
\hat{\psi}_\downarrow^\dagger(\vec{r}\,)+ \mbox{h.c.}\right]
\ee
where $E_k=\hbar^2 k^2/2m$ and
the real non-negative number $\lambda$ 
is now the only variational parameter.
The expectation value of the full Hamiltonian $H$ in the ground state
$|\Psi_\lambda^g\rangle$
of $\mathcal{H}_\lambda$ is readily evaluated from section \ref{subsec:linham}
and from Eqs.(\ref{eq:spectre_bcs},\ref{eq:uiv}),
\be
\langle H\rangle_{\lambda}^{g} = 
\sum_{\vec{k}\in \mathcal{D}} (E_k-\mu)\left(1-
\frac{E_k-\mu}{|E_k-\mu+i\lambda|}\right)
+  g_0 L^3 
\left(\langle \hat{\psi}_\downarrow \hat{\psi}_\uparrow\rangle_\lambda^{g}
\right)^2,
\label{eq:hlg}
\ee
with the anomalous average
$\langle \hat{\psi}_\downarrow \hat{\psi}_\uparrow\rangle_\lambda^{g} =
-L^{-3} \sum_{\vec{k}\in \mathcal{D}}
\lambda/(2|E_k-\mu+i\lambda|).$
In these expressions, the appex $g$ stands for ``ground state".
It remains to require that the first order derivative of
$\langle H\rangle_{\lambda}^{g}$ with respect to $\lambda$ vanishes,
which reduces, after explicit calculation of the derivative,
to the following equation satisfied by the optimal
value $\lambda_g$ of $\lambda$:
\be
g_0 \langle \hat{\psi}_\downarrow \hat{\psi}_\uparrow\rangle_{\lambda_g}^{g} =
\lambda_g.
\label{eq:sca}
\ee
One has recovered, in a slightly different way,
the self-consistency condition Eq.(\ref{eq:sc2}), and one finds that
$\lambda_g$ satisfies the gap equation (with $\tilde{\mu}=\mu$).
This optimal $\lambda_g$ was called earlier $\Delta$.

We can now perform a similar variational calculation, taking as a
variational ansatz an excited state of $\mathcal{H}_\lambda$.
To preserve the symmetry between the two spin components $\uparrow$
and $\downarrow$, we consider the excited state
\be
|\Psi_\lambda^e\rangle = b_{\vec{q}\,\uparrow}^\dagger 
b_{-\vec{q}\downarrow}^\dagger |\Psi_\lambda^g\rangle,
\ee
where $\vec{q}$ is a fixed wavevector. Two Bogoliubov modes
of the Hamiltonian $\mathcal{H}_\lambda$ 
are now in the excited state, with occupation number one, rather than being
in the vacuum state of the corresponding $b$'s.
The mean energy of $H$ in
this excited state is simply, in the thermodynamic limit,
\be
\langle H\rangle_\lambda^e = \langle H\rangle_\lambda^g 
+\frac{2(E_q-\mu)^2+2\lambda g_0\langle \hat{\psi}_\downarrow \hat{\psi}_\uparrow\rangle_\lambda^{g}}{|E_q-\mu+i\lambda|}
+O(L^{-3}).
\label{eq:ener_e}
\ee
It differs from $\langle H\rangle_\lambda^g$ by a term $O(1)$,
whereas the full energy is $O(L^3)$, so that
the optimal value $\lambda_e$ of $\lambda$ for the minimization of 
$\langle H\rangle_\lambda^e$ differs from $\lambda_g=\Delta$ by
a small term $\delta\lambda$.
We can then simply expand Eq.(\ref{eq:ener_e}) in powers of $\delta \lambda$.
The key point now is that the ground state energy varies only to second
order in $\delta\lambda$ since it is minimal in $\Delta$;
since the second order derivative of $\langle H\rangle_\lambda^g$
is $O(L^3)$, $\delta\lambda$ is $O(1/L^3)$ and the contribution 
to the total energy
of the corresponding second order variation of $\langle H\rangle_\lambda^g$
is $O(1/L^3)$ and negligible.
In the remaining terms of Eq.(\ref{eq:ener_e}) we can simply 
set $\lambda=\Delta$.
Using Eq.(\ref{eq:sca}), we are left with
\be
\langle H\rangle_{\lambda_e}^e = \langle H\rangle_{\Delta}^g +
2\epsilon_q + O(1/L^3)
\ee
where $\epsilon_q$ is the BCS spectrum Eq.(\ref{eq:spectre_bcs}).

To summarize, one considers excited states of $\mathcal{H}$ 
as variational states, with a small number of Bogoliubov excitations.
One should then in principle
solve again the self-consistency conditions
Eqs.(\ref{eq:sc1},\ref{eq:sc2}), but 
in the thermodynamic limit the excitation
of a few Bogoliubov modes has a small effect on the density and gap parameter, 
an effect that we have shown to be negligible on the energy. 
One can then directly consider
that the elementary excitations of the quadratic Hamiltonian $\mathcal{H}$ 
associated
to the ground energy BCS state 
actually give the energy of the elementary excitations of BCS theory
\cite{canonique}.

When $\Delta>0$, we see that the minimal value of the BCS spectrum
$\epsilon_k$ with respect to $k$ is non-zero: it has a gap 
$E_{\rm gap}=\mbox{min}_k \epsilon_k$.
When $\tilde{\mu}>0$, which contains the usual regime of the BCS theory,
the regime of condensation of Cooper pairs
(see below), the gap is $E_{\rm gap}=\Delta$, hence the name of $\Delta$.
When $\tilde{\mu}<0$, which contains the regime of a Bose-Einstein
condensate of dimers, the gap is $E_{\rm gap}=
\left(\tilde{\mu}^2+\Delta^2\right)^{1/2}$.
Finally, we get from BCS theory a prediction for the miminal energy
$2 E_{\rm gap}$ required to break a pair,
that is to get from a condensate of $N/2$ pairs,
a condensate of $N/2-1$ pairs and two unpaired atoms \cite{whygap}. 

\subsubsection{BCS predictions in limiting cases}
\label{subsubsec:lc}

We shall not discuss the full solution of the BCS
equations but we briefly review simple limiting cases.
We introduce the Fermi momentum $k_F$ of the ideal Fermi gas,
such that $k_F^3/(6\pi^2)=\rho_\uparrow=\rho_\downarrow$.

\noindent {\bf Limit $k_F a \rightarrow 0^-$:} 
the gap parameter $\Delta$ tends exponentially to zero in this limit,
as we shall see.
For such a small value of $\Delta$, one may set $\Delta=0$ in
Eq.(\ref{eq:dens_homo}), to obtain that
$\tilde{\mu}$ is the Fermi energy of the ideal gas $\hbar^2 k_F^2/2m$, 
so that
\be
\mu \simeq \frac{\hbar^2}{2m} (6\pi^2\rho_\uparrow)^{2/3} + g_0 \rho_\uparrow.
\ee
If one takes the mathematical limit $l\to 0$ the Hartree mean field
term disappears since $g_0\to 0$ in this limit, as already mentioned. 
However in the present weakly
attractive limit one can choose $|a| \ll l \ll k_F^{-1}$ so that
the continuous space limit and the zero-range condition
Eq.(\ref{eq:zero_range_for_model}) for the lattice model
are obtained while $g_0 \simeq g$, see Eq.(\ref{eq:g0_exp}).
The on-site interaction potential
is then treatable in the Born approximation, which makes the BCS
approach more accurate. 
One then indeed recovers the first term $g \rho_\uparrow$ of a
systematic expansion of $\mu$ in powers of $k_F a$ \cite{Heiselberg}.
Now turning to the gap equation, one finds a gap parameter \cite{Randeria}
\be
\Delta \simeq 8 e^{-2} \tilde{\mu}\, e^{-\pi/(2 k_F |a|)}.
\label{eq:equivDelta}
\ee
Since $\tilde{\mu}=\hbar^2 k_F^2/2m>0$ in this weakly attractive limit, 
$|\Gamma_k|$ can assume
extremely large values, for $\hbar^2 k^2/2m < \tilde{\mu}$: the
pairs that are condensed are {\it not} bosons.
Note the non-analytic dependence of the gap on the small parameter
$k_F |a|$,  which indicates that the BCS state in the thermodynamic
limit cannot be obtained by
a perturbative treatment of the interaction potential.
This non-analytic dependence can be readily seen
from Eq.(\ref{eq:gap_homo}), whose integrand diverges 
in $k=\sqrt{2m\tilde{\mu}}/\hbar$, in the limit $\Delta\rightarrow 0$.
Replacing the $k$ integration variable by the ideal gas
mode energy $E=\hbar^2 k^2/2m$,
and approximating the ideal gas density of modes by a constant 
in the energy interval of half-width $\delta\ll \tilde{\mu}$ around $\tilde{\mu}$, one
gets a contribution
\be
\int_{\tilde{\mu}-\delta}^{\tilde{\mu}+\delta} \frac{dE}{\sqrt{\Delta^2 + 
(E-\tilde{\mu})^2}}
\stackrel{\Delta\rightarrow 0}\sim 2\, \ln\frac{2\delta}{\Delta} .
\ee
The same technique applied over e.g.\ the interval $E\in [0,4\tilde{\mu}]$
leads to Eq.(\ref{eq:equivDelta}), when one also takes into account
(to lowest order in $\Delta$) the difference 
between the exact integral in the gap equation and the approximate one.

{\bf Limit $k_F a\rightarrow 0^+$:} the coefficients $\Gamma_k$
tend uniformly to zero in this limit,  because $\tilde{\mu}<0$
and $\Delta \ll |\tilde{\mu}|$,
so that the pair creation operator $C^\dagger$
obeys approximately bosonic commutation relations.
This was expected physically, since the ground state of the gas is 
a condensate of almost bosonic dimers. Since two atoms in a dimer
are at a relative distance $\sim a$, one should ensure $l\ll a$
in the lattice model, so that one takes the mathematical
limit $l\to 0$, which implies $\tilde{\mu}\to \mu$.
Let us check that the BCS theory
correctly predicts a condensate of such dimers. We first simplify the gap equation 
Eq.(\ref{eq:gap_homo}) by using $\epsilon_k \simeq -\mu +\hbar^2 k^2/2m$.
After division by $\Delta$, we get an equation for the chemical potential
\be
\frac{1}{g} \simeq -\int \frac{d^3k}{(2\pi)^3} \left[
\frac{1}{\frac{\hbar^2 k^2}{m}-2\mu}-\frac{m}{\hbar^2 k^2}
\right]
\ee
which leads to
\be
\mu \simeq -\frac{\hbar^2}{2ma^2}.
\ee
This is minus half the binding energy of a dimer, exactly what was expected
(keep in mind that $\mu N = \mu_{\rm mol} N_{\rm mol}$ where $N_{\rm mol}$
is the total number of dimers and is equal to $N/2$, so that the molecular
chemical potential $\mu_{\rm mol}$ is twice the atomic one).
The next step is to expand the integrand of Eq.(\ref{eq:dens_homo})
to leading order in $\Delta$ to calculate the gap parameter \cite{Randeria}:
\be
\Delta  \simeq \frac{2}{\sqrt{3\pi}} (k_F a)^{3/2} \frac{\hbar^2}{m a^2}
\ll \frac{\hbar^2}{m a^2}.
\ee
Note that in this molecular BEC regime, $\Delta$ is {\it not} proportional
to the energy required to break a pair.
One sees from Eq.(\ref{eq:spectre_bcs}) that the gap in $\epsilon_k$ is $\simeq |\mu|$,
since $\mu$ is negative and much larger in absolute value than $\Delta$.
The energy to break a pair is then $2|\mu|$, which is indeed the binding
energy of a dimer.
Finally, by performing the Fourier transform in
 Eq.(\ref{eq:phi_homo}), approximating $V_k/U_k$
to leading order in $\Delta$,
one obtains the pair wavefunction
\be
\phi(\vec{r_1}-\vec{r_2}) \simeq \frac{1}{L^{3/2}} \phi_0(|\vec{r_1}-\vec{r_2}|)
\ee 
where $\phi_0$ is the wavefunction of the
bound state of two atoms given by Eq.(\ref{eq:phi0}), and where we
used $\gamma^2=\sum_{\vec{k}} (V_k/U_k)^2$.

{\bf Limit $k_F |a|=+\infty$}: the numerical solution of the 
gap equation Eq.(\ref{eq:gap_homo})  and of the density equation 
Eq.(\ref{eq:dens_homo}) in the limit $l\to 0$
gives the BCS estimate of the numerical coefficient
$\eta$ of Eq.(\ref{eq:eta}) for the ground branch. 
This estimate is an upper bound \cite{Randeria2}:
\be
\eta \leq \eta_{\rm BCS} = 0.5906\ldots
\ee
A much better estimate was obtained by approximate (fixed node) Monte Carlo 
calculations \cite{Pandharipande,Giorgini},
$\eta \simeq  0.4$,
a value confirmed by recent exact quantum Monte Carlo calculations \cite{Juillet}.
Early measurements of $\eta$ were in contradiction with these
theoretical values \cite{Thomas}, but 
more precise measurement performed in Innsbruck \cite{Grimm_eta}, in Paris
\cite{Salomon_eta} and at Rice University \cite{Hulet_eta} are consistent
with them.

\subsection{Derivation of superfluid hydrodynamic equations from BCS theory}

A key point of the BCS theory is to predict a spectrum of
elementary excitations having a gap, see Eq.(\ref{eq:spectre_bcs}).
One could then be tempted to infer strong predictions on
the thermodynamic properties, e.g.\ that the entropy obeys
an activation law $O(e^{-E_{\rm gap}/k_B T})$ at low temperature,
where $E_{\mathrm{gap}}$ is the energy gap. 
However, in the BEC limit, where the gas is
simply a Bose-Einstein condensate of bosonic dimers at low temperature,
this prediction of an activation law
is obviously wrong, since one knows that the relevant excitation spectrum
is the Bogoliubov spectrum, which has no gap.

In reality, the spatially homogeneous gas, whatever
the considered regime (BEC or BCS), is expected to have a branch of collective
excitations with no gap, behaving as sound waves at low momenta.
These collective excitations correspond to coherent oscillations
of the density of the pairs, which are not gapped, to be distinguished
from the pair breaking excitations which are gapped.
The key point briefly addressed in this section is that the {\it time-dependent}
BCS theory contains such a branch of collective, non-pair-breaking 
oscillations \cite{if_neutral}.
This is similar to what happens for the weakly interacting Bose gas:
the Bogoliubov spectrum is obtained from a linearization of the time
dependent Gross-Pitaevskii equation.

Here, rather than performing an exact linearization of the time-dependent
BCS equations, which leads to the so-called RPA approach
\cite{Blaizot,Griffin_rpa,Bruun_rpa,Minguzzi_rpa,Zwerger_rpa,Combescot_rpa}, we go through
a sequence of simple approximations allowing one to derive superfluid
hydrodynamic equations from the time-dependent BCS theory. This has
several advantages: it is more physical, it easily applies 
to harmonic trapping \cite{Bruun_le_fait}, and it is applicable to
the non-linear time dependent regime, to study analytically
low-energy collective excitations of a trapped superfluid Fermi gas
\cite{Baranov,Anna}, but also the expansion
of the gas after the trap was switched off \cite{Stringari_tof}
and the response of the gas to a rotation of the harmonic trap
\cite{Stringari2,Tonini} even in the non-linear regime \cite{Tonini}.

Coming back to thermodynamical aspects, one may fear that the straightforward
finite temperature extension of the BCS theory, with a variational density
operator which is Gaussian in the field operators \cite{Blaizot}, 
is not able to calculate the critical temperature $T_c$
with a good accuracy, 
because it does not correctly include the collective excitations.
It turns out that this simple finite temperature BCS theory 
correctly gives $T_c$ in the BCS limit $k_F a\to 0^-$ within 
a numerical factor \cite{TCbcs}, but is indeed totally wrong in the
BEC limit $k_F a\to 0^+$ where it does not reproduce at all
Einstein's prediction for the critical temperature of the ideal Bose gas.
Refinements of the BCS theory have been developed to recover
Einstein's prediction and to obtain a calculation of $T_c$ approximately
valid over the whole range of values of $k_F a$ 
\cite{Nozieres,Randeria}.

\subsubsection{Time dependent BCS theory}

This is a direct generalization of the static case 
of section \ref{subsec:emwtbf}. The exact $N$-body Schr\"odinger
equation can be obtained from extremalisation over the time-dependent
state vector $|\Psi(t)\rangle$ of the following action,
\be
S = \int_{t_i}^{t_f} dt\, 
\left\{ \frac{i\hbar}{2} 
\left[\langle \Psi(t)| \frac{d}{dt} |\Psi(t)\rangle
-\mbox{c.c.}\right] - \langle\Psi(t)|H(t)|\Psi(t)\rangle \right\}
\ee
for fixed initial $|\Psi(t_i)\rangle$ and final $|\Psi(t_f)\rangle$
values of the state vector. For an arbitrary variation $|\delta\Psi(t)\rangle$ 
of the ket $|\Psi(t)\rangle$, subject to the condition
\be
|\delta\Psi(t_i)\rangle=|\delta\Psi(t_f)\rangle=0,
\label{eq:dtidtf}
\ee 
we calculate the first order variation of the action,
\be
\delta S= \int_{t_i}^{t_f} dt\, \left\{\langle\delta\Psi(t)|
\left[i\hbar \frac{d}{dt} - H(t)\right]|\Psi(t)\rangle+\mbox{c.c.}\right\}.
\label{eq:dS}
\ee
We have integrated by parts over $t$ to get rid of $(d/dt) |\delta\Psi\rangle$
and we have used Eq.(\ref{eq:dtidtf}) to show that the fully integrated term
vanishes. The condition that $\delta S$ vanishes for all ket variations
indeed leads to Schr\"odinger's equation.

For time dependent BCS theory, one forces the ket to be of the BCS
form Eq.(\ref{eq:psi_bcs}), $|\Psi\rangle = |\Psi_{\rm BCS}\rangle$. 
The variation of the ket is performed within the BCS family, by a variation
of $\Phi=\gamma \phi$ and of $\mathcal{N}$, so that
$|\delta\Psi\rangle = \delta(|\Psi_{\rm BCS}\rangle)$.
We can then introduce the now time dependent quadratic Hamiltonian
$\mathcal{H}(t)$ constructed in section \ref{subsec:emwtbf}
such that, for all variations within BCS family,
\be
(\delta\langle\Psi_{\rm BCS}(t)|) H(t) |\Psi_{\rm BCS}(t)\rangle
=
(\delta\langle\Psi_{\rm BCS}(t)|) \mathcal{H}(t) |\Psi_{\rm BCS}(t)\rangle.
\ee
One then sees that $\delta S$ in Eq.(\ref{eq:dS}) identically vanishes
if the ket evolves with the quadratic Hamiltonian,
\be
i\hbar \frac{d}{dt} |\Psi_{\rm BCS}(t)\rangle  = \mathcal{H}(t)
|\Psi_{\rm BCS}(t)\rangle.
\label{eq:tdbcs}
\ee
It remains to check that the ket evolving this way remains
of the BCS form. To this end, one
shows, with the reasoning having led to Eq.(\ref{eq:com}),
that $\hat{\psi}_\sigma |\Psi_{\rm BCS}(t)\rangle$
is equal to a linear combination of the $\hat{\psi}^\dagger_{\sigma'}
(\vec{r}\,)$ acting on $|\Psi_{\rm BCS}(t)\rangle$.
One then sees that $\mathcal{H} |\Psi_{\rm BCS}\rangle$ is
of the form ``constant plus polynomial of degree exactly two
in $\hat{\psi}^\dagger$" acting on $|\Psi_{\rm BCS}(t)\rangle$,
which can be reproduced by an appropriate time dependence
of $\Phi$.

In practice, the equation of motion for $\Phi$ is not useful.
One rather moves to the Heisenberg picture with respect to the
time dependent Hamiltonian $\mathcal{H}(t)$, as we did in
section \ref{subsec:linham}. Since this Hamiltonian
is quadratic, the equations of motion for the fields are linear, of the form
Eq.(\ref{eq:leom}), where $L^\uparrow$ is now time dependent.
Since these equations are linear, we can solve them by
evolving in Eqs.(\ref{eq:expan0},\ref{eq:expan}) the mode functions $(u_\alpha,v_\alpha)$ 
while keeping constant the quantum coefficients 
$\hat{b}_{\alpha\sigma}$ where $\sigma=\uparrow$ or $\downarrow$:
\be
i\hbar \partial_t 
\left(\begin{array}{c} u_\alpha \\ v_\alpha\end{array}\right) 
= L^\uparrow(t) 
\left(\begin{array}{c} u_\alpha \\ v_\alpha\end{array}\right) .
\label{eq:uvt}
\ee
These equations on the mode functions
are effectively non-linear since coefficients of $L^\uparrow$,
involving the mean density $\langle \hat{\psi}_\uparrow^\dagger
\hat{\psi}_\uparrow \rangle$ and the anomalous average $\langle \hat{\psi}_\downarrow
\hat{\psi}_\uparrow \rangle$,
depend on the mode functions; 
since the $\hat{b}_{\alpha\sigma}$ are constants of motion,
this dependence
is still given by Eqs.(\ref{eq:rho_modal},\ref{eq:delta_modal}), now taking
the time dependent $u_\alpha$'s and $v_\alpha$'s.

\subsubsection{Semi-classical approximation}

The physical situation that we have in mind here is a gas
in the BCS regime ($a<0$, $\mu>0$) in a time dependent
harmonic trap. The trap may be rotating, with an angular velocity $\vec{\Omega}(t)$,
in which case one moves to the rotating frame to eliminate the time dependence
of the trapping potential due to rotation; this introduces an additional term
$-\vec{\Omega}(t)\cdot \vec{L}$ in the one-body Hamiltonian, where
$\vec{L}$ is the angular momentum operator of a particle.
In this case one may expect that quantum vortices form 
\cite{Ketterle_vortices} for a high enough
rotation frequency; the semi-classical approximation
that we present is however restricted to a vortex-free gas.

We wish to treat the equation of motion of $(u_\alpha, v_\alpha)$ 
in the semi-classical approximation.
The general validity condition of a semi-classical approximation is 
that the applied potentials vary slowly over the coherence lengths of the gas,
a coherence length being the typical width of a correlation function of the gas.

A first correlation function here 
is $\langle \psi^\dagger_\uparrow(\vec{r}\,)\psi_\uparrow(\vec{r}\,')\rangle$.
Using the homogeneous gas expression of this function, one sees that it
is the Fourier transform of the function $V_k^2$
of width the Fermi momentum, $\Delta k = k_F$, where $\hbar^2 k_F^2/2m = \mu$.
The associated coherence length is $1/\Delta k= k_F^{-1}$.

A second correlation function is $\langle \hat{\psi}_\downarrow(\vec{r}\,)
\hat{\psi}_\uparrow(\vec{r}\,')\rangle$. For the homogeneous gas this 
is the Fourier transform of the function $U_k V_k$, which according
to Eq.(\ref{eq:uiv}) has a momentum width $\Delta k=m |\Delta|/\hbar^2 k_F$.
The associated coherence length $1/\Delta k$ corresponds in BCS theory
to the length of a Cooper pair, 
\be
l_{\rm BCS} = \frac{\hbar^2 k_F}{m |\Delta|}.
\ee
Since $\Delta < \mu$ in the BCS regime, we keep
$l_{\rm BCS}$ as the relevant coherence length.

A first typical length scale of variation of the matrix elements in 
Eq.(\ref{eq:uvt}) originates from the position dependence of $|\Delta|$: 
in the absence of rotation and for an isotropic trap, 
this is expected to be the Thomas-Fermi radius
$R_{\rm TF}$ of the gas, defined as $\mu = m\omega^2 R_{\rm TF}^2/2$,
where $\omega$ is the atomic oscillation frequency.
This assumes that the scale of variation of the modulus of the gap 
is the same as the one of the density, a point confirmed
in the adiabatic approximation to come.
The condition that the mean field Hartree term and the harmonic potential
have a weak relative variation over $l_{\rm BCS}$ for typical values of
the position also leads to the condition
\be
l_{\rm BCS} \ll R_{\rm TF}.
\ee
For an isotropic harmonic trap this is equivalent to the condition
\be
|\Delta| \gg \hbar \omega.
\label{eq:cond_int}
\ee

In the general time dependent case, however, this is not the whole story, 
since the phase of $\Delta$
can also become position dependent. In the rotating case
for example, with a rotation frequency of the order of $\omega$,
the phase of $\Delta$ may vary as $\simeq m \omega x y/\hbar$; this introduces
a local wavevector $m \omega R_{\rm TF}/\hbar \simeq k_F$, 
making a semi-classical approximation hopeless. We eliminate this problem 
with a gauge transform of the $u$'s and $v$'s \cite{Tonini}:
\bea
\tilde{u}_\alpha(\vec{r},t) &\equiv& u_\alpha(\vec{r},t) e^{-i S(\vec{r},t)/\hbar} \\
\tilde{v}_\alpha(\vec{r},t) &\equiv& v_\alpha(\vec{r},t) e^{+i S(\vec{r},t)/\hbar},
\eea
where $S$ is defined as
\be
\Delta(\vec{r},t) = |\Delta(\vec{r},t)| \,
e^{2 i S(\vec{r},t)/\hbar}.
\ee
The time dependent BCS equations Eq.(\ref{eq:uvt}) are modified,
$\Delta$ being replaced by $|\Delta|$ and the single particle
Hamiltonian $h$ being replaced by the gauge transformed Hamiltonian
\be
\tilde{h}= e^{-i S/\hbar} h e^{+i S/\hbar} + \partial_t S.
\ee

An efficient frame to perform a semi-classical approximation in
a systematic way for the evolution of a wave
in a potential is to use the Wigner representation \cite{Wigner}
of the density operator of the wave.
Here the wave is represented by the two-component spinor
$(\tilde{u}_\alpha,\tilde{v}_\alpha)$ so that we introduce the corresponding
density operator
\be
\hat{\sigma} = 
\left(\begin{array}{cc}
\hat{\sigma}_{11} & \hat{\sigma}_{12} \\
\hat{\sigma}_{21} & \hat{\sigma}_{22}\end{array}\right)
\equiv \sum_{\alpha}
\left(\begin{array}{cc} |\tilde{u}_\alpha\rangle \langle \tilde{u}_\alpha |  &
|\tilde{u}_\alpha\rangle \langle \tilde{v}_\alpha |  \\
|\tilde{v}_\alpha\rangle \langle \tilde{u}_\alpha |  & 
|\tilde{v}_\alpha\rangle \langle \tilde{v}_\alpha | \end{array}\right).
\ee
Note that the matrix elements of $\hat{\sigma}$ in position space are related
to the previously mentioned correlation functions of the gas, up to the gauge
transform, which makes the discussion consistent:
\bea
|\langle\vec{r}\,|\hat{\sigma}_{22}|\vec{r}\,'\rangle| & =&
|\langle \hat{\psi}_\uparrow^\dagger(\vec{r}\,) \hat{\psi}_\uparrow(\vec{r}\,')\rangle | \\
|\langle\vec{r}\,|\hat{\sigma}_{12}|\vec{r}\,'\rangle| & =&
|\langle \hat{\psi}_\downarrow(\vec{r}\,) \hat{\psi}_\uparrow(\vec{r}\,')\rangle |.
\eea

We introduce the Wigner representation of  $\hat{\sigma}$ \cite{Wigner}, assuming
for simplicity a continuous position space:
\begin{equation}
W(\vec{r},\vec{p},t) = \mathrm{Wigner}\{\hat{\sigma}\}\equiv
\int \frac{d^3 x}{(2\pi\hbar)^3} \langle \vec{r}-\vec{x}/2
|\hat{\sigma}| \vec{r}+\vec{x}/2\rangle e^{i\vec{p}\cdot\vec{x}/\hbar}.
\end{equation}
Since $\hat{\sigma}$ is the density operator of a two-component spinor,
the Wigner distribution $W$ is a two by two matrix.
Within this representation, the key quantities of BCS theory, the total density,
the modulus of the gap function and the total matter current have the following
expressions in the general case of a rotating frame:
\bea
\label{eq:rho_wig}
\rho(\vec{r},t) &=& 2\int_{\hbar\mathcal{D}}
d^3{p} \, W_{22}(\vec{r},\vec{p},t) \\
\label{eq:delta_wig}
|\Delta|(\vec{r},t) &=& -g_0 \int_{\hbar\mathcal{D}}
d^3{p}\, W_{12}(\vec{r},\vec{p},t) \\
\label{eq:j_wig}
\vec{\mbox{\j}}\,(\vec{r},t) &=& \rho \, \left[\vec{v}\,(\vec{r},t) -\vec{\Omega}(t)\times\vec{r}\,\right] 
- \frac{2}{m} \int_{\hbar\mathcal{D}}
d^3{p}\ \vec{p}\, W_{22}(\vec{r},\vec{p},t),
\eea
where $\vec{\Omega}$ is the angular velocity of the rotating frame,
and the so-called velocity field is defined as
\be
\vec{v}(\vec{r},t)\equiv \frac{\vec{\rm grad}\,S(\vec{r},t)}{m}.
\ee
We have introduced here the total matter current $\vec{\mbox{\j}}(\vec{r},t)$, 
that obeys by construction
\be
\partial_t \rho + \mathrm{div} \, \vec{\mbox{\j}} =0,
\label{eq:cont}
\ee
In the rotating frame, in a many-body state invariant by exchange of the spin states
$\uparrow$ and $\downarrow$,
it is very generally given by 
\be
\vec{\mbox{\j}}(\vec{r},t)=\frac{\hbar}{im}\left(\langle \hat{\psi}_{\uparrow}^\dagger (\vec{r},t)
\vec{\rm grad}\,\hat{\psi}_{\uparrow}(\vec{r},t)\rangle-
\mbox{c.c.}\right)
- \rho(\vec{r},t)\, \vec{\Omega}(t)\times \vec{r},
\end{equation}
Within BCS theory, these last two relations still hold \cite{Blaizot}
and lead to Eq.(\ref{eq:j_wig}).

The semi-classical expansion then consists e.g.\ in
\be
\mathrm{Wigner}\{V(\hat{\vec{r}}\,)\hat{\sigma}\} =
[V(\vec{r}\,)+\frac{i\hbar}{2}\partial_{\vec{r}\,}V
\cdot \partial_{\vec{p}}+\ldots ] W(\vec{r},\vec{p},t),
\end{equation}
where $V$ is a position-dependent potential.
The successive terms in this expansion
we call zeroth order, first order, etc, in the semi-classical approximation.
Assuming that the momentum derivative of $W$ is $\simeq W/\Delta p$,
where $\Delta p$ is the momentum width of $W$, we recover the fact that
the first order term is small as compared to the zeroth order one
if the variation of $V$ over the coherence length $\hbar/\Delta p$,
which is of the order of the spatial derivative of $V$ times the coherence length,
is small as compared to $V$.

Here we shall need only the equations of motion of 
the Wigner distribution $W$ up to zeroth order in the semi-classical
approximation:
\be
i\hbar\partial_t W(\vec{r},\vec{p},t)
\simeq  \left[L^\uparrow_0(\vec{r},\vec{p},t),W(\vec{r},\vec{p},t)\right]
\label{eq:zeroth}
\ee
where the two by two matrix $L^\uparrow_0$ is equal to
\be
L^\uparrow_0(\vec{r},\vec{p},t)= \left(
\begin{array}{cc}
\frac{p^2}{2m} - \mu_{\rm eff}(\vec{r},t)  & |\Delta|(\vec{r},t) \\
|\Delta|(\vec{r},t) &  - \frac{p^2}{2m}+\mu_{\rm eff}(\vec{r},t) 
\end{array}
 \right).
\label{eq:Lspin}
\ee
We have introduced the position and time dependent function,
\begin{equation}
\mu_{\rm eff}(\vec{r},t) \equiv \mu -U(\vec{r},t) 
-\frac{1}{2} m v^2(\vec{r},t) +
m\vec{v}\,(\vec{r},t)\cdot 
(\vec{\Omega}(t)\times \vec{r}\,) - \partial_t S(\vec{r},t),
\label{eq:mu_eff}
\end{equation}
that may be called effective chemical potential for reasons 
that will become clear later.

At time $t=0$, the gas is at zero temperature. 
By introducing the spectral decomposition
of $L^\uparrow(t=0)$ one can then check that
\be
\hat{\sigma}(t=0) =\theta[L^\uparrow(t=0)]
\ee
where $\theta(x)$ is the Heaviside function. Since $L^\uparrow_0(t=0)$ 
is the classical limit
of the operator ${L}^\uparrow(t=0)$, the leading order semi-classical approximation
for the corresponding Wigner function is, in a standard way, given by
\be
W(\vec{r},\vec{p},t=0) 
\simeq \frac{1}{(2\pi\hbar)^3}\,\theta[L_0^\uparrow(\vec{r},\vec{p},t=0)].
\end{equation}
This implies that the $2\times 2$  matrix $W$ is 
proportional to a pure spin-1/2 state $|\psi\rangle\langle \psi|$ with
\be
|\psi(\vec{r},\vec{p},t=0)\rangle = 
\left(\begin{array}{c} U_0(\vec{r},\vec{p}\,) \\
V_0(\vec{r},\vec{p}\,)
\end{array}
\right)
\ee
where $(U_0,V_0)$ is the eigenvector 
of $L^\uparrow_0(\vec{r},\vec{p},t=0)$ of positive energy
and normalized to unity.
At time $t$, according to the zeroth order evolution Eq.(\ref{eq:zeroth}), 
each two by two
matrix $W$ remains a pure state, with components $U$ and $V$ solving
\be
i\hbar\partial_t \left(\begin{array}{c} U(\vec{r},\vec{p},t) \\
V(\vec{r},\vec{p},t) \end{array} \right)
=L_0^\uparrow(\vec{r},\vec{p},t)
\left(\begin{array}{c} U(\vec{r},\vec{p},t)  \\
V(\vec{r},\vec{p},t) \end{array} \right).
\ee

\subsubsection{Adiabatic approximation}

We then introduce the so-called adiabatic approximation: the vector $(U,V)$, being
initially an eigenstate of $L_0^\uparrow(\vec{r},\vec{p},t=0)$, 
will be an instantaneous
eigenvector of $L^\uparrow_0(\vec{r},\vec{p},t)$ at all later times $t$.
This approximation holds under the validity condition of the adiabaticity
theorem \cite{adiab}, 
discussed for our specific case in \cite{Tonini},
generically requiring that the energy difference 
between the two eigenvalues of $L^\uparrow_0(\vec{r},\vec{p},t)$
be large enough.
As this energy difference can be as small as the gap parameter,
this imposes a minimal value to the gap, not necessarily
coinciding with the one of Eq.(\ref{eq:cond_int}), as shown in \cite{Tonini}.
In this adiabatic approximation, one can take 
\be
W(\vec{r},\vec{p},t) \simeq\frac{1}{(2\pi\hbar)^3}
\,\theta[L_0^\uparrow(\vec{r},\vec{p},t)]=
\frac{1}{(2\pi\hbar)^3}\,|+(\vec{r},\vec{p},t)\rangle
\langle +(\vec{r},\vec{p},t)|
\label{eq:++}
\ee
where $|+(\vec{r},\vec{p},t)\rangle$, 
having real components $(U_{\rm inst},V_{\rm inst})$, is the
instantaneous eigenvector with positive eigenvalue
of the matrix $L_0^\uparrow$ defined in Eq.(\ref{eq:Lspin}).
$(U_{\rm inst},V_{\rm inst})$ are simply the amplitudes 
on the plane wave $\exp(i\vec{p}\cdot\vec{r}/\hbar)$
of the BCS mode functions of a spatially uniform BCS gas 
of chemical potential $\mu_{\rm eff}$
and of gap parameter $|\Delta(\vec{r},t)|$. 
Using Eq.(\ref{eq:rho_wig}) and Eq.(\ref{eq:delta_wig})
with the approximate Wigner distribution Eq.(\ref{eq:++}), 
one further finds that this fictitious spatially uniform
BCS gas is at equilibrium at zero temperature so that the zero
temperature equations of state may be used, relating the
chemical potential to the density, $\mu_0[\rho]$,
and the gap to the density, $\Delta_0[\rho]$, where the functions
$\mu_0$ and $\Delta_0$ may be calculated from the spatially homogeneous
BCS theory.
In particular, the equation of state relating the chemical potential
to the density leads to
\be
\mu_{\rm eff}(\vec{r},t) = \mu_0[\rho(\vec{r},t)]
\label{eq:euler_obtenu}
\ee
which leads, together with Eq.(\ref{eq:mu_eff}), to 
one of the time dependent hydrodynamic equations, the Euler-type one.
Also, $U_{\rm inst}$ and $V_{\rm inst}$ are even functions of
$\vec{p}$, so that the integral in the right hand side of Eq.(\ref{eq:j_wig})
vanishes and Eq.(\ref{eq:cont}) reduces to the 
hydrodynamic continuity equation in a rotating frame,
\be
\partial_t \rho(\vec{r},t) + \mbox{div}\, \left\{\rho(\vec{r},t)
\left[\vec{v}\,(\vec{r},t) - \vec{\Omega}(t)\times \vec{r}\,\right]
\right\}=0.
\ee
Under the adiabatic approximation, the superfluid hydrodynamic equations are
thus derived from BCS theory, remarkably
without having to postulate the expression of the
superfluid current in terms of the gradient of the phase of the order
parameter.
We refer to \cite{Tonini} for a discussion of the validity condition
of the adiabatic approximation.

The presence of superfluid hydrodynamic equations
allows to conclude in the existence of collective modes in
the BCS theory, in a transparent way, by a simple linearization around steady state.
Although we have considered here a trapped system, we note that the formalism
may also be developed in the spatially homogeneous case.
To create a sound wave of wavevector $\vec{q}$, one may apply an external
potential varying with spatial harmonics  $e^{\pm i\vec{q}\cdot\vec{r}}$,
e.g.\ with a Bragg pulse \cite{Grynberg_bragg,Ketterle_bragg}. The semi-classical
approximation holds if this external potential varies slowly
at the scale of the Cooper pair size $l_{\rm BCS}$, that is
$q l_{\rm BCS} \ll 1$. Linearization of hydrodynamic equations in the linear
response regime predicts a linear dispersion relation $\omega_q = c_s q$,
with a sound velocity $c_s$ such that
\be
m c_s^2 = \rho \frac{d\mu_0}{d\rho}[\rho],
\ee
of the order of $\hbar k_F/m$ in the regime $a<0$. 
The semi-classicality condition
$q l_{\rm BCS} \ll 1$ then results in $\hbar \omega_q \ll \Delta$.
This is satisfactory, as for $\hbar\omega_q > 2 \Delta$ one observes in the full RPA
theory a coupling of the sound
wave to the elementary, pair breaking excitations, which may distort 
the dispersion relation and even damp the
sound wave, see e.g.\ \cite{Minguzzi_rpa, Zwerger_rpa, Combescot_rpa}
 and references
therein, a phenomenon not included in superfluid hydrodynamics.

We acknowledge useful suggestions from F\'elix Werner, Mattia Jona-Lasinio,
Christoph Weiss, Alice Sinatra.


\begin{thebibliography}{99}

\bibitem{Houches03}
Castin Y., ``Simple theoretical tools for low dimension Bose gases",
Lecture notes of the 2003 Les Houches Spring School,
{\sl Quantum Gases in Low Dimensions},
Olshanii M., Perrin H., Pricoupenko L. Eds.,
J. Phys. IV France {\bf 116} (2004) 89-132.

\bibitem{Diu}
Diu B., Guthman C., Lederer D., and Roulet B., {\em Physique
Statistique}  (Hermann, Paris, France, 1989).

\bibitem{Efetov}
Efetov K. B. and Larkin A. I., Sov. Phys.  JETP {\bf 42} (1976) 390.

\bibitem{Tracy}
H.G. Vaidya and C.A. Tracy,
Phys. Rev. Lett. {\bf 42}, 3 (1979) and J. Math. Phys. {\bf 20}, 2291 (1979).

\bibitem{bosonisation}
See the paragraph below Eq.(31) in the paper by
Levitov L. S., Lee H.-W., and Lesovik G. B., J. Math. Phys.
{\bf 37} (1996) 4845.

\bibitem{Wigner}
Wigner E. P., Phys. Rev. {\bf 40} (1932) 749.

\bibitem{Minguzzi}
Vignolo P., Minguzzi A., and Tosi M. P.,
Phys. Rev. Lett. {\bf 85} (2000) 2850.

\bibitem{Baranov_pc}
Baranov M., private communication, October 2000.

\bibitem{Minguzzi2}
Akdeniz Z., Vignolo P., Minguzzi A., and Tosi M. P.,
Phys. Rev. A {\bf 66} (2002) 055601.

\bibitem{Petrov}
Petrov D. S., Phys. Rev. Lett. {\bf 93} (2004)  143201.

\bibitem{Moerdijk}
Moerdijk A. J., Verhaar B. J., and Axelsson A., Phys. Rev. A
{\bf 51} (1995) 4852.

\bibitem{Weisskopf}
Blatt J. M. and Weisskopf V. F., in {\it Theoretical Nuclear
Physics} , Wiley, New York (1952).

\bibitem{VarennaJean}
Dalibard J., in {\it Collisional dynamics of ultra-cold atomic gases},
Proceedings of the International School of  Physics Enrico  Fermi, 
Course CXL: Bose -- Einstein condensation in gases, Varenna,  
M. Inguscio, S. Stringari, C. Wieman edts, (1998).

\bibitem{Houches99}
Castin Y., in {\it Coherent atomic matter waves},
 Lecture notes of Les Houches summer school,
 edited by Kaiser R., Westbrook C., and David F.,
 EDP Sciences and Springer-Verlag (2001) 1-136.

\bibitem{CdF}
Cohen-Tannoudji C., 
Course at Coll\`ege de France, Lectures 4, 5 (1998-1999),
available online at  http://www.phys.ens.fr/cours/college-de-france/1998-99/1998-99.htm
                                   
\bibitem{Wilkens}
Busch T., Englert B. G., Rzazewski K., and Wilkens M.,
Found. Phys. {\bf 28} (1998) 549.

\bibitem{Calarco}
Idziaszek Z. and Calarco T., Phys. Rev. A {\bf 71} (2005) 050701(R).


\bibitem{Efimov1}
Efimov V. N., Sov. J. Nucl. Phys. {\bf 12} (1971) 589.

\bibitem{Efimov2} 
Efimov V. N., Nucl. Phys. {\bf A210} (1973) 157.
                                         

\bibitem{Pethick}
Jonsell S., Heiselberg H., and Pethick C. J.,
  Phys. Rev. Lett. {\bf 89} (2002) 250401.

\bibitem{Werner}
Werner F. and  Castin Y.,
Phys. Rev. Lett. {\bf 97} (2006) 150401.

\bibitem{Castin}
Castin Y., Comptes Rendus Physique {\bf 5} (2004) 407.

\bibitem{Tan}
Tan S., cond-mat/0412764.

\bibitem{Werner_v1}
Werner F. and Castin Y., cond-mat/0507399 (v1).

\bibitem{Werner2}
Werner F. and Castin Y.,
Phys. Rev. A {\bf 74} (2006) 053604.

\bibitem{Petrov_fermions}
Petrov D. S., Phys. Rev. A {\bf 67} (2003) 010703.

\bibitem{math_domain}
Reed Michael, and Berry Simon, in {\it Methods of modern mathematical physics 1: functional
analysis}, Academic Press (1980).

\bibitem{doubly_naive}
This is in a way doubly naive for $a\to 0^+$, as far as the pseudo-potential is concerned,
since for $a>0$ the exact problem admits an eigenstate of energy $<3\hbar\omega/2$,
that $\to-\infty$ for $a\to 0^+$.


\bibitem{Heiselberg}
Heiselberg H., Phys. Rev. A {\bf 63} (2001) 043606 and references therein.

\bibitem{Svistunov}
Burovski E., Prokof'ev N., Svistunov B., Troyer M.,
Phys. Rev. Lett. {\bf 96} (2006) 160402;
Burovski E., Prokof'ev N., Svistunov B., Troyer M.,
New J. Phys. {\bf 8} (2006) 153.


\bibitem{Bulgac}
Bulgac A., Drut J. E., Magierski P.,
Phys. Rev. Lett. {\bf 96} (2006) 090404.

\bibitem{Lee}
Lee Dean,  Schaefer Thomas, Phys. Rev. C {\bf 73} (2006) 015202.

\bibitem{Juillet}
Juillet O., cond-mat/0609063.

\bibitem{Mora}
Mora C. and Castin Y., Phys. Rev. A {\bf 67} (2003) 053615.

\bibitem{Pandharipande}
Carlson J., Chang S.-Y., Pandharipande V.R. and  Schmidt K. E.,
Phys. Rev. Lett. {\bf 91} (2003) 050401.

\bibitem{Giorgini}
Astrakharchik G. E.,
Boronat J., Casulleras J., and  Giorgini S.,
Phys. Rev. Lett. {\bf 93} (2004) 200404.

\bibitem{Thomas}
O'Hara K. M., Hemmer S. L., Gehm M. E., Granade S. R., and Thomas J. E.,
Science {\bf 298} (2002) 2179; 
Gehm M. E.,, Hemmer S. L., Granade S. R., O'Hara K. M., and Thomas J. E.,
Phys. Rev. A {\bf 68} (2003) 011401.

\bibitem{Grimm_eta}
M. Bartenstein, A. Altmeyer, S. Riedl, S. Jochim, C. Chin, J. Hecker
Denschlag, R. Grimm,
Phys. Rev. Lett. {\bf 92}, 120401 (2004).

\bibitem{Salomon_eta}
T. Bourdel, L. Khaykovich, J. Cubizolles, J. Zhang, F. Chevy,
 M. Teichmann, L. Tarruell, S. J. J. M. F. Kokkelmans, and C. Salomon,
Phys. Rev. Lett. {\bf 93}, 050401 (2004).

\bibitem{Jin_mol}
Greiner M., Regal C., and Jin D. S., Nature {\bf 426}
(2003) 537.

\bibitem{Grimm_mol}
Jochim S., Bartenstein M., Altmeyer A., Hendl G.,
Riedl S., Chin C., Denschlag J.H., and Grimm R., Science
{\bf 302} (2003) 2101.

\bibitem{Ketterle_mol}
M.W. Zwierlein, C.A. Stan, C.H. Schunck, S.M.F. Raupach,
S. Gupta, Z. Hadzibabic, and W. Ketterle, Phys.
Rev. Lett. {\bf 91}, 250401 (2003).

\bibitem{Hulet_eta}
Partridge G. B., Li W., Kamar R. I., Liao Y. A. and
Hulet R. G., Science {\bf 311} (2006) 503.

\bibitem{Nozieres}
Nozi\`eres P. and Schmitt-Rink S., J. Low Temp. Phys. {\bf 59} (1985) 195.

\bibitem{Randeria}
Randeria M., in {\it Bose-Einstein Condensation},
edited by Griffin A., Snoke D. W., and Stringari S.
(Cambridge University Press, Cambridge) (1995) 355.

\bibitem{boite}
Pricoupenko L., Castin Y., Phys. Rev. A {\bf 69} (2004) 051601(R).

\bibitem{Pandha_bosons}
Cowell S., Heiselberg H., Mazets I.E., Morales J., Pandharipande V.R., 
Pethick C.J., Phys. Rev. Lett. {\bf 88} (2002) 210403.

\bibitem{dim_dim}
Petrov D. S., Salomon C., and Shlyapnikov G. V.,
Phys. Rev. A {\bf 71} (2005) 012708;
Petrov D. S., Salomon C., and Shlyapnikov G. V.,
Phys. Rev. Lett. {\bf 93} (2004) 090404.

\bibitem{LOFF}
Fulde P., Ferrell R. A., Phys. Rev. {\bf 135} (1964) A550;
Larkin A.I., Ovchinnikov Y.N., Sov. Phys. JETP {\bf 20} (1965) 762.

\bibitem{MoraLOFF}
Combescot R., Mora C., Eur. Phys. J. B {\bf 28} (2002) 397 and
references therein.

\bibitem{Ketterle_unba}
Zwierlein M. W., Schirotzek A.,  Schunck C. H., and
Ketterle W., Science {\bf 311} (2006) 492;
Shin Y.,  Zwierlein M. W.,  Schunck C. H., Schirotzek  A., Ketterle  W.,
Phys. Rev. Lett. {\bf 97}, (2006) 030401.

\bibitem{Hulet_unba}
Partridge G. B.,  Li Wenhui,   Liao Y. A.,  Hulet R. G.,  Haque M. and  Stoof H. T. C.,
Phys. Rev. Lett. {\bf 97} (2006) 190407.

\bibitem{Chevy_unba}
There are currently many theoretical papers on this subject.
See references in e.g.\ \cite{Hulet_unba} and in
Chevy F., Phys. Rev. Lett. {\bf 96} (2006) 130401.

\bibitem{Bogoliubov}
Bogoliubov N.~N., Tolmachev V.~V., and Shirkov D.~V., {\em New Method in the
 Theory of Superconductivity} (Academy of the Sciences of the U.S.S.R.,
 Moscow, 1958).

\bibitem{Anderson_rpa}
Anderson P.~W., Phys. Rev. {\bf 112} (1958) 1900.

\bibitem{Zoller}
Jaksch D., Bruder C., Cirac J. I., Gardiner C. W., and Zoller P., Phys. Rev. Lett.
{\bf 81} (1998) 3108.

\bibitem{Bloch}
Greiner M., Mandel O., Esslinger T., H\"ansch T. W., and Bloch I.,
Nature \textbf{415} (2002) 39.

\bibitem{Esslinger_Fermi_lattice}
St\"oferle T., Moritz H., G\"unter K., K\"ohl M., Esslinger T.,
Phys. Rev. Lett. {\bf 96} (2006) 030401.

\bibitem{Ketterle_Fermi_Hubbard}
Chin K.,  Miller D. E., Liu  Y., Stan  C., Setiawan  W., Sanner  C., Xu  K., Ketterle  W.,
Nature {\bf 443} (2006) 961.

\bibitem{BCS}
Bardeen J., Cooper L. N., and Schrieffer J. R.,
Phys. Rev. {\bf 108} (1957) 1175.

\bibitem{ZollerGardiner}
Gardiner C., Zoller P., {\sl Quantum Noise} (Springer, Heidelberg 2004).

\bibitem{Leggett}
Leggett A.~J., Phys. Rev. {\bf 140} (1965) 1869.

\bibitem{Blaizot} 
Blaizot J.-P. and Ripka G., {\it Quantum Theory of Finite
Systems}, The MIT Press (Cambridge, Massachusetts, 1986).

\bibitem{the_curious}
One may wonder if condition Eq.(\ref{eq:cond_ext_cal}) really
implies that $|\Psi_{\mathrm{BCS}}^0\rangle$ is an eigenstate
of $\mathcal{H}$. This is not evident {\it a priori} since the
variation $\delta|\Psi_{\mathrm{BCS}}\rangle$ does not
explore the whole Hilbert space but only the manifold tangent
to the BCS family in $|\Psi_{\mathrm{BCS}}^0\rangle$.
One can show however that the answer is yes:
as sketched below Eq.(\ref{eq:tdbcs}),
$(\mathcal{H}-\mathcal{E}[\Phi_0])|\Psi_{\mathrm{BCS}}^0\rangle$
also belongs to this tangent manifold, so that one may choose
$\delta|\Psi_{\mathrm{BCS}}\rangle$ proportional to it.

\bibitem{cafg}
This expression coincides with Eq.(\ref{eq:gamma_phi}),
since one may check that
$([U]^{-1})_{\vec{k},\vec{r}} = e^{-i\vec{k}\cdot\vec{r}} l^3/(L^{3/2} 
U_k)$.

\bibitem{canonique}
The reasoning was done for a fixed chemical potential and for the excitation
energy of the grand canonical Hamiltonian $H$. What is the corresponding
excitation energy of the canonical Hamiltonian for a fixed number $N$
of particles?
We have a canonical energy in the ground state $E_{\rm can}^g=
\langle H\rangle^g(\mu)+ \mu N$, and in the excited state
$E_{\rm can}^e= \langle H\rangle^e(\mu_e) + \mu_e N$.
Here the chemical potentials $\mu$ and $\mu_e$ differ by $O(1/L^3)$
for a few Bogoliubov excitations, so that one may expand to first
order in $\delta\mu=\mu_e-\mu$. This leads to
$E_{\rm can}^e -E_{\rm can}^g = \langle H\rangle^e(\mu)-
\langle H\rangle^g(\mu) + \delta\mu \left[N+\frac{d}{d\mu} 
\langle H\rangle^g(\mu)\right] + O(1/L^3)$.
Now the quantity in between brackets vanishes, a standard thermodynamic 
result. One can also use $(d/d\mu)
\left[\langle H\rangle^g_{\lambda=\Delta(\mu)}\right]=
\left(\partial_\mu\langle H\rangle^g_{\lambda}\right)_{\lambda=\Delta(\mu)}$
and calculate the partial derivative of Eq.(\ref{eq:hlg}) 
with respect to $\mu$ for a fixed $\lambda$. Setting $\lambda=\Delta$
in the resulting expression leads to the standard thermodynamic result.

\bibitem{whygap}
P. G. de Gennes, in {\it Superconductivity of Metals and Alloys},
section 4.3, Addison-Wesley Publishing Company (Reading, Massachusetts, 
fifth printing 1996).

\bibitem{Randeria2}
Engelbrecht J. R., Randeria M., and S\'a de Melo C., 
Phys. Rev. B {\bf 55} (1997) 15153.

\bibitem{if_neutral}
We consider here the case of neutral particles. For charged
particles the situation is different, because of the long range
Coulomb interaction, see
P. C. Martin, in {\em Superconductivity}, edited by R.~D. Parks
(Dekker, New York, 1969) Vol.~1.

\bibitem{Griffin_rpa}
C{\^o}t{\'e} R. and Griffin A., Phys. Rev. B {\bf 48}  (1993) 10404.

\bibitem{Bruun_rpa}
Bruun G. M. and Mottelson B. R., Phys. Rev. Lett. {\bf 87} (2001) 270403.

\bibitem{Minguzzi_rpa}
Minguzzi A., Ferrari G., Castin Y., Eur. Phys. J. D {\bf 17} (2001) 49.

\bibitem{Zwerger_rpa}
B\"uchler H. P., Zoller P., and Zwerger W., 
Phys. Rev. Lett. 93 (2004) 080401.

\bibitem{Combescot_rpa}
Combescot R., Kagan M. Yu., Stringari S., cond-mat/0607493.

\bibitem{Bruun_le_fait} 
The RPA approach can be applied in an isotropic harmonic
trap, with some non negligible numerical effort, see 
\cite{Bruun_rpa}.

\bibitem{Baranov}
Baranov M.~A.  and Petrov D.~S., Phys. Rev. A {\bf 62} (2000) 041601(R).

\bibitem{Anna}
Minguzzi A. and Tosi M.~P.,  Phys. Rev. A {\bf 63} (2001) 023609.

\bibitem{Stringari_tof}
Menotti C.,  Pedri P., Stringari S.,  Phys. Rev. Lett. {\bf 89} (2002) 250402.

\bibitem{Stringari2}
Cozzini M., Stringari S., Phys. Rev. Lett. {\bf 91} (2003) 070401.

\bibitem{Tonini}
Tonini G., Werner F., Castin Y., Eur. Phys. J. D {\bf 39}  (2006) 283.

\bibitem{TCbcs}
Gor'kov L.~P. and Melik-Barkhudarov T.~K., Sov. Phys. JETP {\bf
13} (1961)  1018.

\bibitem{Ketterle_vortices}
Zwierlein M. W., Abo-Shaeer J. R. , Schirotzek A. , Schunck C.H.,
Ketterle W., Nature {\bf 435} (2005) 1047.

\bibitem{adiab}
Messiah A., in {\sl Quantum Mechanics}, vol.II (North Holland, 1961);
Migdal A.B., in {\sl Qualitative methods in Quantum Theory},
(W.A. Benjamin, Massachusetts, 1977) 155.

\bibitem{Grynberg_bragg}
D. R. Meacher, D. Boiron, H. Metcalf, C. Salomon, and G. Grynberg,
Phys. Rev. A {\bf 50} (1994) R1992 - R1994 .

\bibitem{Ketterle_bragg}
Stamper-Kurn D. M., Chikkatur A.~P., G\"orlitz A., Inouye S.,
Gupta S., Pritchard D.~E. and Ketterle W., Phys. Rev. Lett. {\bf 83}
(1999) 2876.

\end{thebibliography}
\end{document}